\documentclass[nohyper,12pt,letterpaper]{JHEP3}
\pdfoutput=1
\usepackage{graphicx}
\usepackage{amsfonts,amssymb,calc}

\newcommand{\eqn}[1]{(\ref{#1})}
\newcommand{\be}{\begin{equation}}
\newcommand{\ee}{\end{equation}}
\newcommand{\ben}{\begin{displaymath}}
\newcommand{\een}{\end{displaymath}}
\newcommand{\bea}{\begin{eqnarray}}
\newcommand{\eea}{\end{eqnarray}}
\newcommand{\bean}{\begin{eqnarray*}}
\newcommand{\eean}{\end{eqnarray*}}

\newcommand{\ads}[1]{\mbox{${AdS}_{#1}$}}
\newcommand{\adss}[2]{\mbox{$AdS_{#1}\times {S}^{#2}$}}

\newcommand{\bra}[1]{\mbox{$\langle #1 |$}}
\newcommand{\ket}[1]{\mbox{$| #1 \rangle$}}
\newcommand{\braket}[2]{\mbox{$\langle #1  | #2 \rangle$}}

\newcommand{\eg}{{\it e.g.}}
\newcommand{\ie}{{\it i.e.}}

\newcommand{\tr}{\mbox{Tr}}

\newcommand{\commentout}[1]{}

\newcommand{\beq}{\begin{equation}}
\newcommand{\eeq}{\end{equation}}
\newcommand{\beqr}{\begin{displaymath}}
\newcommand{\eeqr}{\end{displaymath}}
\newcommand{\beqa}{\begin{eqnarray}}
\newcommand{\eeqa}{\end{eqnarray}}
\newcommand{\beqar}{\begin{eqnarray*}}
\newcommand{\eeqar}{\end{eqnarray*}}
\newcommand{\cA}{{\cal A}}

\newcommand{\cN}{{\cal N}}
\newcommand{\cD}{{\cal D}}
\newcommand{\cO}{{\cal O}}
\newcommand{\cP}{{\cal P}}

\newcommand{\non}{\nonumber}

\newcommand{\half}{\ensuremath{\frac{1}{2}}}

\newcommand{\N}[1]{\ensuremath{\cN=#1}}

\newcommand{\ps}{\ensuremath{\partial_\sigma}}

\newcommand{\ei}{\ensuremath{\varepsilon_i}}
\newcommand{\sz}{\ensuremath{\sigma_0}}
\newcommand{\NC}{\ensuremath{\mathrm{N}}}

\newcommand{\sst}{\ensuremath{\sin\frac{\sz}{2}}}
\newcommand{\cst}{\ensuremath{\cos\frac{\sz}{2}}}
\newcommand{\hsz}{\ensuremath{\frac{\sz}{2}}}

\newcommand{\Pp}{\ensuremath{\hat{P}}}
\newcommand{\Id}{\ensuremath{\mathbb{I}}}

\newcommand{\pone}{{\tt (I)}}
\newcommand{\ttheta}{\ensuremath{\tilde{\theta}}}
\newcommand{\tlambda}{\ensuremath{\tilde{\lambda}}}
\newcommand{\bbQ}{\ensuremath{\mathbb{Q}}}
\newcommand{\bbH}{\ensuremath{\mathbb{H}}}
\newcommand{\bbR}{\ensuremath{\mathbb{R}}}

\newcommand{\sg}[1]{\ensuremath{\mbox{sg}(#1)}}
\newcommand{\NO}[1]{\ensuremath{\mathbf{:}#1\mathbf{:}}}
\newcommand{\Ima}{\ensuremath{\mathrm{Im}}}
\newlength{\sht}
\newlength{\shtp}
\newlength{\shtm}
\newcommand{\sump}[2]{\ensuremath{
   \settowidth{\shtp}{$\displaystyle \sum_{#1}^{#2}$}
   \settowidth{\shtm}{$\displaystyle \sum$}   
   \setlength{\sht}{(\shtp-\shtm)/2}
   \makebox[0pt][l]{$\displaystyle \sum_{#1}^{#2}$}
   \hspace{\sht} 
   {{{{\phantom{\sum}}}}'}  
                                 }}
\newcommand{\binomial}[2]{\ensuremath{\left(\begin{array}{c}#1 \\ #2  \end{array}\right) }}
\newlength{\qw}
\newlength{\qwt}
\newcommand{\sq}{\ensuremath{
       \settowidth{\qw}{$q$}
       \setlength{\qwt}{-\qw}
       q\hspace{\qwt}/
                           }}
\newlength{\ha}
\newlength{\htwo}

\newcommand{\kets}[4]{
 \settoheight{\ha}{${\scriptstyle | KKII \rangle}$}
 \setlength{\htwo}{\ha/4} 
 \raisebox{\htwo}{${\scriptstyle | #1#2#4#3 \rangle}$}
 }

\newcommand{\cotan}{\ensuremath{\mbox{cotan}}}
\newcommand{\eone}{\ensuremath{\epsilon_1}}
\newcommand{\etwo}{\ensuremath{\epsilon_2}}
\newcommand{\thon}{\theta_1}
\newcommand{\thtw}{\theta_2}
\newcommand{\thth}{\theta_4} 
\newcommand{\thfo}{\theta_3}
\newcommand{\uno}{\thon}
\newcommand{\dos}{\thtw}
\newcommand{\tres}{\thth}
\newcommand{\cuatro}{\thfo}
\newcommand{\aab}[2]{\ensuremath{#1^4\,{#2}\sq{#2}}}
\newcommand{\aad}[3]{\ensuremath{#1^4\, \epsilon#2^2#3^2}}
\newcommand{\aabc}[3]{\ensuremath{{#1}\sq {#1}\,\epsilon #2^2#3^2}}
\newcommand{\aaf}[2]{\ensuremath{{#1}\sq{#1}\,{#2}\sq{#2}}}
\newcommand{\aag}[4]{\ensuremath{\epsilon#1^2#2^2\,\epsilon#3^2#4^2}}
\newcommand{\aah}[4]{\ensuremath{{#1}\sq{#1}\,{#2}\sq{#2}\,\epsilon#3^2#4^2}}
\newcommand{\aaj}[4]{\ensuremath{#1^4#2^4\left(#3\sq#3+#4\sq#4\right)}}
\newcommand{\aak}[4]{\ensuremath{#1^4\,#2\sq#2\,\epsilon#3^2#4^2 }}
\newcommand{\aal}[3]{\ensuremath{#1^4#2^4#3^4 }}
\newcommand{\aam}[4]{\ensuremath{#1^4#2^4\,\epsilon#3^2#4^2 }}
\newcommand{\pola}[4]{\ensuremath{\epsilon^{(1)}_{#1#4} \epsilon^{(2)}_{#2#3} }} 
\newcommand{\polab}[4]{\ensuremath{\epsilon^{(1)}_{#1#3}\epsilon^{(2)}_{#2#4}}}
\newcommand{\vl}{\ket{0}_\lambda}

\title{\LARGE Summing planar diagrams}

\author{Martin Kruczenski \\
        Department of Physics, Purdue University, \\
        525 Northwestern Avenue, W. Lafayette, IN 47907-2036.

E-mail: \email{markru@purdue.edu}}

\abstract{We consider the sum of planar diagrams for open strings propagating on $N$\, D3-branes and show that it can be recast as
the propagation of a closed string with a Hamiltonian $H=H_0-g_s N\, \Pp$ where $H_0$ is the free Hamiltonian and $\Pp$ is the
hole or loop insertion operator. We compute explicitly $\Pp$ and study its properties. When the distance $y$ to the D3-branes
is much larger than the string length, $y\gg \sqrt{\alpha'}$, small holes dominate and $H$ becomes a supersymmetric Hamiltonian 
describing the propagation of a closed string in the full D3-brane supergravity background in a particular gauge that
we call $\sigma$-gauge. At strong coupling, $g_sN\gg1$, there is a region $1\ll y\ll (g_sN)^{\frac{1}{4}}$ where $H$ is a 
supersymmetric Hamiltonian describing the propagation of closed strings in \adss{5}{5}. We emphasize that both results
follow from the open string planar diagrams without any reference to the existence of a D3-brane supergravity background. 
A by-product of our analysis is a closed form for the scattering of a generic closed string state from a D3-brane.
 Finally, we briefly discuss how this method could be applied to a field theory and describe a way to rewrite the planar Feynman 
diagrams as the propagation of a string with a non-local Hamiltonian by identifying the shape of the string with the 
trajectory of the particle.

}

\keywords{string theory, QCD, light-cone frame}

\begin{document}

\section{Introduction}
\label{intro}

 Some years ago, 't Hooft proposed \cite{largeN} the large-N limit as a promising approach to understanding 
the strong coupling regime of gauge theories. In particular, he argued that, when considered in light cone frame, 
a gauge theory looks similar to a string theory and that, by summing the planar diagrams, one could obtain the particular 
effective string theory that describes the strong coupling limit of the gauge theory. The idea, although beautiful and
potentially very useful, was hampered by the fact that summing the planar diagrams appears a difficult task. 
 The situation somewhat changed when Polchinski \cite{Polchinski:1995mt} introduced D-branes. In the low energy limit, 
open strings attached to a D-brane are described by a gauge theory. In particular in the case of a $D3$-brane, the 
gauge theory is \N{4} SYM in 3+1 dimensions. The gauge group is $SU(N)$ where $N$ is the number of D-branes. In the limit
when $N$ is large, the stack of $N$ D-branes becomes very heavy deforming the space around it. In this limit, the D-branes can
be described by a supergravity solution where closed strings propagate giving a novel and interesting interpretation to 
the large N limit. This was understood by Maldacena who proposed the AdS/CFT correspondence \cite{malda}, a precise relation between 
a large N gauge theory, namely \N{4} SYM, and a string theory, IIB on \adss{5}{5}, the near horizon limit of the D3-brane
supergravity solution. This allows to compute various field theory quantities in the strong coupling limit by using the string 
description \cite{WGKP}. Thus, the idea of 't Hooft is realized in the sense that the large-N limit gives rise to a string theory.
It further suggests that it might be possible to realize also the other part, namely, that the planar diagrams can be summed up and 
the string theory dual extracted from the result. 
 In this paper we analyze this possibility elaborating on our previous work \cite{planar}. 

 In \cite{planar} which from now on we call \pone, we considered the one loop amplitude describing the interaction between a stack of $N$
D-branes and a probe brane (see figure \ref{DDint}). When computing the planar corrections in light cone gauge, we found that they
were described by the propagation of a closed string with a Hamiltonian equal to $H=H_0-g_sN \Pp$ where $\Pp$ is the operator that 
describes the insertion of a hole in the world-sheet (or of a loop from a field theory perspective). 
 The operator $\Pp$ was explicitly computed in the bosonic sector and described the scattering of an arbitrary closed string mode
from a D-brane. In the approximation that the holes are small the corrections describe the propagation of a closed string in a
modified supergravity background. Although one should expect this background to be the D-brane supergravity solution, this was not
the case, extra terms appeared in the Hamiltonian. We attributed this to the fact that we only considered the bosonic sector and
expected those extra term to cancel in a full supersymmetric computation.

\FIGURE{
\includegraphics[height=6cm]{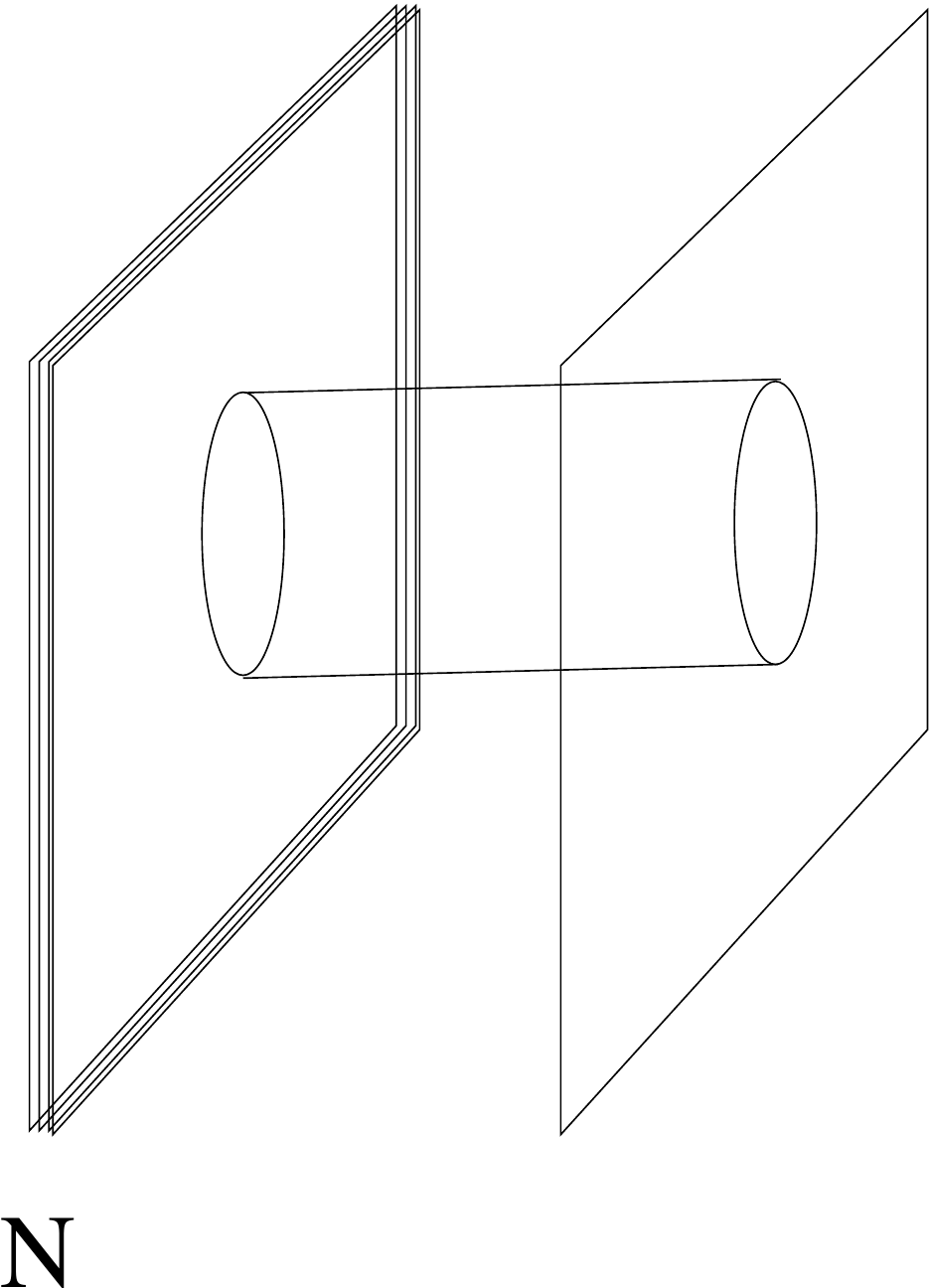}
\caption{The interaction between a stack of $N$ D-branes and a probe brane is given, at lowest order, by a one-loop open string diagram,
or equivalently by a single closed string interchange.}
\label{DDint}
}

 In the present paper we consider D3-branes and find precisely that. Namely, in the limit of small holes the Hamiltonian $H$ 
describes strings propagating in the \emph{full} D3-brane background.  

 Finally let us remark that the emphasis of this paper is in understanding the sum of planar diagrams without any prejudice about 
the result. In particular we do not need that the sum is given in terms of a string theory.  The
Hamiltonian we obtain in the closed string side is non-local and therefore cannot be interpreted as a string Hamiltonian.
This, however would not prevent us from studying planar diagrams since we can study the properties of such Hamiltonian, \eg\ spectrum, 
ground state etc. to derive properties of the open string theory, or eventually field theory, whose planar diagrams we are summing.

 The subject of gauge theories in light cone gauge is a well studied one. For example see the review article \cite{Brodsky:1997de}.
More recent is the work in \cite{Thorn:2005ak} where loop calculations are discussed and \cite{Belitsky:2004sc} where the formulation 
of \N{4} in light cone gauge \cite{Brink:1982pd} is used to compute conformal dimensions of various operators.

 String theory in light cone gauge is also very well studied \cite{GSW}. Earlier work on the subject including the relation to the
large-N limit can be found for example in \cite{Thorn1},  \cite{Orland:1986ve}.

In the case of the superstring light cone gauge was an important method used to construct the theory \cite{Green:1983hw, Green:1984fu}. 
For strings in \adss{5}{5} the light-cone gauge action was described in \cite{Metsaev:2000yu}. In the pp-wave approximation,  
light-cone gauge also was used recently to compare amplitudes with the field theory result\cite{pp-wave}. 
 
 The idea of defining a ``hole'' operator was also already considered for example in \cite{Polyakov:2001af}. A related idea is 
also discussed in \cite{Fischler:1987gz}. There, small holes are studied in the case of the bosonic string. Presumably their
conclusions would be different if the calculation is done for a D3-brane hole on a type IIB world-sheet. In the case of the bosonic string,
an operator similar to the slit insertion operator we discuss here was already computed in \cite{GW} for the case of all Dirichlet 
boundary conditions. 

These previous works indeed suggest that combining the light-cone frame and the introduction of a ``hole'' operator should be useful.

 It should be noted that recently, other approaches to the problem were discussed. In \cite{Bardakci:2001cn,Clark:2003wk} a world-sheet 
description of a gauge theory is derived. The first \footnote{I am grateful to C. Thorn for an explanation of the work in \cite{Bardakci:2001cn}.}, 
finds a representation in terms of a spin system which followed by a mean-field approach gives a world-sheet action and in the second representing 
a free field theory in terms of strings is discussed. In the context of the AdS/CFT correspondence a relation between the Schwinger parametrization 
of Feynman diagrams and particles propagating in \ads{5} space was discussed in \cite{Gopakumar:2003ns}. A more detailed analysis of this proposal 
including various checks can be found in~\cite{Aharony:2006th}. The idea of deriving the AdS/CFT duality using the NSR string (as opposed to 
the GS we use here) is discussed in \cite{Berenstein:1999jq}.  
 
 It is interesting to note that, in the context of topological strings, it was recently observed~\cite{Walcher:2007tp} that the 
open string partition function follows from the closed string partition function by shifting the closed string moduli by terms linear
in the 't Hooft coupling. The Feynman diagram expansion for (topological) open string amplitudes follows in a similar way. It would be 
interesting to understand further if this is related or not to the large-N duality we propose here for ordinary superstrings. Namely, 
that the Feynman diagram expansion of the open string follows from shifting the closed string Hamiltonian by an operator linear in the
't Hooft coupling.

 This paper is organized as follows: in section \ref{planar} we review the main ideas of the previous paper \cite{planar}. 
 In section \ref{PS} we compute the slit insertion operator and study the divergencies of different fields as they approach 
the insertion of a slit. These divergences are the usual divergences that any field has in the presence of an operator insertion
and which determine the operator product expansion between operators. As a result, we find that $\Pp_S$ is not supersymmetric. 
 Defining the correct operator implies multiplying $\Pp_S$ by certain operator insertions at the ends of the slit. In section \ref{Pcomp} 
we compute those insertions and find the final form of the hole insertion operator $\Pp$. This operator $\Pp$ describes the scattering 
of closed strings from a $D3$-brane. When reduced to the massless modes, it reproduces known results providing a useful check as we show 
in section \ref{Dscattering}. In \pone\ it was observed that important information on the background was contained in the limit of $\Pp$ 
for small holes. We compute this limit in section \ref{small holes} and show that it reproduces the propagation of a closed string in the 
full $D3$-brane supergravity background. In section \ref{field theory} we briefly discuss ideas related to the application of the present method to 
field theory planar diagrams. 
Finally we give our conclusions in section \ref{conclusions}. Some calculations and formulas are collected in the appendices.
 In particular a simpler derivation of the Neumann coefficients is described.

\section{Planar diagrams in light cone gauge}
\label{planar}
 In this section we briefly review the results and ideas of paper \pone, \ie\ \cite{planar}. If we consider the diagram in fig.\ref{DDint},
its value can be computed as the regulated sum of the zero point energy of all physical open string oscillators. If, for simplicity, we consider 
a bosonic string and all branes to be $p$-dimensional, the diagram of fig.\ref{DDint} reduces to:
\beq
 Z = \int d^p k \sum_{N^i_n=0..\infty} \omega_k, \ \ \mbox{with} \ \ \omega_k=\sqrt{k^2+m^2}, \ \ \ m^2=\sum_{n\ge1,i}N^i_n-a+L^2\ ,
\eeq
 where $k$ represents the momenta parallel to the brane, $N^i_n$ are the occupation numbers of the oscillators, $L$ is the distance 
between the branes and $a$ is the usual normal ordering constant of the bosonic string ($a=1$). The sum is divergent, to give it a meaning 
we start by doing the following formal manipulations: 
\beq
 Z = \int d^p k \sum_{N^i_n=0}^{\infty} \omega_k \sim 
      \int d^p k \sum_{N^i_n=0}^{\infty} \int_0^{\infty} \frac{d\ell}{\ell^{\frac{3}{2}}} e^{-\ell(k^2+m^2)} \sim
       \int d^{p-1} k_\perp \sum_{N^i_n=0}^{\infty}\int_0^\infty dp^+ e^{-\frac{1}{p^+}(k_\perp^2+m^2)}\ ,
\eeq
 where in the last step we integrated out a spacial coordinate to write the result in a form suggestive of light-cone gauge.  
In fact we can now re-write $Z$ as:
\beq
 Z = \int_0^\infty dp^+ \hat{Z} = \int_0^\infty dp^+ \tr e^{-\beta H_{l.c.}}\ ,
\label{opst}
\eeq
where $\beta=2\pi\alpha'$ is a constant and $\hat{Z}=\tr e^{-\beta H_{l.c.}}$ with $H_{l.c.}$ the light cone Hamiltonian:
\beq
 H_{l.c.} = \frac{1}{4p^+}\left[p_\perp^2+\sum_{n\ge1,i}N^i_n n -\frac{1}{\alpha'} + \frac{L^2}{4\pi^2\alpha'{}^2}\right].
\eeq
 The trace in (\ref{opst}) is over all oscillator states and parallel momenta. In this form the divergence is in the integral over $p^+$
in the limit $p^+\rightarrow\infty$ but now can be physically understood as due to the closed string tachyon propagating along the closed
string channel. For that reason we concentrate on the partition function $\hat{Z}$ which can be computed obtaining the standard result. 
What we are more interested here is that we can rewrite $\hat{Z}$ in a path integral form:
\beq
 \hat{Z} = \int \cD X_{\perp} e^{-\int_{0}^{p^+}d\sigma \int_0^\beta d\tau 
        \left[\left(\partial_\tau X_\perp\right)^2+\left(\partial_\sigma X_\perp\right)^2\right]}.
\eeq   
 We can now interchange $\sigma$ and $\tau$ since they enter equally in the calculation and rewrite the path integral as a computation in the
closed string channel:
\beq
 \hat{Z} =   \bra{B_f}e^{-H_{c.s.}\tau}\ket{B_i}\ ,
\eeq
 where the time of propagation is $\tau=4\alpha'p^+$, namely the length of the open string in the previous calculation, and $\ket{B_{i,f}}$ are
the boundary states corresponding to the branes in the diagram. These states are well known, a good review on how to construct them  
is \cite{DiVecchia:1999rh}. Finally the closed string Hamiltonian is given by:
\beq
 H_{c.s.} = \frac{1}{\sqrt{\alpha'}}\left[\half\alpha'p^2+\sum_{n\ge1,i}n\left(N^{Ii}_n+N^{IIi}_n\right)-2\right]\ ,
\eeq
where $N^{Ii}_n$ ,$N^{IIi}_n$ are the occupation numbers of the left and right moving oscillators. It is also a well-known result that both 
calculations of $\hat{Z}$ coincide \cite{Polchinski}.

\FIGURE{
\includegraphics[height=6cm]{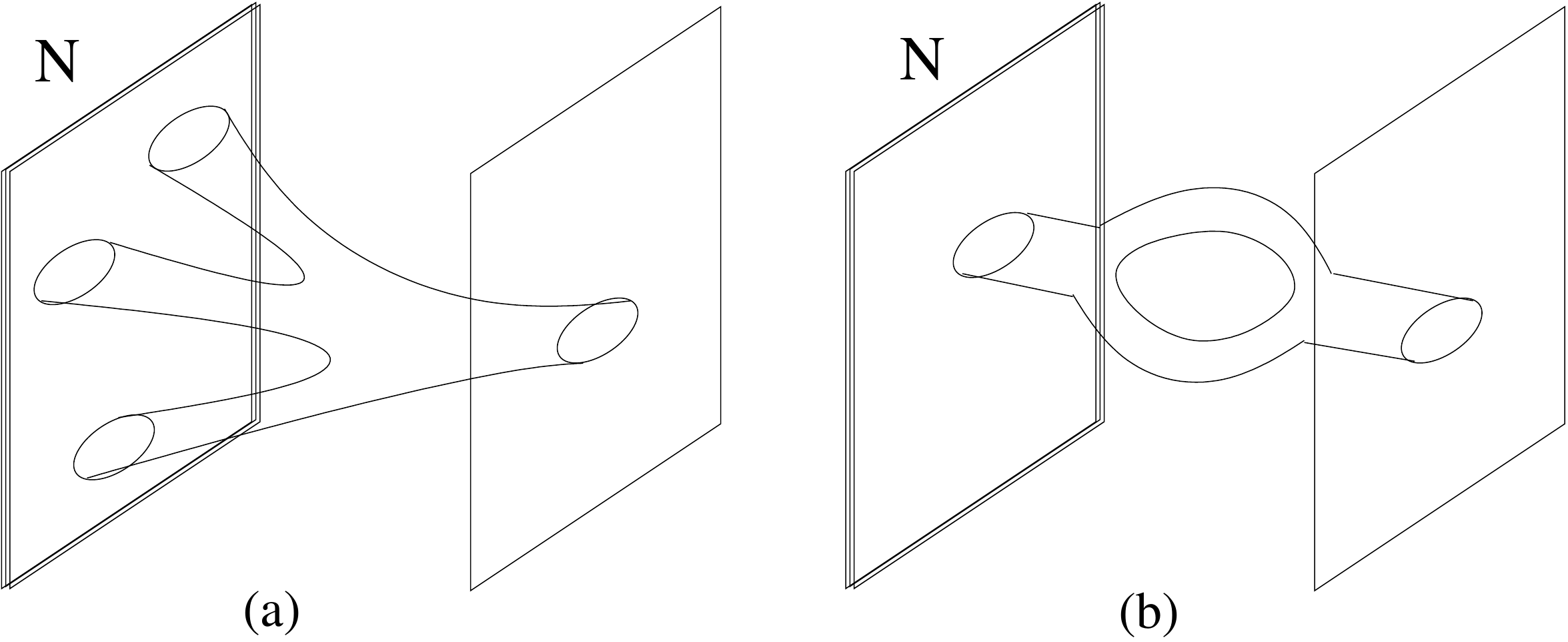}
\caption{Corrections to the diagram of fig.\ref{DDint}. In (a) we depict typical planar corrections and in (b) non-planar ones. 
In the limit $N\rightarrow\infty$ the first ones dominate.}
\label{DDintb}
}

 The purpose of \pone\ was to sum the planar corrections that are obtained from diagrams of the type shown in fig.\ref{DDintb}a 
while discarding those such as the one in fig.\ref{DDintb}b. From the point of view of the closed string we are including all tree level 
corrections including those of the massive modes. From the point of view of open strings, the interaction we should take into account
is the one that splits (or joins) strings as the one depicted in fig.\ref{3vertex}. Notice that the total length of the string is conserved since
it is given by $p^+$. With such vertex we can construct diagrams of the type depicted in fig.\ref{plnpl}a or those as in fig.\ref{plnpl}b. 

\FIGURE{
\includegraphics[height=5cm]{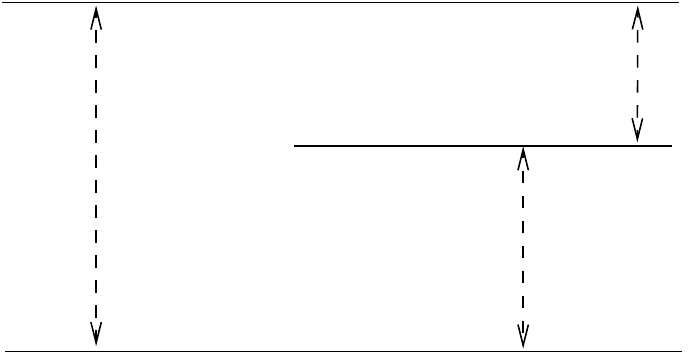}
\caption{The interaction between open strings is given by a three vertex where two strings join or one string splits in two \cite{GSW}. The 
total length of the strings is proportional to $p^+$ and therefore conserved.}
\label{3vertex}
}

 In the planar approximation one can see that those in fig.\ref{plnpl}a dominate. Again, we can now compute, instead, a path integral over such 
world-sheet with appropriate boundary conditions on the slits. The total partition function is
\beq
 \hat{Z} = \sum_{n=0}^{\infty} \frac{(g_sN)^n}{n!} \int \prod_{i=1}^n d\sigma_i^Ld\sigma_i^R d\tau_i \int \cD X_\perp 
  e^{-\int d\sigma d\tau\left[\dot{X}_\perp^2+X_\perp'{}^2\right]} \ ,
\eeq
where the hat indicates that we still have to do the integral on $p^+$. We also have to integrate over all positions of the slits, three
parameters per each. We divide by $n!$ since the slits are identical or, equivalently, we can integrate over the range $0<\tau_1<\ldots<\tau_n<\tau$.  

\FIGURE{
\includegraphics[height=5cm]{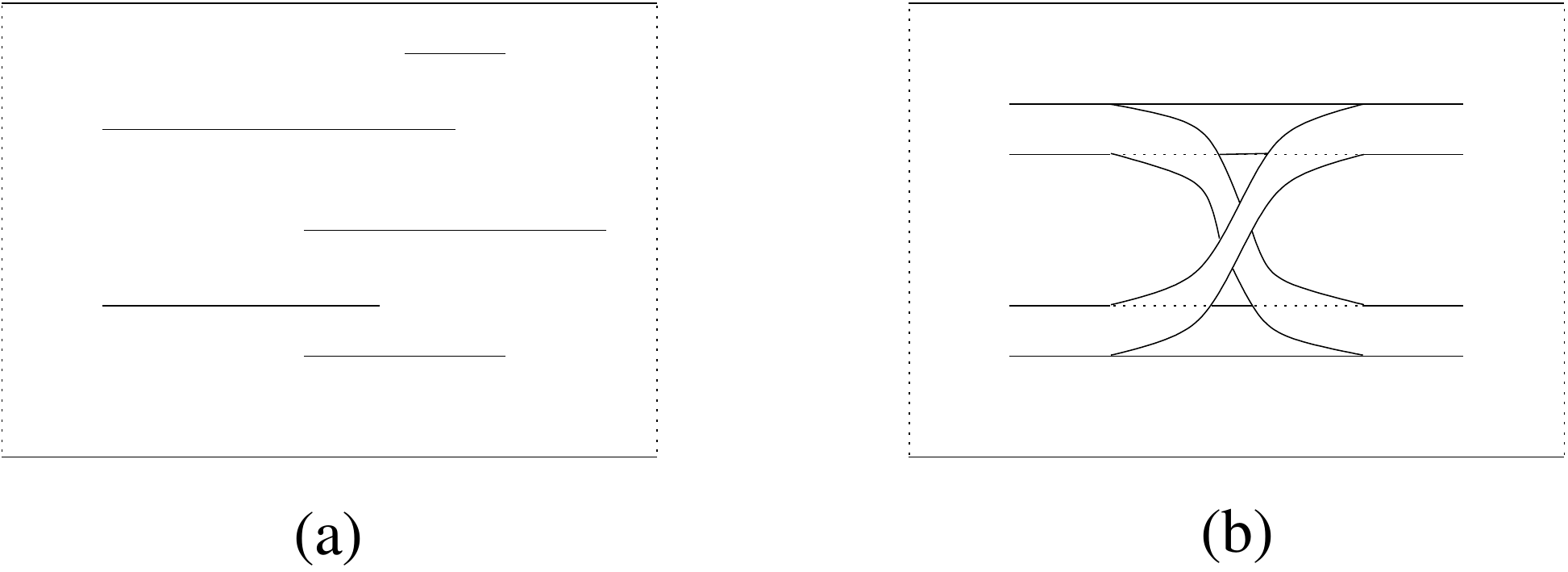}
\caption{Corrections to the diagram of fig.\ref{DDint} as seen in the open string channel. Again we have planar (a) and non-planar (b) 
contributions. }
\label{plnpl}
}

 Again, we can interchange $\sigma$ and $\tau$ to write the diagram in terms of the propagation of a closed
string as shown in fig.\ref{slitP}. It is obvious from the figure that, in this channel, we still have only one closed string. 
In this channel it is convenient to define an operator $P(\sigma_1^L,\sigma_1^R)$ that propagates the closed string from an instant 
before inserting a slit to an instant right after, as depicted in fig.\ref{slitP}. This operator depends on the positions 
$\sigma^L$ and $\sigma^R$ of the slit but not on the time $\tau$ at which it acts. With this operator, we can rewrite $\hat{Z}$ as:
\beqa
 \hat{Z} &=& \sum_{n=0}^\infty(g_sN)^n  \int_{0<\tau_1<\ldots<\tau_n<\tau} \prod_{i=1}^n d\sigma_i^Ld\sigma_i^R d\tau_i \bra{B_f} e^{-H_0(\tau-\tau_n)}
\ldots P(\sigma_2^L,\sigma_2^R)e^{-H_0(\tau_2-\tau_1)}P(\sigma_1^L,\sigma_1^R)e^{-H_0\tau_1}\ket{B_i} \nonumber\\
         &=& \sum_{n=0}^\infty(g_sN)^n  \int_{0<\tau_1<\ldots<\tau_n<\tau} \prod_{i=1}^n d\tau_i \bra{B_f} e^{-H_0(\tau-\tau_n)}
\ldots \Pp e^{-H_0(\tau_2-\tau_1)}\Pp e^{-H_0\tau_1}\ket{B_i}\ ,
\eeqa
where we defined $\Pp=\int d\sigma^Ld\sigma^R P(\sigma^R,\sigma^L)$. If we further define 
\beq
\Pp(\tau) = e^{H_0\tau} \Pp e^{-H_0\tau} \ ,
\eeq
we get
\beqa
 \hat{Z} &=& \sum_{n=0}^\infty(g_sN)^n  \int_{0<\tau_1<\ldots<\tau_n<\tau} \prod_{i=1}^n d\tau_i \bra{B_f} \Pp(\tau_n) \ldots \Pp(\tau_1)\ket{B_i} \\
         &=& _I\bra{B_f} \hat{T} e^{g_sN\int_{0}^\tau \Pp(\tau) d\tau} \ket{B_i}_I \\
         &=& \bra{B_f} e^{-(H_0-\lambda\Pp)\tau}\ket{B_i}\ ,
\eeqa
where $\lambda=g_sN$, the subindex $I$ indicates states in the interaction representation and $\hat{T}$ indicates the time ordered product.
 The last equality is the standard Dyson representation of time dependent perturbation theory if we want to expand the last line
in powers of $\lambda$. Thus, we obtain a closed string Hamiltonian $H=H_0-\lambda\Pp$ which, by definition, is such that
expanding the corresponding evolution operator $U=e^{-H\tau}$ in powers of $\lambda$ recreates, order by order, the perturbative expansion 
in the open string channel. It is clearly important to study such operator and the rest of the paper is devoted to computing $\Pp$ for the 
superstring and analyzing the result. 

\FIGURE{
\includegraphics[height=6cm]{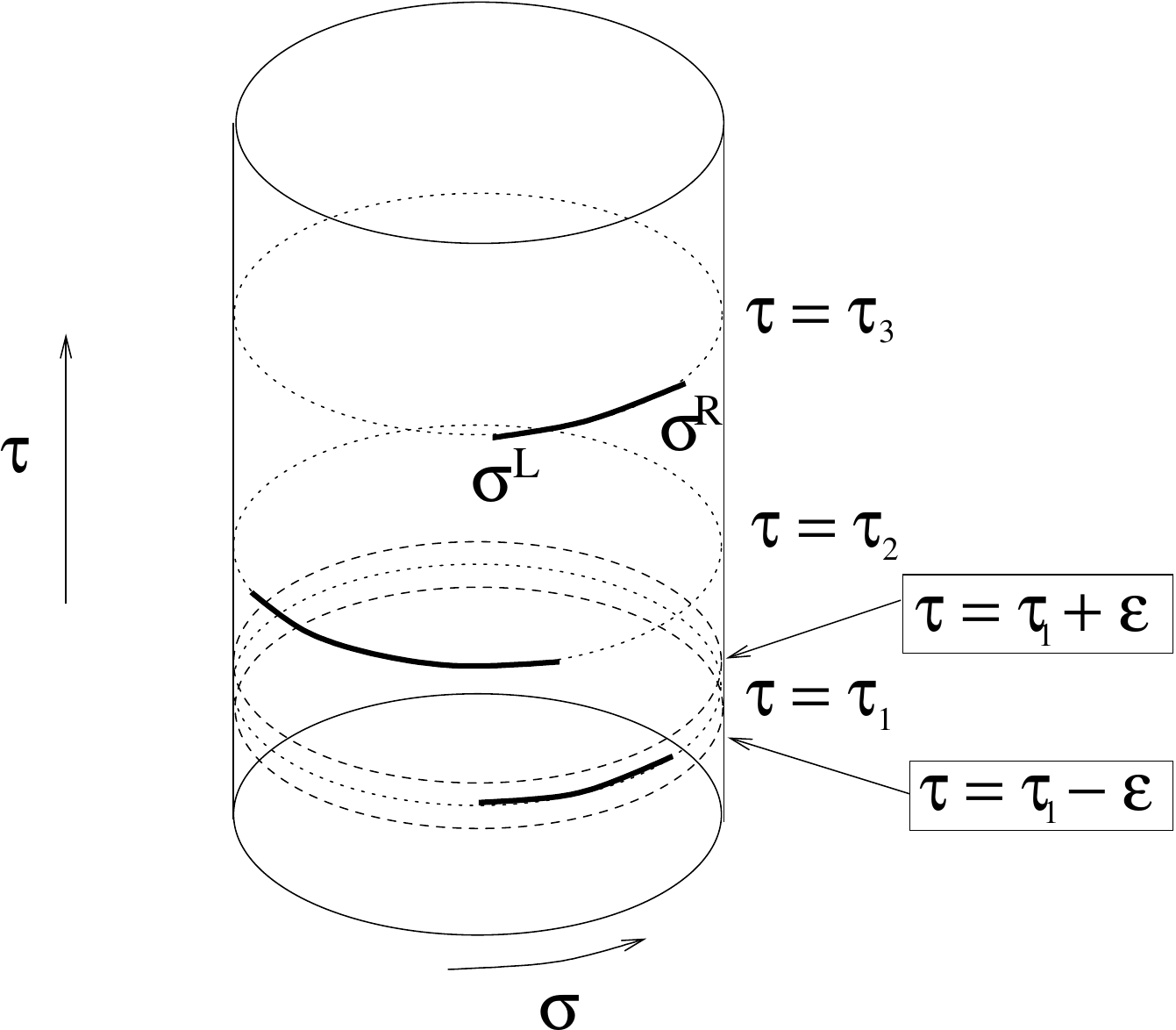}
\caption{In the closed string channel we compute the diagram by defining an operator $\Pp$ that propagates the string across the slit from 
$\tau=\tau_1-\epsilon$ to $\tau=\tau_1+\epsilon$.}
\label{slitP}
}

One caveat is that, if part of the supersymmetry is preserved, the partition function is zero. In the path integral 
method this follows form the fact that there is a fermionic zero mode and that $\int d\theta\,1=0$. In the open string approach follows
from the fact that there are the same number of fermionic and bosonic states at each level and we compute 
\beq
Z = \int_0^\infty dp^+\ \tr\left((-)^Fe^{-\beta H_{l.c.}}\right)\ .
\eeq
 We need $(-)^F$ where $F$ is the fermionic number because the fermions contribute with a minus sign to the zero point energy. From the closed
string point of view we get a zero because both boundary states, initial and final, satisfy the same condition for some given fermionic zero mode.
 If we call the mode $c$ then we should have $c\ket{B_i}=0$ and $\bra{B_f}c=0$ meaning that in $\ket{B_i}$ the mode is empty and in $\ket{B_f}$ it 
is full. Therefore $\bra{B_f} B_i\rangle=0$. For that reason we should take an initial state that breaks supersymmetry. For example the  boundary
state of a D3-brane moving at constant velocity along certain coordinate $Y^I$. In any case, at this stage we are not really concerned on the 
initial and final states since we are interested in the Hamiltonian $H$ that arises and not in actually evaluating the matrix 
element $\bra{B_f}e^{-H\tau}\ket{B_i}$.

\section{The slit operator $\Pp_S$}
\label{PS}

 In this section we compute the slit operator $\Pp_S$ and study its properties. At the end of the section we find that, in the 
case of the superstring, the slit operator is not supersymmetric. The correct operator $\Pp$ is actually a slit with operator insertions 
near the ends of the segment. In the next section we do such computation, which parallels the open string calculations in \cite{GSW}. 

\subsection{Computation of $\Pp_S$}

 In fact, the slit operator for the superstring was computed in \pone. It was written as a two vertex state, namely as a state in the 
tensor product of the space of states of the initial and final strings. Before stating the result let us introduce some notation.
 We consider type IIB superstrings in the $U(1)\times SU(4)$ formalism \cite{GSW}. The spacial coordinates are divided into parallel to the 
brane $X^{\pm}$, $X^{a=1,2}$, and perpendicular $Y^{I=1\ldots 6}$. The coordinates parallel to the brane are divided into light-cone 
coordinates $X^{\pm}$ and transverse. For the transverse ones we sometimes use the redefinition
\beq
 X^R = \frac{1}{\sqrt{2}} \left(X^1+iX^2\right), \ \ \  X^L = \frac{1}{\sqrt{2}} \left(X^1-iX^2\right).
\eeq
 The fermionic coordinates are divided into left movers $\theta^A$, $\lambda_A$, and right movers $\ttheta^A$, $\tlambda_A$. The upper index
$A$ transforms in the fundamental of $SU(4)$ and the lower index $A$ in the antifundamental. At equal time, the anticommutation relations are
\beq
 \{\theta^A(\sigma),\lambda_B(\sigma')\}=\delta^A_B\delta(\sigma-\sigma'), \ \ \ \ \  
 \{\ttheta^A(\sigma),\tlambda_B(\sigma')\}=\delta^A_B\delta(\sigma-\sigma'),
\eeq
 the coordinates are expanded in modes according to
\beqa
X_r^i &=& x_r^i + \sum_{n\neq0} x^i_ne^{in\sigma} = x_r^i + \sum_{n\neq 0} \frac{i}{|n|}\left(a_{irn} - a_{ir,-n}^\dagger\right) e^{in\sigma} ,\\
P_r^i &=&  \frac{1}{2\pi} \left[p_{0r}^i + \sum_{n\neq0} p^i_ne^{in\sigma}\right] = 
           \frac{1}{2\pi} \left[a_{ir0}^\dagger +\half \sum_{n\neq 0}\left(a_{inr} + a_{ir,-n}^\dagger\right)  e^{in\sigma} \right],\\ 
\theta^{A}_r   &=& \sum_{n=-\infty}^{\infty} \theta_{rn}^{A} e^{in\sigma} , \ \ \ \ \ \ \ \ \ \ \ \ \ 
\ttheta^{A}_r  = \sum_{n=-\infty}^{\infty} \ttheta_{rn}^{A} e^{in\sigma} \\
\lambda_{rA}  &=& \frac{1}{2\pi} \sum_{n=-\infty}^{\infty} \lambda_{rnA} e^{in\sigma} , \ \ \ \ \ \ 
\tlambda_{rA} = \frac{1}{2\pi} \sum_{n=-\infty}^{\infty} \tlambda_{rnA} e^{in\sigma} ,
\eeqa
 where the index $r=1,2$ refers to the initial and final strings\footnote{To avoid confusion with the slit operator in later sections we
sometimes use the symbol $\Pi^i=P^i$.}. By convention we defined $p^r_0=a^\dagger_{i0r}$.  The commutation relations are:
\beq
[a_{irn},a_{jsm}^\dagger] = |n|\, \delta_{ij}\delta_{rs}\,\delta_{mn}, \ \ \ \{\theta^{A}_{rn},\lambda_{sBm}\}=\delta_{rs}\delta^A_B\delta_{m+n}
 \ \ \ \{\ttheta^{A}_{rn},\tlambda_{sBm}\}=\delta_{rs}\delta^A_B\delta_{m+n}\ ,
\eeq
and all others zero. The vacuum of the oscillators is defined such that, if $\mathbf{n>0}$, we have
\beqa
&& a_{irn}\ket{0}=0, \ \ a_{ir,-n}\ket{0}=0 \\
&& \theta_{1n}^A\ket{0}=0, \ \ \theta_{2,-n}^A\ket{0}, \ \ \ttheta_{1,-n}^A\ket{0}=0, \ \ \ttheta_{2n}^A\ket{0}=0 \\
&& \lambda_{1nA}\ket{0}=0, \ \ \lambda_{2,-nA}\ket{0}, \ \ \tlambda_{1,-nA}\ket{0}=0, \ \ \tlambda_{2nA}\ket{0}=0 .
\eeqa
 The difference between $r=1,2$ for the fermions is due to the fact that we define the states with time running in opposite direction for the
initial and final strings but we keep the convention that the tilded variables are left moving and the ones with no tilde, right moving.
 We have a set of linearly realized supercharges:
\beq
 Q^+_{A} = \lambda_{0A} , \ \ Q^{+A} = \theta_0^A, \ \ \tilde{Q}^+_{A} = \tlambda_{0A}, \ \ \tilde{Q}^{+A} = \ttheta^A_0 ,
\eeq
and a set of non-linearly realized:
\beqa
Q_{-A} &=& 2\sqrt{2} \int_{-\pi}^{\pi} \rho^I_{AB} \cA^I \theta^B + 8\pi \int_{-\pi}^\pi \cA^L \lambda_A \\
\tilde{Q}_{-A}    &=& 2\sqrt{2} \int_{-\pi}^{\pi} \rho^I_{AB} \tilde{\cA}^I \tilde{\theta}^B + 8\pi \int_{-\pi}^\pi \tilde{\cA}^L \tilde{\lambda}_A \\
Q_{-}^{A} &=& -4\sqrt{2}\pi \int_{-\pi}^{\pi} \rho^{IAB} \cA^I \lambda_B + 4 \int_{-\pi}^\pi \cA^R \theta^A \\
\tilde{Q}_{-}^{A} &=& -4\sqrt{2}\pi \int_{-\pi}^{\pi} \rho^{IAB} \tilde{\cA}^I \tilde{\lambda}_B + 4 \int_{-\pi}^\pi \tilde{\cA}^R \tilde{\theta}^A ,
\eeqa
where
\beq
\cA^I = P^I-\frac{1}{4\pi} \partial_\sigma Y^I, \ \ \ \tilde{\cA}^I = P^I+\frac{1}{4\pi} \partial_\sigma Y^I ,
\eeq
 and the same for $\cA^{R,L}$. They have the commutation relations
\beq
 \left[\cA(\sigma),\cA(\sigma')\right] = -\frac{i}{2\pi}\partial_\sigma \delta(\sigma-\sigma'), \ \ \ 
 \left[\tilde{\cA}(\sigma),\tilde{\cA}(\sigma')\right] = \frac{i}{2\pi}\partial_\sigma \delta(\sigma-\sigma'), \ \ \ 
 \left[\cA(\sigma),\tilde{\cA}(\sigma')\right] = 0 .
\eeq
It is useful to have a list of supersymmetry variations of the different fields:
\beq
 \begin{array}{lclclcl}
\rule{0pt}{18pt} \left[Q_{-A},\cA^I\right] &=& \frac{i\sqrt{2}}{\pi} \rho^I_{AB} \partial_\sigma \theta^B ,& \ \ \ &  
                     \left[\tilde{Q}_{-A},\tilde{\cA}^I\right] &=& -\frac{i\sqrt{2}}{\pi} \rho^I_{AB} \partial_\sigma \tilde{\theta}^B  ,\\  
\rule{0pt}{18pt} \left[Q_{-A},\cA^R\right] &=& 4i \partial_\sigma \lambda_A ,& \ \ \ &  
                     \left[\tilde{Q}_{-A},\tilde{\cA}^R\right] &=& -4i \partial_\sigma \tilde{\lambda}_A             ,\\  
\rule{0pt}{18pt} \left\{Q_{-A},\theta^B\right\} &=& 8\pi \delta_A^B \cA^L   ,& \ \ \ &  
                     \left\{\tilde{Q}_{-A},\tilde{\theta}^B\right\} &=& 8\pi \delta_A^B \tilde{\cA}^L                ,\\  
\rule{0pt}{18pt} \left\{Q_{-A},\lambda_B\right\} &=& 2\sqrt{2} \rho^I_{AB} \cA^I   ,& \ \ \ &  
                     \left\{\tilde{Q}_{-A},\tilde{\lambda}_B\right\} &=& 2\sqrt{2} \rho^I_{AB} \tilde{\cA}^I         ,\\  
\rule{0pt}{18pt} \left[Q_{-}^{A},\cA^I\right] &=& -  2 i\sqrt{2} \rho^{IAB} \partial_\sigma \lambda_B ,& \ \ \ &  
                     \left[\tilde{Q}_{-}^{A},\tilde{\cA}^I\right] &=& 2 i\sqrt{2} \rho^{IAB} \partial_\sigma \tilde{\lambda}_B  ,\\
\rule{0pt}{18pt} \left[Q_{-}^{A},\cA^L\right] &=& \frac{2i}{\pi} \partial_\sigma \theta^A ,& \ \ \ &  
                     \left[\tilde{Q}_{-}^{A},\tilde{\cA}^L\right] &=& -\frac{2i}{\pi} \partial_\sigma \tilde{\theta}^A       ,\\  
\rule{0pt}{18pt} \left\{Q_{-}^{A},\theta^B\right\} &=& -4\sqrt{2}\pi \rho^{IAB} \cA^I   ,& \ \ \ &  
                     \left\{\tilde{Q}_{-}^{A},\tilde{\theta}^B\right\} &=& -4\sqrt{2}\pi \rho^{IAB} \tilde{\cA}^I            ,\\  
\rule{0pt}{18pt} \left\{Q_{-}^{A},\lambda_B\right\} &=& 4 \delta^A_B \cA^R   ,& \ \ \ &  
                     \left\{\tilde{Q}_{-}^{A},\lambda_B\right\} &=& 4 \delta^A_B \tilde{\cA}^R                               ,\\
\rule{0pt}{28pt}\left\{Q^A_+,\lambda_B\right\} &=& \frac{1}{2\pi} \delta^A_B  &&             
                     \left\{\tilde{Q}^A_+,\tilde{\lambda}_B\right\} &=& \frac{1}{2\pi} \delta^A_B    ,\\             
\rule{0pt}{18pt}\left\{Q_{+A},\theta^B\right\} &=& \delta_A^B  &&             
                     \left\{\tilde{Q}_{+A},\tilde{\theta}^B\right\} &=& \delta_A^B  ,
 \end{array}
\label{susyvar}
\eeq   
where the ones not listed vanish.        
Finally, we can define the Hamiltonian $H_0$  and the momentum $P_\sigma$ through
\beqa
H &=& \int d\sigma (H_r+H_l) ,\\
P_\sigma &=& \int d\sigma (H_r-H_l) \label{to}, \\
H_l &=& 2\pi  \left(\cA^L\cA^R+\half\cA^I\cA^I\right) + i \partial_\sigma\lambda_C\theta^C ,\\
H_r &=& 2\pi  \left(\tilde{\cA}^L\tilde{\cA}^R+\half\tilde{\cA}^I\tilde{\cA}^I\right) - i \partial_\sigma\tlambda_C\ttheta^C .
\label{HP}
\eeqa
 A D3-bane boundary state $\ket{B_{D3}}$ was found in \pone\ to be defined by:
\beqa
 \left( \cA^{L,R} + \tilde{\cA}^{L,R} \right)  \ket{B} &=& 0 , \\ 
 \left( \cA^I- \tilde{\cA}^I \right)  \ket{B}&=& 0   , \\
 \left( \theta^A  - \tilde{\theta}^A \right)  \ket{B} &=&0 , \\
 \left( \lambda_A + \tilde{\lambda}_A \right)  \ket{B} &=& 0 ,
\eeqa   
 which preserve
\beq
\bbQ_+^A= Q_+^A - \tilde{Q}^A_+, \ \ \bbQ_{+A}=Q_{+A} + \tilde{Q}_{+A}, \ \ \bbQ_{-A}=Q_{-A}-\tilde{Q}_{-A}, \ \ \bbQ_{-}^A=Q_{-}^A +  \tilde{Q}_{-}^A .
\eeq
This is regarding a boundary state. In the case of the vertex $\ket{V}$ we should impose these conditions on the slit and continuity of
the coordinates in the rest. For Dirichlet boundary conditions this leads to
\beqa
 \left(Y_1^I(\sigma) - Y_2^I(\sigma)\right) \ket{V} &=& 0 , \ \ -\pi\le \sigma\le \pi ,\\
 \left(Y_1^I(\sigma) + Y_2^I(\sigma)\right) \ket{V} &=& 0 , \ \ |\sigma|\le \sz ,\\
 \left(\Pi_1^I(\sigma) + \Pi^I_2(\sigma)\right) \ket{V} &=& 0 , \ \ \sz \le |\sigma| \le \pi ,
\eeqa
and for Neumann to:
\beqa
 \left(\Pi^a_1(\sigma) + \Pi^a_2(\sigma)\right) \ket{V} &=& 0 , \ \ -\pi\le \sigma\le \pi ,\\
 \left(\Pi^a_1(\sigma) - \Pi^a_2(\sigma)\right) \ket{V} &=& 0 , \ \ |\sigma|\le \sz ,\\
 \left(X^a_1(\sigma) - X^a_2(\sigma)\right) \ket{V} &=& 0 , \ \ \sz \le |\sigma| \le \pi ,
\eeqa
 where we understand all operators are evaluated  at $\tau=0$. These conditions are solved by the vertex state:
\beq
 \ket{V} = e^{\sum_{rs,imn} \NC^{rs}_{i,nm} a_{irn}^\dagger a_{ism}^\dagger }   \prod_{i/\ei=+1} \delta(p^i_1+p^i_2) \ket{0}  ,
\label{vertex_state}
\eeq 
 where $i$ runs over all eight bosonic coordinates and the Neumann coefficients $\NC^{rs}_{i,nm}$ where computed in \pone.
 For the fermions the conditions are:
\beqa
 \left(\theta_1^A   - \theta_2^A  - \tilde{\theta}_1^A  + \tilde{\theta}_2^A  \right) \ket{V} &=& 0 ,\ \ \ -\pi\le\sigma\le \pi ,\\   
 \left(\lambda_{1A}  + \lambda_{2A} + \tilde{\lambda}_{1A} + \tilde{\lambda}_{2A} \right) \ket{V} &=& 0 ,\ \ \ -\pi\le\sigma\le \pi ,\\  
 \left(\theta_1^A   - \theta_2^A  + \tilde{\theta}_1^A  - \tilde{\theta}_2^A  \right) \ket{V} &=& 0 ,\ \ \ \sz\le|\sigma|\le \pi ,\\  
 \left(\lambda_{1A}  + \lambda_{2A} - \tilde{\lambda}_{1A} - \tilde{\lambda}_{2A} \right) \ket{V} &=& 0 ,\ \ \ \sz\le|\sigma|\le \pi ,\\  
 \left(\theta_1^A   + \theta_2^A  - \tilde{\theta}_1^A  - \tilde{\theta}_2^A  \right) \ket{V} &=& 0 ,\ \ \ -\sz\le\sigma\le \sz ,\\  
 \left(\lambda_{1A}  - \lambda_{2A} + \tilde{\lambda}_{1A} - \tilde{\lambda}_{2A} \right) \ket{V} &=& 0 ,\ \ \ -\sz\le\sigma\le \sz . 
\eeqa
 To construct the vertex state it is useful to define new fermionic variables:
\beq
\begin{array}{lclclcl}
\rule{0pt}{18pt} \Xi^A  &=& \frac{1}{\sqrt{2}} \left(\theta_1^A+\tilde{\theta}_2^A\right)     , &\ \ & \bar{\Xi}_A  &=& \frac{1}{\sqrt{2}} \left( \lambda_{1A} + \tilde{\lambda}_{2A} \right) ,\\
\rule{0pt}{18pt} \chi_A &=& \frac{1}{\sqrt{2}} \left(\lambda_{2A}+\tilde{\lambda}_{1A}\right)  , &&     \bar{\chi}^A &=& \frac{1}{\sqrt{2}} \left( \theta_2^A  + \tilde{\theta}_1^A  \right) ,\\
\rule{0pt}{18pt} c_A    &=& \frac{1}{\sqrt{2}} \left(\tilde{\lambda}_{1A}- \lambda_{2A} \right), &&     \bar{c}^A    &=& \frac{1}{\sqrt{2}} \left( \tilde{\theta}_1^A - \theta_2^A     \right) ,\\
\rule{0pt}{18pt} d^A    &=& \frac{1}{\sqrt{2}} \left(\theta_1^A - \tilde{\theta}_2^A\right)  , &&     \bar{d}_A    &=& \frac{1}{\sqrt{2}} \left( \lambda_{1A} - \tilde{\lambda}_{2A} \right) , 
\end{array} \label{cddef}
\eeq
 in terms of which the conditions are
\beqa
 \left(\chi_A  + \bar{\Xi}_A  \right) \ket{V} &=& 0 ,\ \ \ -\pi\le\sigma\le \pi   ,\non\\   
 \left(\Xi^A   - \bar{\chi}^A \right) \ket{V} &=& 0 ,\ \ \ -\pi\le\sigma\le \pi   ,\non\\  
 \left(\bar{c}^A +      d^A   \right) \ket{V} &=& 0 ,\ \ \  \sz\le|\sigma|\le \pi ,\label{veq1} \\  
 \left(\bar{d}_A -      c_A   \right) \ket{V} &=& 0 ,\ \ \  \sz\le|\sigma|\le \pi ,\non\\  
 \left(     d^A  - \bar{c}^A  \right) \ket{V} &=& 0 ,\ \ \ -\sz\le\sigma\le \sz   ,\non\\  
 \left(\bar{d}_A +      c_A   \right) \ket{V} &=& 0 ,\ \ \ -\sz\le\sigma\le \sz . \non
\eeqa
The first two conditions are solved by the state
\beq
 e^{\sum_{m\ge 1} \left(\chi_{mA} \Xi^{A}_{-m} + \bar{\Xi}_{A,-m} \bar{\chi}^A_{m}\right) } \prod_B (\chi_{0B}+\bar{\Xi}_{0B}) \ket{0} .
\label{stf1}
\eeq
The other four conditions can be solved by introducing yet another set of fermionic modes
\beq
\begin{array}{lclcclcll}
 a_{nA}^\dagger &=& c_{nA}          &\ , \mbox{if}\ (n>0)  &\ \ \ & b_n^{A\dagger} &=&\bar{c}_n^A  &\ , \mbox{if}\ (n>0)    ,\\
 b_n^{A\dagger} &=& d_{n}^{A}       &\ , \mbox{if}\ (n<0)  &\ \ \ & a_{nA}^\dagger &=&\bar{d}_{nA} &\ , \mbox{if}\ (n<0)    ,\\
 a^A_n          &=& \bar{c}_{n}^{A} &\ , \mbox{if}\ (n<0)  &\ \ \ & b_{nA}         &=&{c}_{nA}     &\ , \mbox{if}\ (n<0)    ,\\
 b_{nA}         &=& \bar{d}_{nA}    &\ , \mbox{if}\ (n>0)  &\ \ \ & a_{n}^{A}      &=&{d}_{n}^{A}  &\ , \mbox{if}\ (n>0)    ,
\end{array}
\eeq  
 and defining the state
\beq
 \ket{V} = e^{\sum_{m,n\neq 0} V_{nm} |m| b^{A\dagger}_{n} a^{\dagger}_{mA} 
             +\sum_{m\neq 0} (\bar{b}^A_{0} \alpha_m + \bar{a}^A_{0} \beta_m) a^{\dagger}_{mA}} \ket{0} ,
\label{stf2}
\eeq
where
\beqa
 V_{nm} &=& -2\left(\NC^{11}_{nm}(\ei=-1) + \NC^{12}_{nm}(\ei=-1)\right) ,\\
 \alpha_m &=& - |m| \left(\NC^{11}_{0m}(\ei=-1) + \NC^{12}_{0m}(\ei=-1)\right) ,\label{alphadef}\\
 \beta_m &=& - m \left(\NC^{12}_{m0}(\ei=1) - \NC^{11}_{m0}(\ei=1)\right) \label{betamdef}.
\eeqa 
 The zero modes were defined as 
\beqa
{a}_{0A}&=&\bar{d}_{0A}-c_{0A}  = \frac{1}{\sqrt{2}}\left(\lambda_{1A0}-\tilde{\lambda}_{2A0}-\tilde{\lambda}_{1A0}+\lambda_{2A0}\right),\non  \\
\bar{a}_{0}^{A}&=&{d}_{0}^{A}-\bar{c}_{0}^{A}= \frac{1}{\sqrt{2}}
   \left(\theta^A_{10}-\tilde{\theta}^A_{20}-\tilde{\theta}^A_{10}+\theta^A_{20}\right),\label{abdef}\\
 {b}_{0A}&=& c_{0A}+\bar{d}_{0A}= \frac{1}{\sqrt{2}}\left(\tilde{\lambda}_{1A0}-\lambda_{2A0}+\lambda_{1A0}-\tilde{\lambda}_{2A0}\right),   \non\\
\bar{b}_{0}^{A}&=&\bar{c}_{0}^{A}+{d}_{0}^{A}= \frac{1}{\sqrt{2}}\left(\tilde{\theta}^A_{10}-\theta^A_{20}+\theta^A_{10}-\tilde{\theta}^A_{20}\right),\non
\eeqa
 and obey
\beq
 \{a_{0A},\bar{a}_0^B\} = 2 \delta_A^B, \ \ \  \{b_{0A},\bar{b}_0^B\} = 2 \delta_A^B .
\eeq
 The vacuum obeys
\beq
 a_{0A} \ket{0} =0, \ \ \ \ b_{0A}\ket{0} =0 .
\label{veq2}
\eeq
 The meaning of the representation in terms of a vertex state is better understood by writing the vertex state corresponding to the 
identity operator which is\footnote{See the discussion in appendix \ref{scattering}.}
\beqa
 \ket{\Id} &=& 4 \bar{b}_0^4 \int \frac{d^6q}{(2\pi)^6} e^{iq^I(y^I_1+y^I_2)} e^{\Delta_0} \prod_{i/\ei=+1} \delta(p^i_1+p^i_2) \ket{0} ,\\
\Delta_0 &=& -\sum_{i,m\neq0}\frac{1}{|m|}a^\dagger_{i1m}a^\dagger_{i2,-m}
       +\sum_{n>0} \left(\lambda_{2nA}\theta_{1,-n}^A + \tlambda_{1nA}\ttheta_{2,-n}^A 
       + \lambda_{1,-nA}\theta_{2n}^A+\tlambda_{2,-nA}\ttheta_{1n}^A\right) \nonumber \\
        &=&  -\sum_{i,m\neq0}\frac{1}{|m|}a^\dagger_{i1m}a^\dagger_{i2,-m} 
             +\sum_{m\ge1}\left( \chi_{mA}\Xi^A_{-m}+\bar{\Xi}^A_m \bar{\chi}^A_m\right)+\sum_{n\neq0} b^{\dagger A}_na^\dagger_{A,-n}   .
\eeqa
 Acting on this state we can replace:
\beq
 a^\dagger_{i2m}\rightarrow -a_{i1,-m}, \ \ 
 \theta_{2n}^A\rightarrow \theta_{1n}^A, \ \ \ttheta_{2n}^A\rightarrow\ttheta_{1n}^A, \ \ 
 \lambda_{2nA}\rightarrow-\lambda_{1nA}, \ \ \tlambda_{2nA}\rightarrow-\tlambda_{1nA} .
\eeq
 If we have an operator which is a function of only creation operators, after doing the replacement we get an operator acting only
on string 1 and in normal ordered form. This shows that the vertex state is a way to write the operator normally ordered. In 
particular we can rewrite the operator $\Pp_S$ as an operator rather than vertex state as:
\beqa
\Pp_S &=& \NO{ e^{\Delta^{(D)}_B+\Delta^{(N)}_B+\Delta_F}  } ,\label{P_S}\\ \nonumber\\
\Delta_B^{(D)} &=& -\sum_{m,n\neq0} |mn| \NC^{11D}_{mn} y^I_m y^I_n 
          + 2 i q^I \sum_{n\neq0} \NC^{11D}_{0n} |n| y^I_n + q^2 N^{11D}_{00}  ,\\    
\Delta_B^{(N)} &=& 4\sum_{m,n\neq0} |mn| \NC^{11N}_{mn} p^a_m p^a_n + 4 p_0^a \sum_{n\neq0} \frac{1}{n}\beta_n p^a_n + 4 k^2\ln\cos\frac{\sz}{2} ,\\    
\Delta_F&=&4\sum_{m,n\neq0}|m|\sg{n}N^{11D}_{mn} \Theta_{n} \bar{\Lambda}_m + 
          2\Theta_0 \sum_{m\neq0}\beta_m\bar{\Lambda}_m+\bar{b}_0^A\sum_{m\neq0}\alpha_m\bar{\Lambda}_m\ ,
\label{P2}
\eeqa
 where the colons indicate normal ordering and the upper index $D$ in $\NC^{11D}_{mn}$ means that we evaluate the Neumann coefficient for 
Dirichlet boundary conditions (\ie\ $\varepsilon =-1$), and in the case of $\NC^{11N}_{mn}$ for $\varepsilon=+1$. We also introduced the 
notation
\beq
 \bar{b}_0^4 = \frac{1}{24} \epsilon_{ABCD} \bar{b}_0^A \bar{b}_0^B \bar{b}_0^C \bar{b}_0^D ,
\eeq
and defined the fields:
\beqa
 \Theta^A &=& \frac{1}{\sqrt{2}}\left(\theta^A-\ttheta^A\right), \ \ \  \Lambda_A = \frac{1}{\sqrt{2}}\left(\lambda_A-\tlambda_A\right), \\
 \bar{\Theta}^A &=& \frac{1}{\sqrt{2}}\left(\theta^A+\ttheta^A\right), \ \ \ \bar{\Lambda}_A = \frac{1}{\sqrt{2}}\left(\lambda_A+\tlambda_A\right),
\label{Tdef}
\eeqa
whose mode expansions we used in writing $\Pp_S$. In the form (\ref{P2}) the oscillator part is written as an operator but not the zero modes.
We get the final expression by doing a Fourier transform:
\beq
 \Pp_S =  \int \frac{1}{4} d^4\bar{b}_0 \int \frac{d^6q}{(2\pi)^6} e^{-b_0^A\bar{\Lambda}_{0A}-iq^Iy^I} \Pp_S ,
\eeq
where we use the same symbol $\Pp_S$ to denote different representations of the same operator. The factor $\frac{1}{4}$ is from the fact that
$\bar{b}_0^4 \rightarrow \frac{1}{4}\mathbb{I}$ as discussed in appendix \ref{scattering}. 

To do the $q$ integral we have to note that
$N^{11}_{00}=\ln\sin\frac{\sz}{2}<0$. The result is:
\beqa
\Pp_S &=& \frac{1}{2^8\pi^6}
           \frac{1}{\left|N^{11D}_{00}\right|^{3}} \int d^4\bar{b}_0 \NO{e^{\bar{\Delta}^{(D)}_B+\Delta_B^{(N)} +\bar{\Delta}_F}} , \label{P3} \\ \\
\bar{\Delta}_B^{(D)} &=& -\sum_{m,n\neq0} |mn| \bar{\NC}^{11D}_{mn} y^I_{-m} y^I_{-n} 
          + \frac{y^I}{N^{11D}_{00}} \sum_{n\neq0} \NC^{11D}_{0n} |n| y^I_{-n} + \frac{y^2}{4 N^{11D}_{00}}  ,\nonumber\\    
\Delta_B^{(N)} &=& 4\sum_{m,n\neq0} |mn| \NC^{11N}_{mn} p^a_m p^a_n + 4 p_0^a \sum_{n\neq0} \frac{1}{n}\beta_n p^a_n + 4 k^2\ln\cos\frac{\sz}{2} ,\nonumber\\    
\bar{\Delta}_F&=&4\sum_{m,n\neq0}|m|\sg{n}N^{11D}_{mn} \Theta_{n}^A \bar{\Lambda}_{mA} + 
          2\Theta_0^A \sum_{m\neq0}\beta_m\bar{\Lambda}_{mA}+\bar{b}_0^A\left(-\bar{\Lambda}_{0A}+\sum_{m\neq0}\alpha_m\bar{\Lambda}_{mA}\right) \nonumber,
\eeqa
 where we defined
\beq
 \bar{\NC}^{11D}_{mn} = \NC^{11D}_{mn}- \frac{ \NC^{11D}_{m0} \NC^{11D}_{n0}}{ \NC^{11D}_{00}}.
\eeq
From the properties of the Neumann coefficients we can derive:
\beqa
 \sum_{n\neq0} \bar{\NC}^{11D}_{nm}e^{in\sigma} &=& \frac{1}{2|m|} e^{-im\sigma} , \ \ \ \mbox{if} \ \ \ |\sigma|<\sz ,\\
 \sum_{n\neq0} |n| \bar{\NC}^{11D}_{nm}e^{in\sigma} &=&\half \frac{\NC^{11D}_{m0}}{\NC^{11D}_{00}}, \ \ \ \mbox{if} \ \ \ |\sigma|>\sz .
\eeqa
 Using this together with the properties of the Neumann coefficients listed in \pone, we readily find that $\Pp_S$ as defined in eq.(\ref{P3}) satisfies:
\beqa
 {}[Y^I(\sigma),\Pp_S] &=& 0 ,\ \ \mbox{for}\ \ -\pi<\sigma<\pi, \\
 {}[\Pi^I(\sigma),\Pp_S] &=& 0 ,\ \ \mbox{for}\ \ \sz<|\sigma|<\pi, \\
 {}Y^I(\sigma)\Pp_S &=& 0 ,\ \ \mbox{for}\ \  |\sigma|<\sz,\\
 {}[\Pi^a(\sigma),\Pp_S] &=& 0 ,\ \ \mbox{for}\ \ -\pi<\sigma<\pi, \\
 {}[X^a(\sigma),\Pp_S] &=& 0 ,\ \ \mbox{for}\ \ \sz<|\sigma|<\pi, \\
 {}\Pi^a(\sigma)\Pp_S &=& 0 ,\ \ \mbox{for}\ \  |\sigma|<\sz,\\
 {}[\Theta(\sigma),\Pp_S] &=& 0 ,\ \ \mbox{for}\ \ -\pi<\sigma<\pi, \\
 {}[\bar{\Lambda}(\sigma),\Pp_S] &=& 0 ,\ \ \mbox{for}\ \ -\pi<\sigma<\pi, \\
 {}[\bar{\Theta}(\sigma),\Pp_S] &=& 0 ,\ \ \mbox{for}\ \ \sz<|\sigma|, \\
 {}[\Lambda(\sigma),\Pp_S] &=& 0 ,\ \ \mbox{for}\ \ \sz<|\sigma|, \\
 {}\Theta(\sigma) \Pp_S &=& 0 ,\ \ \mbox{for}\ \ |\sigma|<\sz, \\
 {}\bar{\Lambda}(\sigma)\Pp_S &=& 0 ,\ \ \mbox{for}\ \ |\sigma|<\sz, 
\eeqa
which imply that indeed $\Pp_S$ projects over the right boundary conditions on the region $|\sigma|<\sz$ and does nothing for $\sz<|\sigma|$.
 In doing these calculations it is useful to note that 
\beq
{}[\cO, \NO{e^{\Delta}}] = \NO{[\cO,\Delta]e^\Delta},
\eeq
whenever $\cO$ is an operator linear in oscillators and $\Delta$ is quadratic in oscillators. 
 
Having found different useful representations of the operator $\Pp_S$ we proceed to study its properties.
     
\subsection{Divergences of operators near $\Pp_S$} 

 Whenever one inserts an  operator in the world-sheet, other field becomes singular near the insertion. For example if one inserts the operator
$X^a(z_0)$ then the (world-sheet) energy momentum tensor has a pole at $z=z_0$ whose residue is $\partial_z X^a(z_0)$. This simply means that 
the energy momentum tensor generates translations on the world-sheet. If we insert a slit the situation is no different. For example the 
energy momentum tensor should also have a singularity representing a translation of the slit. Of particular importance for us are translations 
in $\sigma$. It is clear that the slit is ``almost'' invariant under such translations. Indeed under an infinitesimal translation in $\sigma$ 
the only variation occurs at the ends of the slit, in the region $|\sigma|<\sz$ no change is observed. Therefore we expect the translation 
operator to have pole singularities localized at the ends of the segment.  

 With this in mind we proceed now to study different fields and see what singularities they have at the end points of the slit. The analysis
is the same as the one in \cite{GSW}.
Consider the field $\cA^i$ whose mode expansion is:
\beq
\cA^i_r(\sigma) = \frac{1}{2\pi} a^\dagger_{ir0} + \frac{1}{2\pi} \sum_{n>0} \left(a_{inr}e^{in\sigma}+a^\dagger_{irn}e^{-in\sigma}\right).
\eeq
Now we compute
\beqa
\lefteqn{\cA^i_r(\sigma) e^{\sum_{rsimn}N^{rs(i)}_{nm} a^\dagger_{irn}a^\dagger_{ism}} \ket{0} =} && \\
&& \ \ \ \ = e^{\sum_{rsimn}N^{rs(i)}_{nm} a^\dagger_{irn}a^\dagger_{ism}} 
 \left[\frac{1}{2\pi} a^\dagger_{ir0} + \frac{1}{2\pi} \sum_{n>0} \left(2\sum_{sm}|n| N^{rs(i)}_{nm}a^\dagger_{ism} 
  e^{in\sigma}+a^\dagger_{irn}e^{-in\sigma}\right)\right] \ket{0} \nonumber.
\eeqa
 There is a singularity coming from the double sum which we express as
\beq
\cA^i_r(\sigma) \sim \frac{1}{\pi} \sum_{n>0,sm}|n| N^{rs(i)}_{nm}a^\dagger_{ism} e^{in\sigma}.
\eeq
 The behavior of the Neumann coefficients for large value of the arguments was derived in \pone. This allows us to obtain, for example,
\beqa
\sum_{n>0} e^{in\sigma} N^{1s(i)}_{nm} |n| &\simeq& 
    -\frac{1}{\sqrt{2\pi\sin\sz}} \sum_{n>0} \Ima\left(\frac{e^{i\frac{\pi}{4}-in\sz}}{\sqrt{n}}f^{s(i)}_m\right)e^{in\sigma}\\
 &\simeq& -\frac{1}{2\sqrt{2\sin\sz}} \left(\frac{f^{s(i)}_m}{\sqrt{\sigma-\sz}}+\frac{\bar{f}^{s(i)}_m}{\sqrt{-\sigma-\sz}}\right),
\eeqa
where the approximation refers to the leading behavior near $\sigma=\pm\sz$. In this way we can do a lengthy but straight-forward study of all 
the fields and obtain the leading singularities as:
\beqa
\cA^i_1 &\sim& \ei\tilde{\cA}^i_1  \sim -\tilde{\cA}^2_i \sim -\ei \cA_2^i \sim 
       \frac{ Z^i}{\sqrt{\sigma-\sz}} + \frac{\bar{Z}^i}{\sqrt{-\sigma-\sz}} \nonumber ,\\
\frac{1}{\sqrt{2}}\,d^A &\sim& -\frac{1}{\sqrt{2}}\, \bar{c}^A \sim \theta_1^A  \sim -\ttheta^A_1  
  \sim \frac{Y^A}{\sqrt{\sigma-\sz}} + \frac{\bar{Y}^A}{\sqrt{-\sigma-\sz}}  ,\\ 
\frac{1}{\sqrt{2}}\, \partial_\sigma c_A &\sim& \frac{1}{\sqrt{2}}\, \partial_\sigma \bar{d}_A \sim \partial_\sigma\lambda_{1A} 
 \sim \partial_\sigma\tlambda_{1A} \sim  i\left(\frac{V_A}{\sqrt{\sigma-\sz}}  + \frac{\bar{V}_A}{\sqrt{-\sigma-\sz}}\right) ,
\nonumber
\eeqa
where we defined the operators:
\beqa
 Z^i &=& -\frac{\sqrt{2}}{4\pi\sqrt{\sin\sz}}\sum_{sm} f^{s(i)}_{m}a^\dagger_{ism} ,\\
 \bar{Z}^i &=& -\frac{\sqrt{2}}{4\pi\sqrt{\sin\sz}}\sum_{sm} \bar{f}^{s(i)}_{m}a^\dagger_{ism}, \\
 Y^A &=& \frac{1}{\sqrt{\sin\sz}}\left\{\half\bar{a}^A_0\sin\frac{\sz}{2}+\frac{i}{2}\bar{b}_0^A \cos\frac{\sz}{2} 
        + \sum_{n\neq0}\bar{f}^{1(D)}_nb^{\dagger A}_n\right\} ,\\
 \bar{Y}^A &=& \frac{1}{\sqrt{\sin\sz}}\left\{\half\bar{a}^A_0\sin\frac{\sz}{2}-\frac{i}{2}\bar{b}_0^A \cos\frac{\sz}{2} 
        + \sum_{n\neq0}f^{1(D)}_nb^{\dagger A}_n\right\} ,\\
 V^A &=& -\frac{1}{2\pi\sqrt{\sin\sz}} \sum_{m} |m|\bar{f}^{1(D)}_m a^\dagger_{mA} ,\\  
 \bar{V}^A &=& -\frac{1}{2\pi\sqrt{\sin\sz}} \sum_{m} |m|\bar{f}^{1(D)}_m a^\dagger_{mA} .
\eeqa  
 A very useful check is to use the singularities of the translation operator (\ref{to}) to compute the commutator:
\beq
{}[P_\sigma,\Pp_S] = -i \partial_\sigma \Pp_S ,
\label{Pder}
\eeq
which we expect to give the sigma derivative of the operator we commute it with. To verify that, we use, as shown in fig.\ref{Pvar} 
that the commutator is 
\beq
 {}[P_\sigma,\Pp_S] =  \oint (H_r-H_l) \Pp_S ,
\eeq
where the integral is over the contour in the figure. It is equal to the commutator because it precisely represents the difference between
applying first $\Pp_S$ and then $P_\sigma$ and doing the same in opposite order. The first observation is that the integral outside the 
slit cancel each other. On the slit, both sides are independent but the boundary conditions imply $H_r-H_l=0$ so the integral vanishes there.

\FIGURE{
\includegraphics[height=5cm]{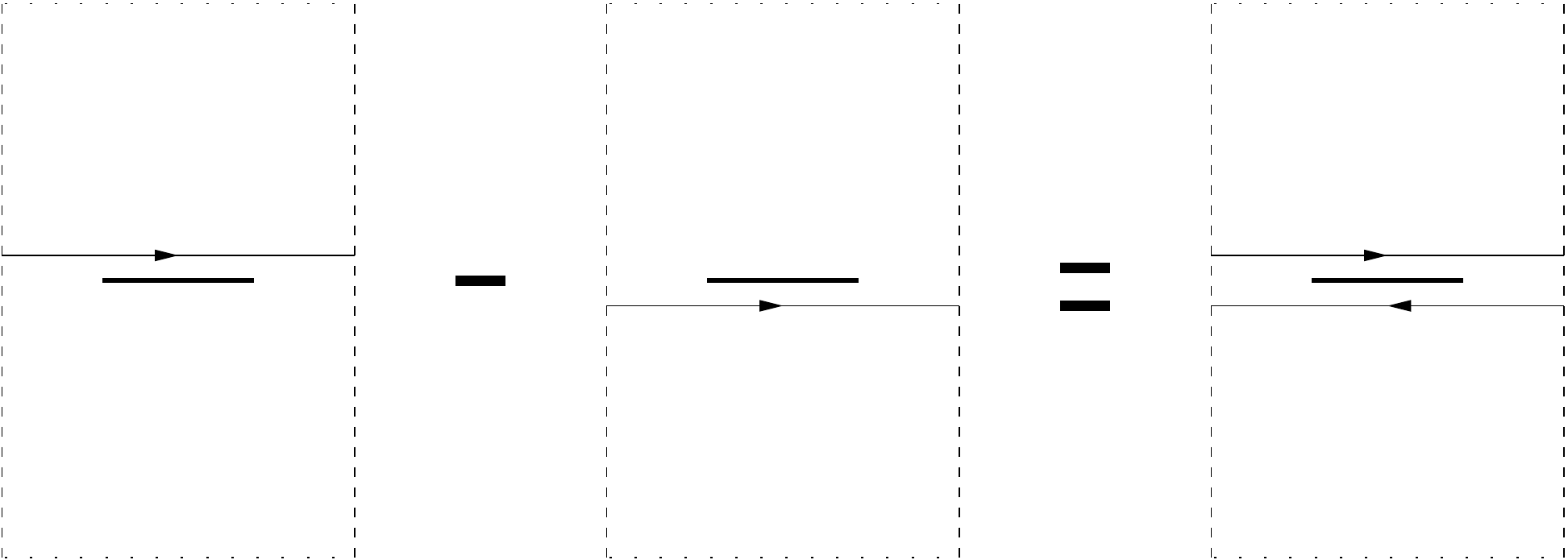}
\caption{To compute the commutator between the slit and the integral over sigma of an operator we apply them in different order and subtract.
The result is a closed contour integral around the slit.}
\label{Pvar}
}

 The only contribution comes from the singularities at the end points of the string. Deforming the contour we get two integrals along circles
centered at $\rho=\pm\sz$. If we write the two circles as $\rho=\sigma_0+\epsilon$, $\rho=-\sigma_0+\epsilon$ we obtain
\beqa
 {}[P_\sigma,\Pp_S] =  \oint (H_r-H_l) &=& \left(\oint\frac{d\bar{\epsilon}}{\bar{\epsilon}}-\oint\frac{d\epsilon}{\epsilon}\right)
                                           \left[2\pi Z^LZ^R+\pi Z^IZ^I-V_AY^A\right]  \\
                                       & & -\left(\oint\frac{d\bar{\epsilon}}{\bar{\epsilon}}-\oint\frac{d\epsilon}{\epsilon}\right)
                                           \left[2\pi \bar{Z}^L\bar{Z}^R+\pi \bar{Z}^I\bar{Z}^I-\bar{V}_A\bar{Y}^A\right] \nonumber  ,
\eeqa
where the minus sign comes from the fact that \eg\ $\cA^i\sim Z^i/\sqrt{\epsilon}$ near $\sz$ but $\cA^i\sim \bar{Z}^i/\sqrt{-\epsilon}$ near $-\sz$.
Remembering that the contours are oriented counterclockwise we get
\beqa
{}[P_\sigma,\Pp_S] &=& -i \left[ \rule{0pt}{18pt}  4\pi^2 \bar{Z}^I\bar{Z}^I+8\pi^2\bar{Z}^L\bar{Z}^R+4\pi\bar{Y}^A\bar{V}_A  \right. \\
                   &&     \left. \rule{0pt}{18pt} -4\pi^2 Z^IZ^I-8\pi^2Z^LZ^R-4\pi Y^AV_A \right] .
\eeqa
At the same time a straightforward computation using the properties of the Neumann coefficients gives:
\beqa
 \partial_\sigma(\Delta_B+\Delta_F) &=& 4\pi^2 \bar{Z}^I\bar{Z}^I+8\pi^2\bar{Z}^L\bar{Z}^R+4\pi\bar{Y}^A\bar{V}_A   \\
                              &&      -4\pi^2 Z^IZ^I-8\pi^2Z^LZ^R-4\pi Y^AV_A ,
\eeqa
which proves the identity (\ref{Pder}). To perform the sigma derivative we introduced the $\sigma$ dependence in $\Pp_S$ through
(\eg\ in the vertex representation):
\beqa
 \ket{\Pp_S} &=& e^{\Delta_B+\Delta_F} \prod_{i/\ei=+1} \delta(p^i_1+p^i_2) \prod_B (\chi_{0B}+\bar{\Xi}_{0B}) \ket{0} ,\\
 \Delta_B &=& \sum_{rs,imn} \NC^{rs}_{i,nm}e^{- i(n+m)\sigma} a_{irn}^\dagger a_{ism}^\dagger  ,\\
 \Delta_F &=&  \sum_{m,n\neq 0} V_{nm} |m|e^{i(n+m)\sigma} b^{A\dagger}_{n} a^{\dagger}_{mA}+\sum_{m\neq 0} (\bar{b}^A_{0} \alpha_m 
                     + \bar{a}^A_{0} \beta_m) e^{im\sigma}  a^\dagger_{mA} \\
           &&   +  \sum_{m\ge 1} \left(\chi_{mA} \Xi^{A}_{-m} + \bar{\Xi}_{A,-m} \bar{\chi}^A_{m}\right) .
\eeqa  
 It is instructive also to use the divergencies and write:
\beqa
{}[P_\sigma,\Pp_S] &=& \epsilon \cP(\sz) + \epsilon \cP(-\sz) \ \ \ \mbox{with} \\
       \cP(\sigma) &=& 8\pi^2 \Pi^L\Pi^R + \frac{1}{4} \partial_\sigma Y^I\partial_\sigma Y^I + 2\pi i \partial_\sigma\bar{\Lambda} \Theta,
\eeqa
which has the following meaning: $\epsilon\cP(\sz)$ means to evaluate $\epsilon\cP(\sz+\epsilon)$ in the limit where $\epsilon \rightarrow0$,\ie\
keeping the divergent piece of $\cP(\sz)$. The same for $\epsilon\cP(\sz)=\lim_{\epsilon\rightarrow0} \cP(-\sz+\epsilon)$. 
Notice that the minus sign we discussed before reappears and we get the same operator evaluated at the two points.  
 
 Recall now that the operator $\Pp_S$ is a function of $\sigma_L$ and $\sigma_R$, the positions of the two extreme points. Since in our
variables we have $\sigma_L=\sigma-\sz$ and $\sigma_R=\sigma+\sz$ we get
\beq
 \partial_\sigma \Pp_S = \partial_{\sigma_L} \Pp_S + \partial_{\sigma_R} \Pp_S .
\eeq
 If we change $\sigma_L$ the only variation in $\Pp_S$ occurs precisely at that end-point, the rest of the slit is unmodified. The same 
if we change $\sigma_R$. Thus we conclude that:
\beqa
\partial_{\sigma_L} \Pp_S &=&  \epsilon \cP(-\sz) ,\\
\partial_{\sigma_R} \Pp_S &=&  \epsilon \cP(\sz) ,
\label{derLR}
\eeqa
that we are going to find useful later on. Without this trick we should have evaluated explicitly $\partial_{\sz} \Pp_S$ which seems a
very difficult task. 
 
\subsection{Supersymmetric transformation of $\Pp_S$}

 The conserved supersymmetric charges commute according to
\beqa
\{\bbQ_+^A,\bbQ^B_-\} &=& -2\sqrt{2}P^I\rho^{IBA} ,\\
\{ \bbQ_{+A} , \bbQ_{-B}\} &=&  2\sqrt{2}P^I\rho^I_{BA} ,\\
\{\bbQ_{-A},\bbQ^B_-\} &=& 2(H_l-H_r) \delta_A^B = -16 P_\sigma \delta_A^B .
\eeqa
 One is used to the fact that the supercharges commute to the Hamiltonian but, after interchanging $\sigma\leftrightarrow\tau$ they
commute to translations in $\sigma$. This is rather interesting since the Hamiltonian has a correction of order $\lambda$ but $P_\sigma$ does not.
If the supercharges had anti-commuted to $H$ then they should have had terms of order $\lambda$  but, since they not, there is no reason for them 
to be corrected. In fact as we see below they are not. On the other hand, we can use the Jacobi identity and obtain
\beqa
&& {}[ \{ \bbQ_A^- , \bbQ^{-B}\},\Pp_S]+ \{[\Pp_S,\bbQ_A^-],\bbQ^{-B}\} + \{[\bbQ^{-B},\Pp_S],\bbQ_A^-\} =0 \ \ \ \ \Rightarrow \\
&& {}-16\delta_A^B [P_\sigma,\Pp_S]+ \{[\Pp_S,\bbQ_A^-],\bbQ^{-B}\} + \{[\bbQ^{-B},\Pp_S],\bbQ_A^-\} =0 .
\eeqa
Since $\Pp_S$ is not invariant under translations it cannot be invariant under supersymmetry, \ie\ we cannot have $[\Pp_S,\bbQ_A^-]=0$ and
$[\bbQ^{-B},\Pp_S]=0$ since $[P_\sigma,\Pp_S]\neq0$. In fact using the same ideas as the previous subsection it is very simple to find out that
\beq
{}[\bbQ_{-A},\Pp_S] = \left\{ 2i \left. \rho^I_{AB} \epsilon \partial_\sigma Y^I \Theta^B \right|_{\sigma=\sz} +
                        2i \left. \rho^I_{AB} \epsilon \partial_\sigma Y^I \Theta^B \right|_{\sigma=-\sz}  \right\} \Pp_S ,
\eeq
and 
\beq
{}[\bbQ^A_{-},\Pp_S] = \left\{-\left. 8\pi i \sqrt{2} \epsilon \Pi^R \Theta^A \right|_{\sigma=\sz} 
                       -\left. 8\pi i \sqrt{2} \epsilon \Pi^R \Theta^A \right|_{\sigma=-\sz} \right\} \Pp_S .
\eeq
Since $\Pp_S$ does not commute with the supersymmetries that are preserved by the D3-brane it cannot be the Hamiltonian. In fact, 
as is well-known \cite{GSW}, one has to insert operators at the end of the slit such that the supersymmetric current has new singularities 
canceling the ones coming from the slit. We discuss this in the next section.

\subsection{$U(1)$ rotational symmetry}

 In light cone-gauge, there is a manifest $SO(2)=U(1)$ symmetry that rotates the coordinates parallel to the brane but transverse to the light-cone,
namely $X^{a=1,2}$. The fields transform according to:
\beq
\begin{array}{ccccccc}
X^R\rightarrow e^{i\phi}\,X^R,  & \ \ \ & \Pi^R\rightarrow e^{i\phi}\, \Pi^R, 
   & \ \ \ & X^L\rightarrow  e^{-i\phi}\, X^L,& \ \ \ & \Pi^L\rightarrow e^{-i\phi}\,\Pi^L \\ &&&&&&\\ 
\Theta\rightarrow e^{-\frac{i}{2}\phi}\,\Theta, && \Lambda\rightarrow e^{\frac{i}{2}\phi}\,\Lambda,
  &&\bar{\Theta}\rightarrow e^{-\frac{i}{2}\phi}\,\bar{\Theta} && \bar{\Lambda} \rightarrow e^{\frac{i}{2}\phi}\,\bar{\Lambda} .
\end{array}
\eeq
 It is clear that, in (\ref{P3}), $\bar{\Delta}^{(D)}_B$ and $\bar{\Delta}^{(N)}_B$ are invariant under the $U(1)$. However, 
$\bar{\Delta}_F$ has a term proportional to $\bar{b}_0^A$ which is not invariant unless we rotate 
$\bar{b}_0^A\rightarrow e^{-\frac{i}{2}\phi} \bar{b}_0^A$. If we do that, the integral $\int d\bar{b}_0^A$ rotates as (recall this
is a fermionic integral):
\beq
  \int d\bar{b}_0^A \rightarrow e^{2i\phi} \int d\bar{b}_0^A .
\eeq
 Therefore the slit operator transforms as
\beq
 \Pp_S \rightarrow e^{2i\phi} \Pp_S ,
\eeq
under rotations. One way to confirm this is to compute, from eq. ($\ref{P3}$) the limit of $\Pp_S$ as $\sigma_0\rightarrow 0$ which results in
\beq
 \Pp_S \simeq_{\sz\rightarrow 0} \frac{1}{\left|N_{00}^{11D}\right|^3} \bar{\Lambda}_0^4 ,
\label{P0s}
\eeq 
where we used the properties of the Neumann coefficients derived in \pone\ and $\bar{\Lambda}_{0A}$ is the zero mode of $\bar{\Lambda}_A$.
 Now, it is obvious that for small $\sz$, $\Pp_S$ has charge $+2$ which, since it is an integer, should be independent of $\sz$.  
This is another reason why we cannot think of $\Pp_S$ as a Hamiltonian which should preserve the $U(1)$ rotational symmetry. 
Again, the same insertions that make $\Pp$ supersymmetric make it invariant under the $U(1)$.  Note that for $\sz\rightarrow 0$, the operator $\Pp_S$
actually vanishes since $\left|N_{00}^{11D}\right| = \left|\ln\sin\frac{\sz}{2}\right| \rightarrow\infty$. In eq.(\ref{P0s}) we kept the
leading contribution.    
 
\subsection{Algebra of the $\Pp_S$}

In this subsection we make some comments about the operator $\Pp_s$ for the case of the bosonic strings. They are outside the main line
of development of the paper and we include them for future reference. The point we want to make is that, since the operator $\Pp_S$ imposes
the boundary state boundary conditions on the slit they should obey the relations:
\beqa
{}[\Pp_S(\sigma_L,\sigma_R),\Pp_S(\sigma'_L,\sigma'_R)] &=& 0, \ \ \ \forall\ \sigma_{L,R}, \sigma'_{L,R}, \\
 \Pp_S(\sigma_L,\sigma_R) \Pp_S(\sigma'_L,\sigma'_R) &=& \Pp_S(\sigma_L,\sigma'_R) \ \ \ \forall\ \sigma_L<\sigma'_L<\sigma_R<\sigma'_R  , \\
\Pp_S(\sigma_L,\sigma_R) \Pp_S(\sigma'_L,\sigma'_R) &=& \Pp_S(\sigma_L,\sigma_R) \ \ \ \forall\ \sigma_L<\sigma'_L<\sigma'_R<\sigma_R  , \\
\Pp_S(\sigma_L,\sigma_R) \ket{B} &=& \ket{B} \ \ \ \forall\ \sigma_L,\sigma_R ,
\label{Pproj}
\eeqa
 where $\ket{B}$ is the boundary state. These relations establish the idea that $\Pp_S$ is a projector. For the superstring we expect similar relations
but we have not investigated the issue. 
 
\section{Operator insertions: computation of $\Pp$}
\label{Pcomp}

 We have to insert operators at the end of the slit in such a way that the resulting operator commutes with the supercharges and is invariant
under the transverse $U(1)$. We propose the ansatz
\beq
 \Pp = \alpha_3 \int_{-\pi}^{\pi} d\sigma \int_0^{\pi} d\sigma_0 H_1(\sigma_L) H_1(\sigma_R) \Pp_S(\sigma_L,\sigma_R) .
\label{Pdef}
\eeq
where the slit extends from $\sigma_l$ to $\sigma_R$ with $\sigma_{R,L}\sigma \pm\sigma_0$, $\sigma$ being the position of the center of 
the slit and $\sigma_0$ its half-width. The constant $\alpha_3$ is inserted to provide an overall normalization and is going to be determine
later by comparison with previously know results from scattering of closed strings from D-branes.  
 In the open string channel it is known which operators to insert \cite{GSW} and we expect them to be essentially the same here since
we are only doing a $\sigma\leftrightarrow\tau$ interchange. Nevertheless let us reason what we can have. As we discuss later it is convenient
to have operators that commute with $\Pp_S$. As we saw in the previous section, $\Pi^{L,R}$, $Y^I$ and $\Theta^A$, $\bar{\Lambda}_A$ commute
with $\Pp_S$ independently of the position in which they are inserted. We also have to add up operators with the same charge under the $U(1)$ that
rotates the transverse Neumann coordinates (transverse to the light-cone directions, not the D3-brane).  This leads to a solution analogous
to the one in \cite{GSW}:
\beq
 H_1 = \sqrt{\epsilon} \left\{\Pi^L-\frac{i}{8\pi\sqrt{2}}\, \epsilon\,\partial_\sigma Y^I \rho^I_{CD}\Theta^C\Theta^D-\epsilon^2\Pi^R\Theta^4\right\}.
\label{H1insertion}
\eeq
 Of course the precise coefficients follow from the calculation but we anticipated the result. We would like to compute the commutator of
the supercharges with $H_1$. To do that it is better to rewrite (\ref{susyvar}) in terms of the fields and supercharges we are using now.  
The result is
\beq
\begin{array}{lclclcl}
\rule{0pt}{18pt} \left[\bbQ_{-A},X^L\right] &=& 0 ,& \ \ \ &  
                     \left[\bbQ_{-A},X^R\right] &=& -8i\sqrt{2}\Lambda_A  ,\\  
\rule{0pt}{18pt} \left[\bbQ_{-A},\Pi^L\right] &=& 0 ,& \ \ \ &  
                     \left[\bbQ_{-A},\Pi^R\right] &=& 2\sqrt{2}i\partial_\sigma\bar{\Lambda}_A             ,\\  
\rule{0pt}{18pt} \left[\bbQ_{-A},Y^I\right] &=& -4i\rho^I_{AB}\Theta^B   ,& \ \ \ &  
                     \left[\bbQ_{-A},P^I\right] &=& \frac{i}{\pi} \rho^I_{AB}\partial_\sigma\bar{\Theta}^B               ,\\  
\rule{0pt}{18pt} \left\{\bbQ_{-A},\bar{\Lambda}_B\right\} &=& -\frac{1}{\pi} \rho^I_{AB} \partial_\sigma Y^I   ,& \ \ \ &  
                     \left\{\bbQ_{-A},\Lambda_B\right\} &=& 4 \rho^I_{AB} P^I         ,\\  
\rule{0pt}{18pt} \left\{\bbQ_{-A},\Theta^B\right\} &=& 8\pi\sqrt{2}\delta_A^B \Pi^L   ,& \ \ \ &  
                     \left\{\bbQ_{-A},\bar{\Theta}^B\right\} &=& -2\sqrt{2}\delta_A^B \partial_\sigma X^L         ,\\ \\
\rule{0pt}{18pt} \left[\bbQ_{-}^{A},\partial_\sigma X^L\right] &=& -4i\sqrt{2}\partial_\sigma\bar{\Theta}^A   ,& \ \ \ &  
                     \left[\bbQ_{-}^{A},\partial_\sigma X^R\right] &=&    0    ,\\  
\rule{0pt}{18pt} \left[\bbQ_{-}^{A},\Pi^L\right] &=& i\frac{\sqrt{2}}{\pi} \partial_\sigma \Theta^A ,& \ \ \ &  
                     \left[\bbQ_{-}^{A},\Pi^R\right] &=& 0  ,\\
\rule{0pt}{18pt} \left[\bbQ_{-}^{A},\partial_\sigma Y^I\right] &=& 8\pi i \rho^{IAB} \partial_\sigma\bar{\Lambda}_B ,& \ \ \ &  
                     \left[\bbQ_{-}^{A},P^I\right] &=& -2i\rho^{IAB} \partial_\sigma\Lambda_B         ,\\  
\rule{0pt}{18pt} \left\{\bbQ_{-}^{A},\bar{\Lambda}_B\right\} &=& 4\sqrt{2}\delta_B^A \Pi^R   ,& \ \ \ &  
                     \left\{\bbQ_{-}^{A},\Lambda_B\right\} &=& -\frac{\sqrt{2}}{\pi} \delta_B^A\partial_\sigma X^R       ,\\  
\rule{0pt}{18pt} \left\{\bbQ_{-}^{A},\Theta^B\right\} &=& 2\rho^{IAB} \partial_\sigma Y^I   ,& \ \ \ &  
                     \left\{\bbQ_{-}^{A},\bar{\Theta}^B\right\} &=& -8\pi\rho^{IAB} P^I                               .\\
 \end{array}
\label{susyvarb}
\eeq          
With this table it is a simple task to compute:
\beqa
{}[\bbQ_{-A},H_1] &=&  -2i \epsilon^{\frac{3}{2}} \rho^I_{AB}\ps Y^I \Pi^L\theta^B 
                    + \frac{\epsilon^{\frac{3}{2}}}{3\pi\sqrt{2}} \ps\left(\epsilon_{ABCD} \Theta^B\Theta^C\Theta^D\right) \\
                  &&  + \frac{\epsilon^{\frac{3}{2}}}{3\pi\sqrt{2}} \left\{-8\pi^2\epsilon\Pi^L\Pi^R-2\pi i \ps\bar{\Lambda}_F\Theta^F \right\}
                       \epsilon_{ABCD} \Theta^B\Theta^C\Theta^D ,
\eeqa
which implies
\beq
H_1 [\bbQ_{-A},\Pp_S] + [\bbQ,H_1] \Pp_S = \frac{\epsilon^{\frac{3}{2}}}{3\pi\sqrt{2}} \partial_{\sigma_R} 
                                           \left(\epsilon_{ABCD} \Theta^B\Theta^C\Theta^D \Pp_S\right) ,
\eeq
where $H_1$ is evaluated at $\sigma_R$ and we used eq.(\ref{derLR}). The same is valid at $\sigma_L$. If we define the 
operator:
\beq
 \hat{\bbQ}_{-A} = \frac{\epsilon^{\frac{3}{2}}}{3\pi\sqrt{2}} \epsilon_{ABCD} \Theta^B\Theta^C\Theta^D ,
\eeq
we can write:
\beqa
\lefteqn{{}[\bbQ_{-A},\int d\sigma_L d\sigma_R H_1(\sigma_L) H_1(\sigma_R)\Pp_S(\sigma_L,\sigma_R)]\ =} && \\
&& \ \ = \int d\sigma_L d\sigma_R H_1(\sigma_R) \partial_{\sigma_L} \left( \hat{\bbQ}_{-A}(\sigma_L)\Pp_S\right) 
 + \int d\sigma_L d\sigma_R H_1(\sigma_L) \partial_{\sigma_R} \left( \hat{\bbQ}_{-A}(\sigma_R)\Pp_S\right) \nonumber \\
  &&\ \ = - \int d\sigma_L d\sigma_R \hat{\bbQ}_{-A}(\sigma_L) \partial_{\sigma_R}\left(H_1(\sigma_R)\Pp_S\right) +
    \int d\sigma_L d\sigma_R H_1(\sigma_L) \partial_{\sigma_R} \left( \hat{\bbQ}_{-A}(\sigma_R)\Pp_S\right) \nonumber,
\eeqa
where we replaced $\partial_{\sigma_L}=\partial_\sigma-\partial_{\sigma_R}$ and integrated by parts in $\sigma$. Also, all the operators are made
out of the same commuting fields so the order is not important. Finally we can integrate in $\sigma_R$ to get:
\beq
 \int_{\sigma_L}^{\sigma_L+2\pi} d\sigma_R \partial_{\sigma_R}\left(H_1(\sigma_R)\Pp_S\right) = H_1(\sigma_L)\left(\Pp_{2\pi}-\Pp_0\right),
\eeq
where $\Pp_0$ is the operator corresponding to a slit of zero size and $\Pp_{2\pi}$ the operator corresponding to a slit of size 
$2\pi$. Doing the same with the other integral we get
\beqa
\lefteqn{{}[\bbQ_{-A},\Pp] = [\bbQ_{-A},\int d\sigma_L d\sigma_R H_1(\sigma_L) H_1(\sigma_R)\Pp_S] \ = }\\
&&  = - \int d\sigma_L \hat{\bbQ}_{-A}(\sigma_L) H_1(\sigma_L) \left(\Pp_{2\pi}-\Pp_0\right) 
      + \int d\sigma_L \hat{\bbQ}_{-A}(\sigma_L) H_1(\sigma_L) \left(\Pp_{2\pi}-\Pp_0\right) =0 \nonumber .
\eeqa 
We conclude that $\Pp$ defined in (\ref{Pdef}) is supersymmetric under this charge. The other charge $\bbQ^{A}_-$ works the same with 
\beq
\hat{\bbQ}^A_{-} = \frac{i\sqrt{2\epsilon}}{\pi} \Theta^A .
\eeq
 It is worth mentioning that, in a later section, we compute $\Pp$ in the limit of small holes obtaining a local operator invariant under 
supersymmetry, providing an independent check of supersymmetry.  Therefore the operator $\Pp$ is the correct operator to represent a hole 
or loop insertion in the superstring. It is useful to write it in normal ordered form. That amounts essentially to replacing every field by 
its divergent part. 
 However, an important point is that there is an extra contribution from the contraction between $P^a$'s and also between $\ps Y^I$'s coming 
from  $H_1(\sigma_L)$ and $H_1(\sigma_R)$. If we think of them as vertex insertions this is the propagator in the presence of the slit which 
has singularities. In the two vertex state formalism what we want to compute is for example
\beq
 \cA^i_s \cA^i_r e^{\Delta_B} \ket{0} .
\eeq
 We can commute the annihilation operators in the $\cA$'s through $e^{\Delta_B}$ which is, in fact, the calculation we did to obtain the divergencies.
However when we apply the second $\cA$, there are creation operators acting on $\ket{0}$ coming from applying the first $\cA$. The result
 is that the divergence is in fact:
\beq
 \cA^i_1 \cA^i_1 \sim \frac{1}{\sqrt{\sigma-\sz}}\frac{1}{\sqrt{-\sigma-\sz}}\left\{Z^I\bar{Z}^I -\frac{1}{32\pi^2}\frac{1}{\sin\sz}\right\} .
\eeq
Except for this subtlety, the rest amounts simply to replacing the operators by their divergencies to obtain:
\beq
\Pp = \NO{\hat{H} \Pp_S} ,
\label{PNOa}
\eeq
with
\beqa
 \hat{H} &=& \left( Z^L + \frac{i}{\sqrt{2}} \rho^I_{AB}Z^IY^AY^B-4 Z^RY^4\right)
             \left( \bar{Z}^L - \frac{i}{\sqrt{2}} \rho^J_{AB}\bar{Z}^I\bar{Y}^A\bar{Y}^B-4 \bar{Z}^R\bar{Y}^4\right) \nonumber\\
         &&  +\frac{1}{8\pi^2} \frac{1}{\sin\sz} \left(Y^4+\bar{Y}^4+\frac{1}{4}\epsilon_{ABCD}Y^AY^B\bar{Y}^C\bar{Y}^D\right) ,
\label{PNO}
\eeqa 
which is a very useful form of $\Pp$. We remind the reader of the notation $Y^4=\frac{1}{24}\epsilon_{ABCD}Y^AY^BY^CY^D$.
 
 As a final point, for later use, we emphasize that all the ideas described in this section fix $\Pp$ up to an overall constant that we 
are not able to compute. 

\section{Scattering of massless strings from D-branes, a check of $\Pp$}
\label{Dscattering}

 The operator $\Pp_S$ has the physical interpretation of describing the scattering of a closed string in an arbitrary state from a D3-brane.
This is a by product of our computation, namely a closed form for the scattering of a generic closed string state from a D3-brane.
Usually, one is interested only in the scattering of massless modes which has been computed 
in \cite{Klebanov:1995ni,MG}\footnote{See also \cite{Chan:2006qf} 
for some recent work on the subject.}. The relation between $\Pp$ and scattering off D-branes follows from the diagram in fig.\ref{Dscat}. 
It describes the free propagation of a closed string from $\tau=-\infty$ to $\tau=0$ at which time, the operator $\Pp$ is applied. After 
that, the closed string propagates freely again. Therefore, if the initial and final states are eigenstates of the Hamiltonian, the diagram 
is proportional to the matrix element of $\Pp$ between those two states. On the other hand, the diagram can be conformally mapped to an annulus 
with two closed string vertex insertions which is the more standard way of computing scattering from D-branes. Since the scattering of 
massless states is known, it is useful to recompute it with the operator $\Pp$, as a check. In the vertex representation, we should 
sandwich the vertex state with the vacuum of the oscillators. If we do that all terms containing creation operator cancel. In particular, 
in the exponent only the bosonic part gives a contribution which reduces to (see also \pone):
\beq
 \Delta_B = \sum_{rs,imn} N^{rs}_{i,00} a_{ir0}^\dagger a_{is0}^\dagger  = q^2 \ln\sin\frac{\sz}{2}+ 4k^2\ln\cos\frac{\sz}{2}   .
\eeq 
 The operator insertions also reduce to their zero modes namely:
\beqa
Z^I & \rightarrow & Z^I_0 = \frac{i\sqrt{2}}{4\pi} \frac{\cos\frac{\sz}{2}}{\sqrt{\sin\sz}} q^I ,\\
Z^{L,R} & \rightarrow & Z^{L,R}_0 = \frac{2\sqrt{2}}{4\pi} \frac{\sin\frac{\sz}{2}}{\sqrt{\sin\sz}} k^{L,R} ,\\
Y^A &\rightarrow & Y^A_0 = \frac{1}{2\sqrt{\sin\sz}} y^A, \ \ \ \mbox{with} \ \ \ y^A=\bar{a}_0^A\sin\frac{\sz}{2}+i\bar{b}_0^A\cos\frac{\sz}{2}.
\label{ydef}
\eeqa
 With that, the operator insertion (\ref{PNO}) reduces to:
\beqa
\hat{H}_{\scriptstyle \mbox{zero modes}} &=& \frac{1}{\sin\sz}\left(\frac{1}{\pi\sqrt{2}}\sin\frac{\sz}{2} k^L 
  -\frac{1}{16\pi}\frac{\cos\frac{\sz}{2}}{\sin\sz}q^I\rho^I_{AB} y^Ay^B
  -\frac{1}{4\pi\sqrt{2}}\frac{\sin\frac{\sz}{2}}{\sin^2\sz}k^R y^4 \right) \nonumber\\
 && \times \left(\frac{1}{\pi\sqrt{2}}\sin\frac{\sz}{2} k^L 
  -\frac{1}{16\pi}\frac{\cos\frac{\sz}{2}}{\sin\sz}q^I\rho^I_{AB} \bar{y}^A\bar{y}^B
 -\frac{1}{4\pi\sqrt{2}}\frac{\sin\frac{\sz}{2}}{\sin^2\sz}k^R \bar{y}^4 \right) \nonumber\\
&&  +\frac{1}{2^7\pi^2\sin^3\sz}\left(y^4+\bar{y}^4+\frac{1}{4}\epsilon_{ABCD}y^Ay^B\bar{y}^C\bar{y}^D\right) \label{HZM} .
\eeqa
 Now we should expand in terms of $\bar{a}_0^A$, $\bar{b}_0^A$ and do the integrals over $\sz$. This is a lengthy calculation that uses the
identities listed in the appendix for the $\rho^I_{AB}$ matrices. The result is better written classified by the number of fermionic operators contained:
\beqa
H^{(0)}_{[0]} &=& -\frac{1}{8\pi^2}\,s\, k^Lk^L\, A(s,t), \label{HZM0}\\
H^{(0)}_{[2]} &=& -\frac{1}{2^6\pi^2\sqrt{2}} k^Lq^I\rho^I_{AB}\left(t\,\bar{b}_0^A\bar{b}_0^B-s\,\bar{a}_0^A\bar{a}_0^B\right)\,A(s,t), \label{HZM2}\\
H^{(0)}_{[4]} &=& -\frac{1}{2^{8}\pi^2}\left\{\bar{a}_0^4s^2+\bar{b}_0^4 t^2
                  -\frac{t}{4}q^I\rho^I_{AB}\bar{a}_0^A\bar{a}_0^Bq^J\rho^I_{CD}\bar{b}_0^C\bar{b}_0^D\right\}\,A(s,t), \label{HZM4}\\
H^{(0)}_{[6]} &=& -\frac{1}{2^8\pi^2\sqrt{2}} k^Rq^I\rho^I_{AB}
                   \left(t\,\bar{a}_0^A\bar{a}_0^B\,\bar{b}_0^4-s\,\bar{b}_0^A\bar{b}_0^B\,\bar{a}_0^4\right)\,A(s,t), \label{HZM6}\\
H^{(0)}_{[8]} &=&  -\frac{1}{2^7\pi^2}\,s\,k^Rk^R\,\bar{a}_0^4\,\bar{b}_0^4\,A(s,t) \label{HZM8} ,
\eeqa
 which summarizes the scattering of massless modes from the D3-brane. Let us however explain the notation: $q^I=q_1+q_2$ is the total momentum 
transfer to the D3-brane, $k_1=-k_2$ is the conserved parallel momentum. We defined $s=-q^2$ and $t=-4k^2=-8k^Lk^R$ and also introduced the function
\beq
A(s,t)=A(q^2,k^2) = \frac{\Gamma\left(2k^2\right)\Gamma\left(\frac{q^2}{2}\right)}{\Gamma\left(2k^2+\frac{q^2}{2}+1\right)} ,
\eeq  
which comes from the integral in $\sz$. The result, particularly for $H^{(0)}_{[4]}$ was simplified using identities between Euler's $\Gamma$ functions.  
To compare with other calculations we should insert polarizations states. For example for transverse polarizations we should compute:
\beq
\varepsilon^{(1)}_{IJ}\,  \varepsilon^{(2)}_{KL}\, 
\bra{0}  \rho^{IAB}\, \rho^{JCD} \lambda_{1A0}\lambda_{1B0} \tlambda_{1C0}\tlambda_{1D0} \,
         \rho^{IEF} \rho^{JGH} \lambda_{2E0}\lambda_{2F0} \tlambda_{2G0}\tlambda_{2H0}\,  H^{(0)}_{[4]} \,\beta^4\, \ket{0} ,
\eeq
remembering that $\bar{a}^A_0=\frac{1}{\sqrt{2}}\left(\theta^A_{10}-\ttheta^A_{20}-\ttheta^A_{10}+\theta_{20}\right)$ and
$\bar{b}^A_0=\frac{1}{\sqrt{2}}\left(\theta^A_{10}-\ttheta^A_{20}+\ttheta^A_{10}-\theta_{20}\right)$.  We also defined the vacuum of the zero modes as:
\beq
 \lambda_{rA0} \ket{0}=0, \ \ \tlambda_{rA0}\ket{0}=0 ,
\eeq
so that massless polarization states are created from the vacuum by the $\lambda$'s. This is not the same zero mode state that enters in the 
definition of $\Pp$. The only difference is that the latter is annihilated by $\beta=\bar{\chi}_0-\Xi_0$ whereas the vacuum of the $\lambda$'s is not.
For that reason, the factor $\beta^4$ appears mapping one vacuum to the other. For the calculation we should use 
\beq
\beta = \frac{1}{2}\left(-\theta^A_{10}-\ttheta^A_{20}+\ttheta^A_{10}+\theta_{20}\right) ,
\eeq
and expand everything in powers of the $\theta$'s.  Contracting each term with the corresponding $\lambda$'s we obtain a 
result that agrees perfectly with \cite{Klebanov:1995ni,MG}. In fact, it is a very useful check since it depends on many 
details of the previous calculations. The full calculation for the scattering of NSNS massless modes is done in appendix \ref{scattering}. 

\FIGURE{
\includegraphics[height=6cm]{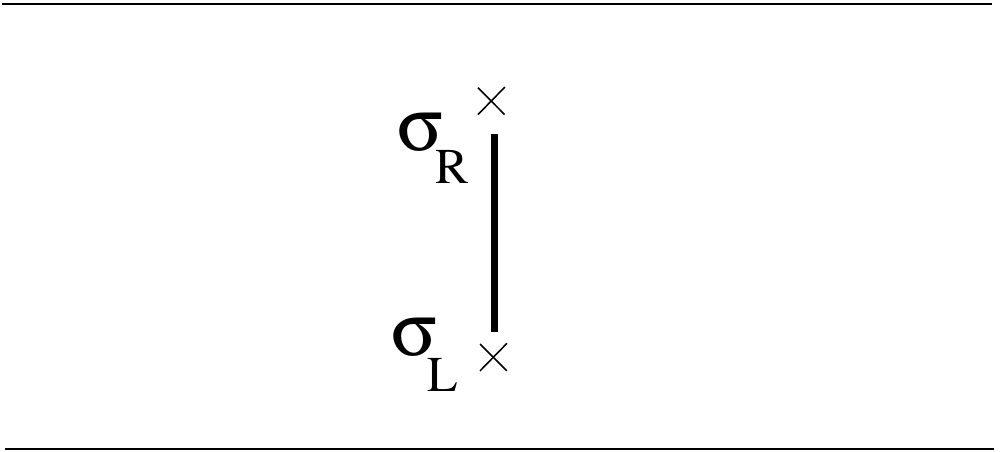}
\caption{The slit insertion operator can be used to compute scattering of closed strings from D-branes in the Green-Schwarz formalism.}
\label{Dscat}
}

\section{The limit of small holes}
\label{small holes}

When the holes become small they can be replaced by an insertion of a local operator that we compute here. In order to do so, we use
the properties of the Neumann coefficients to obtain the small $\sz$ expansions of the operators:
\beqa
Z^I &=& \frac{i\sqrt{2}}{4\pi\sqrt{\sin\sz}} \left\{q^I+\frac{i\sz}{2}\ps Y^I - \frac{\sz^2}{8} q^I + \frac{i\sz^2}{4}\ps^2Y^I
              +\frac{i\sz^3}{8}\ps^3Y^I - \frac{i\sz^3}{48} \ps Y^I \ldots\right\} \nonumber ,\\
Z^{L,R} &=& \frac{\sz}{\sqrt{2\sin\sz}}\left\{ \Pi^{L,R}+\frac{\sz}{2}\ps \Pi^{L,R} + \frac{\sz^2}{4} \ps^2 \Pi^{L,R} 
              - \frac{\sz^2}{24}\Pi^{L,R} +\ldots \right\} ,\\
Y^A &=& \frac{1}{\sqrt{\sin\sz}} \left\{ \frac{i}{2}\bar{b}^A_0 + \frac{\sz}{2} \Theta^A - \frac{i}{16}\sz^2 \bar{b}_0^A 
        +\frac{\sz^2}{4} \ps \Theta^A - \frac{\sz^3}{48} \Theta^A + \frac{\sz^3}{8}\ps^2\Theta^A + \ldots \right\} \nonumber .
\eeqa
Replacing in $\hat{H}$ and keeping the most singular terms as $\sz\rightarrow 0$, we get:
\beqa
 \hat{H} &\simeq_{\sz\rightarrow 0}& 
             - \frac{1}{32\pi^2} \frac{1}{\sz^3} (q^2-2) \bar{b}_0^4 - \frac{1}{8\sz} \Pi^a\Pi^a \bar{b}_0^4 
             - \frac{i}{2^6\pi^2\sz} \bar{b}_0^4 q^I \ps^2 Y^I - \frac{1}{2^7\pi^2\sz}\bar{b}^4_0 \ps Y^I\ps Y^I  \nonumber \\
          && +\frac{1}{3\,2^7\pi^2\sz}\rho^{IFA}\rho^J_{AB} \epsilon_{PQRF}\bar{b}_0^P\bar{b}_0^Q\bar{b}_0^R \Theta^B
                \left(\ps Y^I q^J- \ps Y^J q^I \right) \nonumber\\
          && -\frac{1}{2^6\pi^2\sz} q^Iq^J\rho^I_{AB}\rho^J_{CD} \bar{b}^A_0 \bar{b}^B_0 \Theta^C \Theta^D 
             - \frac{i}{3\,2^7\pi^2\sz} q^I q^J \rho^{IAF} \rho^J_{AB} \epsilon_{PQRF} \bar{b}_0^P\bar{b}_0^Q\bar{b}_0^R \ps\Theta^B \nonumber \\
          && -\frac{i}{3\,2^6\pi^2\sz} \epsilon_{ABCD} \bar{b}_0^A\bar{b}_0^B\bar{b}_0^C \ps\Theta^D 
             + \frac{1}{\sz}\frac{\sqrt{2}}{16\pi} \left[ \Pi^L q^I \rho^I_{AB} \bar{b}^A_0 \bar{b}^B_0 
                                   +  \Pi^R q^I \rho^I_{AB} \Theta^A\Theta^B \bar{b}_0^4 \right] \nonumber . \\
\eeqa
 Note that the first term has a cubic divergence $\frac{1}{\sz^3}$ typical of the tachyon since it gives rise to a pole at $q^2=2$. However
here it is precisely multiplied by $(q^2-2)$ so the residue of the pole is zero and the tachyon poles cancels. Note however that near $q^2=0$,
the term $\frac{1}{\sz^3}$ combines with order $\sz^2$ terms coming from the exponent to give a $\frac{1}{\sz}$ pole. Some of these terms are
precisely the spurious terms that were present in the bosonic calculation of \pone\ and that, as we will see cancel against contributions 
from the insertions. 
 To check that, we need to expand the exponent to order $\sz^2$. The result is
\beqa
\Delta_B+\Delta_F &\simeq_{\sz\rightarrow 0}& q^2 \ln\sz + i q^I Y^I_{NZ} -\frac{\sz^2}{8} \ps Y^I \ps Y^I + \frac{i\sz^2}{4} q^I \ps^2 Y^I \nonumber\\
                 &&  -2\pi^2\sz^2 \Pi^a\Pi^a + 2\pi\left(\frac{i\sz^2}{2}\Theta\ps\bar{\Lambda} 
                   -\bar{b}_0^4 \bar{\Lambda}_{NZ}-\frac{\sz^2}{4} \bar{b}_0^A\ps^2\bar{\Lambda}_A \right) .
\eeqa
where the subindex $NZ$ in $Y^I_{NZ}$ and $\bar{\Lambda}_{NZ}$ indicates the oscillator part of the corresponding operator, \ie\ without
the zero mode. Expanding the exponential and combining with the expansion of $\hat{H}$ we get a pole $\frac{1}{\sz}$ in sigma. 
The integral of which is
\beq
\int_0^\epsilon \sz^{q^2-1} = \frac{\epsilon^{q^2}}{q^2} \sim \frac{1}{q^2} , \ \ \ (q^2\rightarrow 0).
\eeq
For this reason, small holes dominate as $q^2\rightarrow 0$, namely $q^2\ll 1$ in string units. 
As discussed before we still have to do the Fourier integrals:
\beq
 \int \frac{d^6 q}{(2\pi)^6}  e^{iq^I y^I} \int \frac{1}{4} d^4 \bar{b}_0 e^{-\bar{b}_0^A \bar{\Lambda}_{A0}} \times F(q^J,\bar{b}_0^A),
\eeq
where $F$ represents the result of the calculation we just did and the factors $\frac{1}{4}$ and $\frac{1}{(2\pi)^6}$ are the same we already
discussed (see appendix \ref{scattering}). The integrals in $q$ are straight-forward: 
\beq
 \int \frac{d^6 q}{(2\pi)^6} \frac{1}{q^2} e^{iqy} = \frac{1}{4\pi^3} \frac{1}{y^4}, \ \ \ \ 
 \int \frac{d^6 q}{(2\pi)^6} \frac{q_J}{q^2} e^{iqy} = \frac{1}{\pi^3} \frac{iy_J}{y^6}, \ \ \ \  
\eeq
The integral in $\bar{b}_0^A$ is done according to the formulas:
\beqa
\int d^4 \bar{b}_0\, e^{\bar{b}_0^A\xi_A } &=& \xi^4 ,\\ 
\int d^4 \bar{b}_0\, e^{\bar{b}_0^A\xi_A }\, \bar{b}_0^A &=& -\frac{1}{6} \epsilon^{ABCD} \xi_B\xi_C\xi_D ,\\ 
\int d^4 \bar{b}_0\, e^{\bar{b}_0^A\xi_A }\, \bar{b}_0^A \bar{b}_0^B&=& -\half \epsilon^{ABCD} \xi_C\xi_D   ,\\ 
\int d^4 \bar{b}_0\, e^{\bar{b}_0^A\xi_A }\, \bar{b}_0^A \bar{b}_0^B \bar{b}_0^C&=& \epsilon^{ABCD} \xi_D  ,\\ 
\int d^4 \bar{b}_0\, e^{\bar{b}_0^A\xi_A }\, \bar{b}_0^A\bar{b}_0^B\bar{b}_0^C\bar{b}_0^D&=& \epsilon^{ABCD} .
\eeqa
In particular $\int d^4 \bar{b}_0\, \bar{b}_0^4 =1$. After a straight-forward and not so lengthy calculation we obtain for $\Pp$
in the limit $\sz\rightarrow 0$:
\beqa
  \frac{16\pi^3}{\alpha_3} \Pp \simeq \bbH &=&  -\frac{1}{4}\frac{1}{Y^4} \Pi^a\Pi^a - \frac{1}{2^6\pi^2}\frac{1}{Y^4}\ps Y^I\ps Y^I
                      + \frac{i}{16\pi Y^4} \left(\Theta^A\ps\bar{\Lambda}_A+\bar{\Lambda}_A\ps\Theta^A\right) \nonumber \\
   && -i\sqrt{2}\pi\frac{Y^I}{Y^6}\Pi^L\rho^{ICD} \bar{\Lambda}_C\bar{\Lambda}_D + \frac{i\sqrt{2}}{4\pi} \frac{Y^I}{Y^6}  
        \Pi^R \rho^I_{AB}\Theta^A\Theta^B  \nonumber \\
   && -\frac{i}{8\pi} \rho^{IAC}\rho^J_{CB} \bar{\Lambda}_A\Theta^B \frac{1}{Y^6} \left(\ps Y^I Y^J-\ps Y^J Y^I\right) \nonumber\\
   && +\frac{1}{4Y^6} \left(\delta^{IJ}-6\frac{Y^IY^J}{Y^2}\right) \rho^{IAB}\rho^J_{CD} \bar{\Lambda}_A\bar{\Lambda}_B \Theta^C\Theta^D . 
\label{Hfinal}
\eeqa
As a check of the calculation we can compute for example $[\bbQ_{-A},\bbH]=0$. It is also useful to check that $\bbH$ is hermitian.
We define the following hermiticity relations:
\beq
\left(\Pi^L\right)^\dagger = \Pi^R, \ \ \ \left(\bar{\Lambda}_D\right)^\dagger= \frac{1}{2\pi}\Theta^D ,
\eeq
in a basis where the matrices $\rho^I_{AB}$ are unitary, \ie\ $\left(\rho^{IAB}\right)^*=\rho^I_{BA}$ (which means that the corresponding 
Dirac matrices are hermitian). If we write now the full Hamiltonian describing the propagation of the closed string 
in the $\sigma\leftrightarrow\tau$ channel we have:
\beq
H_{\left[\scriptstyle D3\ bkg.\right]} = H_0 -\lambda \frac{\alpha_3}{16\pi^3} \bbH = H_0-2^7\pi^2\lambda \bbH,
\label{Htotal}
\eeq
where we used that $\alpha_3=2^{11}\pi^5$ as determined in appendix \ref{scattering}, eq.\eqn{a31}.  
 The value of $H_0$ was given in eq.(\ref{HP}) but it is convenient to rewrite it in terms of the variables we are using now:
\beqa
H_0 &=& 2\pi\int d\sigma\, \left(\Pi_X^2+\Pi_Y^2+\frac{1}{16\pi^2}\left(\ps X\right)^2+\frac{1}{16\pi^2}\left(\ps Y\right)^2\right) \nonumber\\
    & & +i\int d\sigma\,\left(\ps\Lambda\bar{\Theta}+\ps\bar{\Lambda}\Theta\right) ,
\label{H0b}
\eeqa
where we use the definitions (\ref{Tdef}). The bosonic part of $H$, including the first two terms of $\bbH$, is the the Hamiltonian describing the 
propagation of a closed string in the full D3-brane background in $\sigma$-gauge as computed in \pone. We propose that the full $H$ is the
Hamiltonian for closed strings in the D3-background in this particular gauge. To our knowledge, the fermionic part was not known. It might
seem strange that $H$ is linear in $\lambda$ but that is a feature of the $\sigma$ gauge as explained in \pone.  
 
Thus, we see that the full supergravity background has emerged from the open string calculation. 
We also emphasize that the operator $H$ we found is a full quantum operators which should be understood in normal ordered form. 

It is interesting now to take the near horizon limit. Formally we rescale:
\beq
\begin{array}{ccccccc}
X^a\rightarrow\frac{1}{\xi}X^a,  & \ \ \ & \Pi_X^a\rightarrow \xi\, \Pi_X^a, 
   & \ \ \ & Y^I\rightarrow \xi\, Y^I,& \ \ \ & \Pi^I\rightarrow \frac{1}{\xi}\Pi^I \\ &&&&&&\\ 
\Theta\rightarrow \xi\,\Theta, && \Lambda\rightarrow\frac{1}{\xi}\,\Lambda,&&\bar{\Theta}\rightarrow \frac{1}{\xi}\,\bar{\Theta}
&& \bar{\Lambda} \rightarrow \xi\,\bar{\Lambda} ,
\end{array}
\eeq
preserving the canonical commutation relations. Quite interestingly, under this rescaling, all the terms in $\bbH$ scale as $\frac{1}{\xi^2}$. 
However for $H_0$ we get:
\beqa
H_0&\rightarrow& 2\pi\int d\sigma\, \left(\xi^2\Pi_X^2+\frac{1}{\xi^2}\Pi_Y^2
                   +\frac{1}{16\pi^2}\frac{1}{\xi^2}\left(\ps X\right)^2+\frac{\xi^2}{16\pi^2}\left(\ps Y\right)^2\right) \nonumber\\
    & & +i\int d\sigma\,\left(\frac{1}{\xi^2}\ps\Lambda\bar{\Theta}+\xi^2\ps\bar{\Lambda}\Theta\right) .
\eeqa
 Now we take the limit $\xi\rightarrow 0$. Naively, in this limit we would drop terms such as $\frac{\xi^2}{16\pi^2}\left(\ps Y\right)^2$ but 
in fact the derivative can be as large as we want so that would not be correct. If we look more carefully, however, there is also a term
$(\ps Y)^2$ in $\bbH$ that goes as $\frac{1}{\xi^2}$. Therefore in the limit we keep the term in $\bbH$ and discard the one in $H_0$. 
 The result is that in the near horizon limit the Hamiltonian reduces to:
\beqa
H_{\left[\scriptstyle AdS_5\times S^5\right]} 
   &=& 2\pi\int d\sigma\, \left(\Pi_Y^2+\frac{1}{16\pi^2}\left(\ps X\right)^2\right)+i\int d\sigma\, \ps\Lambda\bar{\Theta} \nonumber \\
   &&+ 32\pi^2 \lambda \int d\sigma\, \left\{\frac{1}{Y^4} \Pi^a\Pi^a + \frac{1}{16\pi^2}\frac{1}{Y^4}\ps Y^I\ps Y^I
                      - \frac{i}{4\pi Y^4} \left(\Theta^A\ps\bar{\Lambda}_A+\bar{\Lambda}_A\ps\Theta^A\right) \right. \nonumber \\
   && +i4\sqrt{2}\pi\frac{Y^I}{Y^6}\Pi^L\rho^{ICD} \bar{\Lambda}_C\bar{\Lambda}_D - \frac{i\sqrt{2}}{\pi} \frac{Y^I}{Y^6}  
        \Pi^R \rho^I_{AB}\Theta^A\Theta^B \nonumber\\
   && +\frac{i}{2\pi} \rho^{IAC}\rho^J_{CB} \bar{\Lambda}_A\Theta^B \frac{1}{Y^6} \left(\ps Y^I Y^J-\ps Y^J Y^I\right) \nonumber \\
   && \left.-\frac{1}{Y^6} \left(\delta^{IJ}-6\frac{Y^IY^J}{Y^2}\right) \rho^{IAB}\rho^J_{CD} \bar{\Lambda}_A\bar{\Lambda}_B \Theta^C\Theta^D \right\} .
\label{ads55}
\eeqa
 The bosonic part of this Hamiltonian, including the overall normalization\footnote{As we recall, the normalization was determined by comparing with 
the supergravity background so it is a consistency check rather than a prediction from the open string side.}, exactly agrees with the Hamiltonian 
of closed strings in \adss{5}{5}. Again we propose that the complete $H$ describes strings in \adss{5}{5}. Although we have not checked it, we expect 
the result to agree with the Hamiltonian derived from the Metsaev-Tseytlin action \cite{Metsaev:1998it} after some appropriate 
$\kappa$-symmetry fixing and after taking $\sigma$-gauge.  Note however that here we derived the result by analyzing planar diagrams 
in the open string theory without any reference to \adss{5}{5} or any supergravity background for that matter. 

 Note also that when taking the limit $\xi\rightarrow 0$ the final Hamiltonian scaled as $\xi^{-2}$. This is fine because, in the evolution operator
$U=e^{-H\tau}$, $\tau$ scales as $\xi^2$ and therefore $U$ is invariant. To see that, recall that $\tau\sim p^+$ in the closed string channel
and we should rescale $p^+\rightarrow \xi p^+$ since $p^+$ is a momentum parallel to the brane.  On top of that we have that, in $\sigma$-gauge,
$X^+=\sigma$ so we should rescale $\sigma \rightarrow \frac{1}{\xi}\sigma$. Since we want $\sigma$ to run from $0$ to $2\pi$ we do a conformal
transformation $(\sigma,\tau)\rightarrow (\xi\sigma,\xi\tau)$ so that $\sigma$ remains invariant and $\tau\rightarrow \xi^2\tau$ as mentioned before.   

 We can also be more precise in the region of validity of our result.
 When deriving the Hamiltonian we consider small holes which dominate in the limit $q^2\rightarrow 0$. More precisely, we require 
$q^2\ll 1$ in string units which is equivalent to $Y^2\gg 1$. After that we want some of the terms in $\bbH$ to dominate those in $H_0$. 
This happens if  $Y^2 \ll \sqrt{\lambda}$, therefore we need
\beq
 1 \ll Y^2 \ll \sqrt{\lambda} ,
\eeq
to recover strings in \adss{5}{5}. This implies $\lambda\gg 1$, namely a strong coupling limit. This, however, is not the decoupling limit
of Maldacena which is taken at $Y^2\ll 1$. In fact the throat region is the relevant region for the double scaling limit proposed by I. Klebanov
and further studied in \cite{Klebanov:1997kc}. The work presented here might help to illuminate that.  On the other hand, if one tries to derive 
the AdS/CFT correspondence with this approach, further work would be needed to understand the region $Y^2\ll 1$. In that region, both, small and
large holes appear to be important. From our perspective, the AdS/CFT correspondence indicates that this should not be true, namely the fact that
we have the same \adss{5}{5} background in the region $Y^2\ll1$ suggest that even there, all the contribution comes from small holes.

\section{Comments on applications to field theory}
\label{field theory}

 We have discussed how to sum planar diagrams for open superstrings. It would be interesting to apply the same ideas to sum the planar diagrams
of a field theory with fields in the adjoint. We gave some ideas to that respect in paper \pone\ and here we continue to study such matter. 
 However this section is mainly speculative and outside the main line of development of the paper. 

\FIGURE{
\includegraphics[height=11cm]{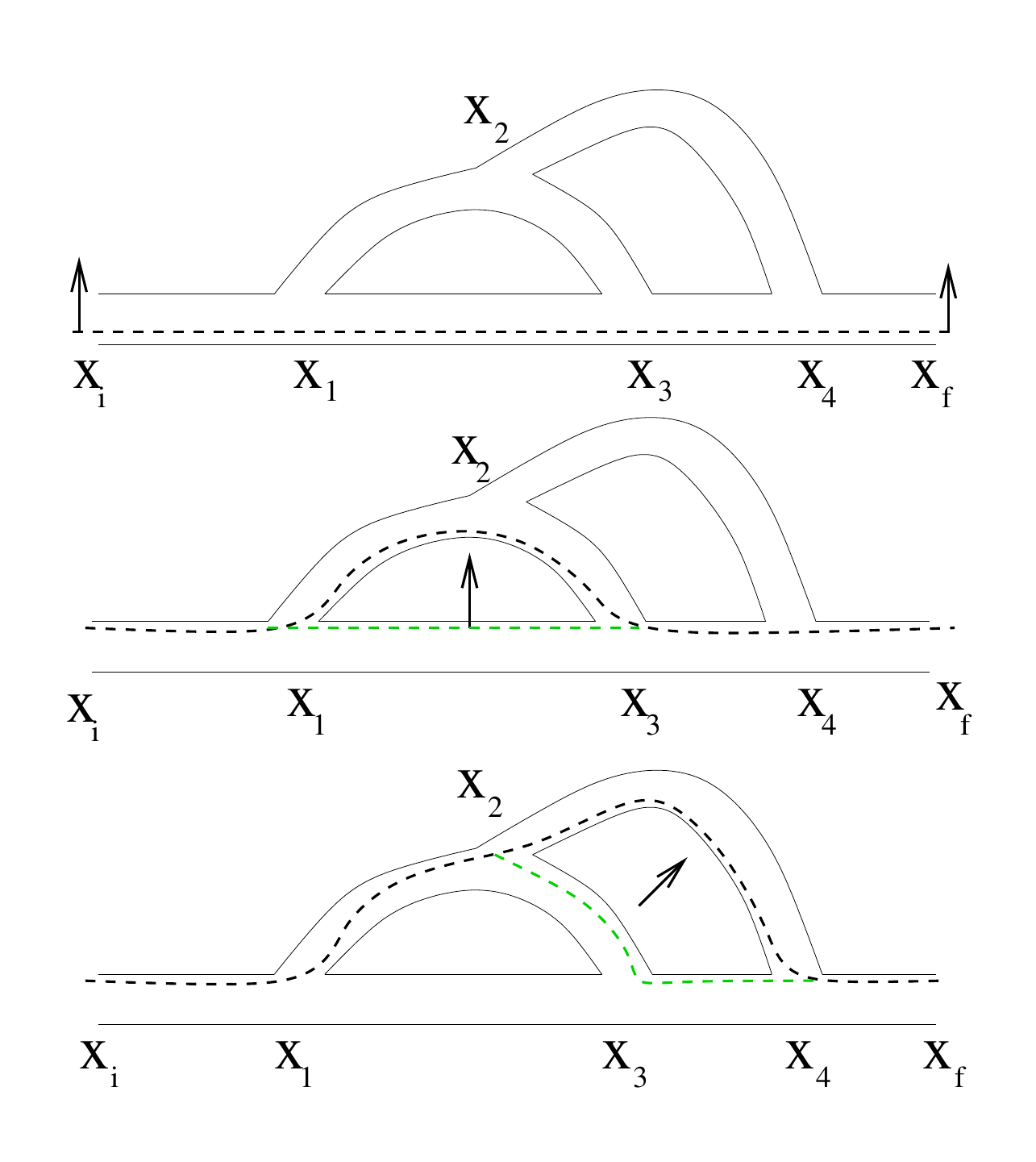}
\caption{Two loop planar Feynman diagram in coordinate representation and double line notation. The dashed line indicates a string 
whose shape is the same as the trajectory of the particle. The state of the string changes suddenly every time we cross a loop. 
The change is equivalent to applying the loop insertion operator $\Pp$ to the string state.}
\label{fd1}
}

 Consider we want to compute a Feynman diagram such as the one in fig.\ref{fd1} which is in the usual coordinate representation, not 
in light-cone frame. We argue that it can be computed by considering a string whose shape is the trajectory of the particle and which evolves
in discreet steps across the diagram. The evolution acts whenever the string crosses a loop as is indicated in the figure. Note that such
description is only possible if the diagram is planar, otherwise we cannot get unique intermediate states for the shape of the string. 

To be more specific, let us look at the simpler case of fig.\ref{fd2}. That diagram is given by   
\beq
A =  \int d^d x_2 d^dx_3 \frac{1}{|x_1-x_2|^{d-2}}\frac{1}{|x_2-x_3|^{d-2}}\frac{1}{|x_2-x_3|^{d-2}}\frac{1}{|x_3-x_4|^{d-2}} .
\eeq
 An alternative expression for the propagators is obtained through
\beq
\int_0^\infty d\bar{\sigma} \frac{1}{\bar{\sigma}^{\frac{d}{2}}} e^{-\half\frac{(x_1-x_2)^2}{\bar{\sigma}}}  
   = \frac{2^{\frac{d}{2}-1}\,\Gamma\left(\frac{d}{2}-1\right)}{|x_1-x_2|^{d-2}} .
\eeq
The integrand can be written as a path integral using 
\beq
 \int_{\begin{array}{l} \scriptstyle X(0)=x_1 \\ \scriptstyle X(\bar{\sigma})=x_2 \end{array} } \cD X(\sigma) \ 
      e^{-\half\int_{0}^{\bar{\sigma}} (\ps X)^2 d\sigma} = 
      \frac{1}{\bar{\sigma}^{\frac{d}{2}}} e^{-\frac{(x_1-x_2)^2}{2\bar{\sigma}}} .
\eeq
 Suppose we now consider an open string whose states are given by its shape in a given parameterization: $\ket{X(\sigma),\bar{\sigma}}$, namely 
the shape is characterized by a function $X(\sigma)$ with $0\le\sigma\le\bar{\sigma}$.  The states are orthogonal, namely
\beq
\bra{X_1(\sigma),\bar{\sigma}_1} X_2(\sigma), \bar{\sigma}_2\rangle = \delta(\bar{\sigma}_1-\bar{\sigma}_2)
                    \prod_{0<\sigma<\bar{\sigma}_1} \delta\left(X_1(\sigma)-X_2(\sigma)\right).
\eeq
Define now the ``boundary'' state:
\beq
 \ket{x_1,x_2,\bar{\sigma}} = \int_{\begin{array}{l} \scriptstyle X(0)=x_1 \\ \scriptstyle X(\bar{\sigma})=x_2 \end{array} }  
          \cD X(\sigma) e^{-\frac{1}{4}\int_{0}^{\bar{\sigma}} (\ps X)^2 d\sigma} \ket{X(\sigma),\bar{\sigma}},
\eeq
 which is not normalized, in fact its norm is 
\beq
 \bra{x_1,x_2,\bar{\sigma}} x_1,x_2,\bar{\sigma}\rangle =  
   \int_{\begin{array}{l} \scriptstyle X(0)=x_1 \\ \scriptstyle X(\bar{\sigma})=x_2 \end{array} }
   \cD X(\sigma) \ e^{-\half \int_{0}^{\bar{\sigma}} (\ps X)^2 d\sigma} = 
      \frac{1}{\bar{\sigma}^{\frac{d}{2}}} e^{-\frac{(x_1-x_2)^2}{2\bar{\sigma}}} ,
\eeq
in such a way that
\beq
 \int_0^{\infty} d\bar{\sigma} \bra{x_1,x_2,\bar{\sigma}} x_1,x_2,\bar{\sigma}\rangle = 
   \frac{2^{\frac{d}{2}-1}\,\Gamma\left(\frac{d}{2}-1\right)}{|x_1-x_2|^{d-2}} .
\eeq
 Let us further define a tensor product between the states of the string such that
\beq
 \int d^d x \ket{x_1,x,\sigma_1} \otimes \ket{x,x_2,\bar{\sigma}-\sigma_1} = \ket{x_1,x_2,\bar{\sigma}},
\label{tp}
\eeq
 that is, we glue the two paths, using that the actions add up. We can now write the diagram as
\beqa
 A &=& \lambda^2 \int d^d x_2d^dx_3 \int d\bar{\sigma}_1  d\bar{\sigma}_2 d\bar{\sigma}_3 d\bar{\sigma}_4
  \bra{x_1,x_2,\bar{\sigma}_1} x_1,x_2,\bar{\sigma}_1\rangle \bra{x_2,x_3,\bar{\sigma}_2} x_2,x_3,\bar{\sigma}_2\rangle \nonumber \\
  &&\ \ \ \ \ \ \ \  \times\ \bra{x_2,x_3,\bar{\sigma}} x_2,x_3,\bar{\sigma}_3\rangle \bra{x_3,x_4,\bar{\sigma}_4} x_1,x_2,\bar{\sigma}\rangle  \\
  &=& \lambda^2 \int d^d x_2d^dx_3 \int d\bar{\sigma}_1 d\bar{\sigma}_2 d\bar{\sigma}_3 d\bar{\sigma}_4
        \left(\rule{0pt}{18pt} \bra{x_1,x_2,\bar{\sigma}_1}\otimes\bra{x_2,x_3,\bar{\sigma}_3}\otimes\bra{x_3,x_4,\bar{\sigma}_4}\right) \nonumber \\
     && \left(\rule{0pt}{18pt} \Id \otimes \ket{x_2,x_3,\bar{\sigma}_3}\bra{x_2,x_3,\bar{\sigma}_2}\otimes\Id\right) 
        \left(\rule{0pt}{18pt} \ket{x_1,x_2,\bar{\sigma}_1}\otimes\ket{x_2,x_3,\bar{\sigma}_2}\otimes\ket{x_3,x_4,\bar{\sigma}_4}\right) , \nonumber
\eeqa
where we considered the initial and final strings divided in three pieces of which we should glue the pieces at both ends as
indicated by the identities in the intermediate operator and, for the middle piece, we should project both sides over the boundary state
as also indicated. Note that the pieces in the middle can have different lengths in $\sigma$. 

\FIGURE{
\includegraphics[width=12cm]{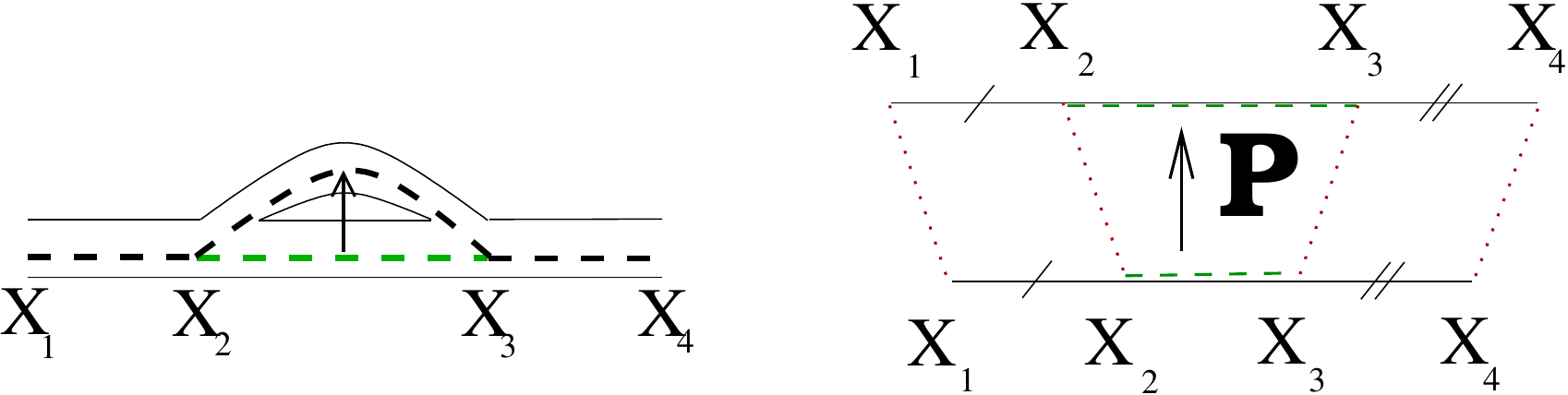}
\caption{One loop planar Feynman diagram in coordinate representation and double line notation. The dashed line indicates the string
as in fig.\ref{fd1}. On the right we draw the diagram as the propagation of a string with a discreet step given by $\Pp$. The left
and right pieces of the string are identified and the middle one is projected over the boundary state. The result is the same as the diagram
on the left.}
\label{fd2}
}

 As a last step, using the tensor product (\ref{tp}) we can write $A$ as:
\beq 
A =  \lambda^2  \int d\bar{\sigma}_1  d\bar{\sigma}_2 d\bar{\sigma}_3 d\bar{\sigma}_4
      \bra{x_1,x_4,\bar{\sigma}_f=\bar{\sigma}_1+\bar{\sigma}_3+\bar{\sigma}_4} \Pp(\bar{\sigma}_1,\bar{\sigma}_2,\bar{\sigma}_3) 
         \ket{x_1,x_4,\bar{\sigma}_i=\bar{\sigma}_4+\bar{\sigma}_1+\bar{\sigma}_2} ,
\eeq
 with 
\beq
 \Pp(\bar{\sigma}_1,\bar{\sigma}_2,\bar{\sigma}_3) = \Id 
                \otimes \ket{X(\bar{\sigma}_1),X(\bar{\sigma}_2),\bar{\sigma}_3}\bra{X(\bar{\sigma}_1),X(\bar{\sigma}_2),\bar{\sigma}_2}\otimes\Id .
\eeq
 Perhaps the notation is not very precise but the meaning is: we cut the string at the points $\sigma=\bar{\sigma}_1$ and $\sigma=\bar{\sigma}_2$.
 We get three pieces.  We leave the left and right pieces as they are but to the middle one we apply the operator 
$\ket{X(\bar{\sigma}_1),X(\bar{\sigma}_2),\bar{\sigma}_3}\bra{X(\bar{\sigma}_1),X(\bar{\sigma}_2),\bar{\sigma}_2}$. It is clear that
the result is the Feynman diagram that we want. If we define the operator 
\beq
 \Pp = \int d\bar{\sigma}_1  d\bar{\sigma}_2 d\bar{\sigma}_3 \Pp(\bar{\sigma}_1,\bar{\sigma}_2,\bar{\sigma}_3) ,
\eeq
then we have
\beqa
 A &=& \lambda^2  \int d\bar{\sigma}_1  d\bar{\sigma}_2 d\bar{\sigma}_3 d\bar{\sigma}_4
      \bra{x_1,x_4,\bar{\sigma}_f=\bar{\sigma}_1+\bar{\sigma}_3+\bar{\sigma}_4} \Pp(\bar{\sigma}_1,\bar{\sigma}_2,\bar{\sigma}_3) 
         \ket{x_1,x_4,\bar{\sigma}_i=\bar{\sigma}_4+\bar{\sigma}_1+\bar{\sigma}_2} \nonumber \\
   &=& \lambda^2  \int d\bar{\sigma}_1  d\bar{\sigma}_2 d\bar{\sigma}_3 d\bar{\sigma}_i 
      \bra{x_1,x_4,\bar{\sigma}_f=\bar{\sigma}_i+\bar{\sigma}_3-\bar{\sigma}_2} \Pp(\bar{\sigma}_1,\bar{\sigma}_2,\bar{\sigma}_3) 
         \ket{x_1,x_4,\bar{\sigma}_i} \\
   &=& \lambda^2 \int d\bar{\sigma}_f d\bar{\sigma}_i 
        \bra{x_1,x_4,\bar{\sigma}_f}  \Pp \ket{x_1,x_4,\bar{\sigma}_i} \nonumber .
\eeqa
 In this way we can write any planar Feynman diagram for the cubic theory in terms of multiple $\Pp$ insertions. We hope this representation is useful and
can be used to sum the planar diagrams of the theory but we leave the issue for future investigation. Here we just want to emphasize that similar
methods as the ones employed for open strings can also be discussed within a field theory. As mentioned before, they use in an essential way that 
the diagrams are planar so they capture an important property that they have, namely, that one can think of them as a string going across the 
diagram always in a well defined state.

\section{Conclusions}
\label{conclusions}

 In this paper we apply the method described in \pone\ (\ie\ \cite{planar}) to the planar diagrams of open superstrings propagating on a 
stack of $N$ D3-branes. We find that the sum of planar diagrams is described by the propagation of a closed string with a non-local Hamiltonian
$H$ which includes a hole insertion operator $\Pp$ that can be explicitly computed. The result is given in eqs.(\ref{Pdef}) and (\ref{H1insertion}), 
or equivalently, eqs.(\ref{PNOa}) and (\ref{PNO}). At distances from the D3-brane larger than a string length, $H$ reduces to the propagation 
of strings in the full D3-brane supergravity background in a particular gauge that we call $\sigma$-gauge and which was defined in \pone.  
 To our knowledge, this Hamiltonian, which is shown in eqs.(\ref{Htotal}), (\ref{H0b}) and (\ref{Hfinal}), is new since only the bosonic part 
was known before. In the near-horizon limit it reduces to the propagation of a closed string in \adss{5}{5} as shown in eq.(\ref{ads55}). 
This last Hamiltonian has a novel form although it should be equivalently derived from the Metsaev-Tseytlin action \cite{Metsaev:1998it}. We emphasize 
however that in both cases the important point is that we derived these Hamiltonians from the analysis of the open string planar diagrams without 
any reference whatsoever to the existence of the D3-brane supergravity
background. We also stress the fact that we can study the full non-local operator $H$ even when it does not have the nice interpretation
of a string in an external background. Properties of the planar diagrams are contained in properties of $H$ such as the spectrum, ground state 
existence of gap etc. Presumably a non-local $H$ is the general situation even for a field theory. 
 In the previous paper \pone\ some doubts were raised regarding possible higher order corrections in $\lambda$ to $\Pp$. In the 
supersymmetric case we saw no indication of such corrections. Divergences due to the tachyon are absent in the superstring. Furthermore,
at low energy the theory reduces to \N{4} SYM which is finite in light-cone gauge~\cite{Brink:1982pd}. Also, the supersymmetry algebra is such 
that no first order corrections in $\lambda$ are required for the conserved supercharges suggesting that no higher order corrections are needed 
for the Hamiltonian. The usual reasoning is that, since the supercharges anticommute to $H$ and $H$ has a term of order $\lambda$, then the
supercharges also should have terms of order $\lambda$ which, when anticommuted, will contribute to $H$ at order $\lambda^2$. In our case,
the supercharges anticommute to a translation in the world-sheet spacial direction and not to $H$. Therefore this reasoning does not apply. 
 To complement these ideas one should compute explicitly, for example, two loop diagrams and check that divergences are indeed absent. This
is outside the scope of the present paper but seems a feasible calculation. 

 One other thing we should emphasize is that scattering amplitudes can also be computed as discussed in \pone. In that case we have an infinitely
long open string that propagates. The hole insertion operator should work similarly. In particular for small holes there should be no
difference. 

 It should be interesting to understand the small holes in conformal gauge which might give a simpler way to compute $\Pp$. In that
gauge, however, we do not know how to argue that the sum of planar diagrams exponentiate as it does in light-cone gauge.  
 
 Note also that we map the open and closed string in a very precise way such that any calculation done with planar light-cone diagrams in the open 
string theory can be equivalently understood as a closed string calculation which obviously gives the same result.  

 The sum of planar diagrams for the open string includes the sum of planar diagrams for \N{4}. In this paper we do not study how to extract
such sum from the open strings although it can be argued that, after deriving the supergravity background, one can use the same reasoning as
Maldacena to take the decoupling limit. The improvement being that we do not assume the existence of a supergravity description and
consider the sum of planar diagrams instead. In any case a more direct approach to the field theory should be desirable.

\section{Acknowledgments}

I am grateful to A. Guijosa, J. Maldacena, C. Thorn, A. Tseytlin, T. ter Veldhuis, J. Walcher, Chi Xiong and L. Urba for comments and discussions. 
This work was supported in part by NSF under grant PHY-0653357 and PHY-0847322.

\appendix

\section{Useful formulas}
\label{formulas}

\subsection{Formulas involving the matrices $\rho^I_{AB}$}

 The matrices $\rho^I_{AB}$ and their inverses $\rho^{I\,AB}$ are defined in \cite{GSW}. Some useful properties are:
\beqa
\rho^I_{AB} \rho^{J\,BC} + \rho^J_{AB} \rho^{I\,BC} &=& 2 \delta^{IJ}\delta^C_A ,\non\\
\rho^I_{AB}\rho^I_{CD} &=& -2\epsilon_{ABCD} ,\non\\
\rho^I_{AB} \rho^J_{CD} \epsilon^{ABCD} &=& -8\delta^{IJ} , \label{rhof}\\
\rho^I_{AB}\rho^{I\,CD} &=& -2\left(\delta^C_A\delta^D_B-\delta_B^C\delta_A^D\right) , \non\\
\rho^I_{AB} &=& \half \epsilon_{ABCD} \rho^{I\,CD}, \non\\
\tr \left(\rho^K\rho^M\rho^L\rho^N\right) &=& 4\left(\delta^{KM}\delta^{LN}-\delta^{KL}\delta^{MN}+\delta^{KN}\delta^{ML}\right) ,\non
\eeqa
and the Fierz identity:
\beq
q^I q^J \rho^I_{AB}\rho^J_{CD}\, a^A b^B a^C b^D = -\half q^Iq^J \rho^I_{AB}\rho^J_{CD}\, a^Aa^B b^Cb^D -\half q^2\epsilon_{ABCD}\, a^Aa^Bb^Cb^D ,
\eeq
where $a^A$, $b^A$ are anticommuting variables and $q^I$ is a six vector. In fact, many other properties can be easily found by noting that
\beq
\gamma^I = \left(\begin{array}{cc}0&\rho^{I\,AB}\\\rho^I_{AB} &0  \end{array}  \right) ,
\eeq
are $SO(6)$ Dirac gamma matrices. Note that in a basis where the $\gamma^I$ are hermitian ($\left(\gamma^I\right)^\dagger=\gamma^I$), 
we have that the $\rho$'s are unitary: $\left(\rho^I\right)^\dagger=\left(\rho^I\right)^{-1}$.

\section{Simplifying the Neumann coefficients}
\label{simpN}

 The Neumann coefficients $N^{rs}_{mn}$ appearing in the expression (\ref{vertex_state}) for the vertex state defining $\Pp_S$ were computed in \pone. 
In this appendix we revisit such computation and show that, actually, equivalent but much simpler expressions can be found for them. We also compare 
the results with previous computations in \pone\ and \cite{GW}.

 The idea of the computation can be found in \cite{GSW} (see also \pone). The main steps are, briefly, as follows. First we consider, 
as in fig.\ref{Dscat}, the cylinder with a slit and parameterize it with two real coordinates $-\infty<\tau<\infty$ and $-\pi<\sigma<\pi$. 
The slit is parallel to the $\sigma$ axis, sits at $\tau=0$ and extends from $-\sigma_0<\sigma<\sigma_0$. It is convenient to introduce a 
complex coordinate $\rho=\tau+i\sigma$ and also $u=e^\rho=e^{\tau+i\sigma}$.  Then, we find a conformal transformation $z(u)$ to the upper 
half-plane parameterized by $z$, $\Ima z>0$. Such transformation is unique up to a choice of two points $z_1=z(u=0)$ and $z_2=z(u=\infty)$ 
corresponding to where the end points of the cylinder map to. Using the $SL(2,\bbR)$ invariance of the half-plane we can fix three parameters 
and then only one parameter or moduli remains. This remaining parameter is the width of the slit, namely $2\sigma_0$. The Green function on
the upper half-plane $G(z,z')$ is given by
\beq
G(z,z') = \ln|z-z'| + \ei \ln|z-\bar{z}'|
\eeq
where $\ei=-1$ or $\ei=+1$ according if we want $G(z,z')$ to satisfy Dirichlet or Neumann boundary conditions on the real axis.
 Using the conformal transformation we just described we can map it back to the Green function $G(u,u')$ on the cylinder. To proceed we need 
to consider four different cases according if $u$ and $u'$ are in the initial $\tau\rightarrow-\infty$ or final string $\tau\rightarrow\infty$. 
In each case we expand $G(u,u')$ appropriately in a series expansion in $u=e^{\tau+i\sigma}$ and $u'=e^{\tau'+i\sigma'}$ (or equivalently 
Fourier expansion in $\sigma$, $\sigma'$). 
 The coefficients are precisely $N^{rs}_{mn}$ where the index $r,s=1,2$ indicate the initial and final strings. The logarithmic singularities
when the two points coincide (which can only happen if they are on the same (initial or final) string) should be subtracted. More concretely, 
we write
\beqa
 G(u,u') &=& \ln|u-u'| +\sum_{n=-\infty}^{\infty} \sum_{m=-\infty}^{\infty} N^{11}_{mn} e^{|m|\tau+|n|\tau'} e^{im\sigma+in\sigma'}, \ \ 
             \tau,\tau'\rightarrow -\infty \\
 G(u,u') &=& \sum_{n=-\infty}^{\infty} \sum_{m=-\infty}^{\infty} N^{12}_{mn} e^{|m|\tau-|n|\tau'} e^{im\sigma+in\sigma'}, \ \ 
             \tau\rightarrow -\infty ,\tau'\rightarrow +\infty \\
 G(u,u') &=& \sum_{n=-\infty}^{\infty} \sum_{m=-\infty}^{\infty} N^{21}_{mn} e^{-|m|\tau+|n|\tau'} e^{im\sigma+in\sigma'}, \ \ 
             \tau\rightarrow+\infty,\tau'\rightarrow -\infty \\
 G(u,u') &=& \ln\left|\frac{1}{u}-\frac{1}{u'}\right| 
             +\sum_{n=-\infty}^{\infty} \sum_{m=-\infty}^{\infty} N^{22}_{mn} e^{-|m|\tau-|n|\tau'} e^{im\sigma+in\sigma'}, \ \ 
             \tau,\tau'\rightarrow +\infty 
\eeqa
In this way, $N^{11}_{mn}$ for $m,n>0$ is determined by the terms depending on $u,u'$ whereas $N^{11}_{m,-n}$ for $m,n>0$ is determined 
by those depending on $u,\bar{u}'$ and so on. Although what we describe has the character of a recipe, it can be verified, a posteriori, that
the coefficients so computed obey all the properties necessary for the vertex state to satisfy the appropriate conditions explained in section \ref{PS}.
In the rest of the appendix we show how to find explicit expressions for $N^{rs}_{mn}$ by computing $G(u,u')$ and then doing the power series expansion. 
This seems straight-forward and indeed it is. However different (but equivalent) expressions can be found according on how we choose to do the 
Taylor expansion of $G(u,u')$ and also on how we choose the points $z_{1,2}$. In \pone\ we chose $z_1=i$, $z_2=iy$, $0<y<1$ where 
$y=\frac{1-\sin\frac{\sigma_0}{2}}{1+\sin\frac{\sigma_0}{2}}$.
In a similar computation, Green and Wai \cite{GW} chose $z_1=i$, $z_2=i-a$ with $0<a<\infty$ and related to $\sigma_0$ by $a=2\tan\frac{\sigma_0}{2}$. 
Here we show that choosing $z_{1,2}= i e^{\pm i\frac{\sigma_0}{2}} $ leads to much simpler expressions.   

 We start by proving that an $SL(2,\bbR)$ transformation leaves the Neumann coefficient appearing in the operator $\Pp$ invariant. Then we
show how to find new, simpler expressions for $N^{rs}_{nm}$ and finally we compare the results with \pone\ and \cite{GW}.
 
\subsection{$SL(2,\bbR)$ invariance of the Neumann coefficients}

 Suppose that we define a new complex variable $w$ spanning the upper half plane and related to $z$ by 
\beq
z=\frac{a w + b}{c w +d}
\eeq 
where $a$,$b$,$c$,$d$ are real numbers and $ad-bc=1$. 
The Green function in the upper half plane is taken to be
\beq
G(z,z') = \ln|z-z'| + \ei \ln|z-\bar{z}'|
\eeq
where $\ei=-1$ for Dirichlet and $\ei=+1$ for Neumann boundary conditions.
 Under the $SL(2,\bbR)$ transformation it maps to
\beq
 G(w,w') = \ln|w-w'| + \ei \ln |w-\bar{w}'| - (1+\ei) \ln|c w+d| - (1+\ei) \ln |cw' - d|
\eeq
 whereas, having we used $w$ directly, we would have taken just the first two terms. The difference vanishes for Dirichlet boundary
conditions, namely $\ei=-1$ and there is nothing to prove. In the case of Neumann boundary conditions we observe that each extra
terms depends on $w$ or $w'$ but not on both. After mapping back, each term will be power expanded in $u$ or $u'$ but not both 
so it will modify only Neumann coefficients with at least one zero subindex:
\beq
 N^{rs}_{m0} \rightarrow N^{rs}_{m0} + C^r_m
\label{shift}
\eeq
Moreover, if the coefficient depends on $u$, it will not matter if we expand for $u'\rightarrow 0$ or $u'\rightarrow\infty$ so the
contribution $C^r_m$ is independent of $s$ as indicated. The bosonic part of the vertex state in eq.(\ref{vertex_state}) 
will be modified as
\beq
\sum_{rs,imn} N^{rs}_{mn} a^\dagger_{irm} a^\dagger_{isn} \rightarrow 
\sum_{rs,imn} N^{rs}_{mn} a^\dagger_{irm} a^\dagger_{isn} + \sum_{r,m,i/\ei=+1} C^r_m a^\dagger_{irm} (a^\dagger_{i,s=1,n=0}+a^\dagger_{i,s=2,n=0}) 
\eeq
However, by definition $a^\dagger_{s=1,n=0}=p_1$ and $a^\dagger_{s=2,n=0}=p_2$ are the momentum of the initial and final strings. 
Since we are  considering a Neumann direction, momentum conservation implies $p_1+p_2=0$ and therefore $\ket{V}$ is not modified. 
Perhaps, more easily, when writing the operator $\Pp_S$ as in eqs.(\ref{P_S})--(\ref{P2}), the Neumann coefficients with subindex zero do not appear
unless they correspond to Dirichlet boundary conditions. 

In the fermionic part, the coefficient $N^{rs}_{m0}(\ei=+1)$ only enters in the definition of $\beta_m$ in eq.(\ref{betamdef}) and in such a 
form that is not affected by the shift ($\ref{shift}$). The insertion operators that also appear in $\Pp$ do not depend on the Neumann coefficients
and therefore are not affected. We have then proved that the operator $\Pp$ is invariant under the $SL(2,\bbR)$ transformation, as it should.
We are therefore free to choose, up to one moduli, the points $z_{1,2}$ at our convenience.

\subsection{Simplified expressions for $N^{rs}_{mn}$}

Now we choose $z_{1,2}=ie^{\pm i\frac{\sigma_0}{2}}$ which defines the conformal transformation
\beq
z = i e^{-i\frac{\sigma_0}{2}} \sqrt{\frac{u-e^{i\sigma_0}}{u-e^{-i\sigma_0}}}, \ \ \ \ u=e^{-i\sigma_0} \frac{z^2+e^{i\sigma_0}}{z^2+e^{-i\sigma_0}}
\eeq
The Green function 
\beq
G(z,z') = \ln|z-z'| + \ei \ln|z-\bar{z}'|
\eeq
maps to 
\beqa
G(u,u') &=&        \ln\left|\sqrt{(u-\mu)(\mu u'-1)}-\sqrt{(u'-\mu)(\mu u -1)}\right|  \\
  && + \ei \ln\left|\sqrt{(\mu u -1)(\mu \bar{u}'-1)}+\sqrt{(u-\mu)(\bar{u}'-\mu)}\right| \\
  && -\frac{1}{2} (1+\ei) \ln \left|u-\mu^{-1}\right| - \frac{1}{2} (1+\ei) \ln \left|u'-\mu^{-1}\right|
\eeqa
where we defined $\mu=e^{i\sigma_0}$. The terms in the last line are of the same type as we proved, in the previous subsection, 
to be irrelevant in the computation of $\Pp$. Namely, they obviously go away for $\ei=-1$ and, as they only depend on one 
or the other variable, when $\ei=+1$ they also go away in $\Pp$ because of momentum conservation. For that reason we can
define a new equivalent Green function
\beqa
\tilde{G}(u,u') &=&        \ln\left|\sqrt{(u-\mu)(\mu u'-1)}-\sqrt{(u'-\mu)(\mu u -1)}\right|  \\
  && + \ei \ln\left|\sqrt{(\mu u -1)(\mu \bar{u}'-1)}+\sqrt{(u-\mu)(\bar{u}'-\mu)}\right| 
\label{tildeG}
\eeqa
 Expanding in powers of $u$, $u'$ for small $u$, $u'$ we get
\beqa
 \tilde{G}(u,u') &\simeq& \ln|u'-u| + \ln\left|\frac{\mu}{2}-\frac{2}{\mu}\right| +\ei \ln|1+\mu| + \sump{nm}{} N^{11}_{nm} u^n u'{}^m \\
           &\simeq& \ln|u'-u| + \ln \sin \sigma_0 + \ei \ln\left[2 \cos\frac{\sigma_0}{2}\right]+ \sump{nm}{} N^{11}_{nm} u^n u'{}^m 
\eeqa
where we used $\mu=e^{i\sigma_0}$. From here we find (see below however) 
$N^{11}_{00}=\ln \sin \sigma_0 + \ei \ln\left[2 \cos\frac{\sigma_0}{2}\right]$ and the rest of the coefficients are obtained by continuing 
the expansion. 
To compute them explicitly we consider the term in $\tilde{G}$ that depends only on $u$, $u'$ (and not their conjugates) and 
observe that, by a straight-forward calculation, we get:
\beqa
\left[ u\partial_u + u'\partial_{u'}\right] \lefteqn{ \half \ln\left[\frac{\sqrt{(u-\mu)(\mu u'-1)}-\sqrt{(u'-\mu)(\mu u -1)}}{u'-u}\right]=} && \\
 &=& \frac{\mu}{4} \frac{1-uu'}{\sqrt{(u-\mu)(u'-\mu)(\mu u-1)(\mu u'-1)}} -\frac{1}{4} \\
 &=& \frac{1}{4}\sump{m,n=0}{\infty} \left(P_nP_m-P_{n-1}P_{m-1}\right) u^n u'{}^m 
\eeqa 
where the prime in the summation sign indicates that we omit the term with $m=n=0$. The functions $P_n=P_n(\cos\sigma_0)$ are the usual 
Legendre polynomials and we defined by convenience $P_{-1}(\cos\sigma_0)=0$. The Legendre Polynomials appear in virtue of the identity \cite{Grad}
\beq
\frac{1}{\sqrt{(u-\mu)(\mu u-1)}} = \frac{1}{\sqrt{\mu}}\frac{1}{\sqrt{u^2-2u\cos\sigma_0+1}} 
  = \frac{1}{\sqrt{\mu}} \sum_{n=0}^\infty P_n(\cos \sigma_0) u^n
\eeq
We then identify 
\beq
 N^{11}_{nm} = \frac{1}{4}\frac{P_nP_m-P_{n-1}P_{m-1}}{n+m} , \ \ \ \ n,m>0, \ \ n+m\neq 0
\eeq
 To get the coefficients $N^{11}_{n,-m}$ , $(n,m>0)$ we need to consider the term depending on $u$ and $\bar{u}'$ for which we have 
\beqa
\left[ u\partial_u - \bar{u}'\partial_{\bar{u}'}\right] \lefteqn{ \frac{\ei}{2}
        \ln\left[\sqrt{(\mu u -1)(\mu \bar{u}'-1)}+\sqrt{(u-\mu)(\bar{u}'-\mu)}\right]=} && \\
 &=& -\frac{\mu\ei}{4} \frac{u-\bar{u}'}{\sqrt{(u-\mu)(u'-\mu)(\mu u-1)(\mu u'-1)}}  \\
 &=& \frac{\ei}{4}\sum_{m,n=0}^\infty \left(P_nP_{m-1}-P_{n-1}P_{m}\right) u^n u'{}^m
\eeqa 
 where again, the Legendre polynomials are evaluated at $\cos\sigma_0$ and we use the convention $P_{-1}(\cos\sigma_0)=0$.
Thus,
\beq
 N^{11}_{n,-m}=\frac{\ei}{4}\frac{P_nP_{m-1}-P_{n-1}P_{m}}{n-m} , \ \ \ \ n,m>0, \ \ n-m\neq 0
\eeq
The case $N^{11}_{n,-n}$ has to be considered separately. In this case we can take just the derivative with respect to $u$ for example and,
after a tedious but simple calculation, we obtain
\beq
N^{11}_{n,-n}=-\frac{\ei}{2n}\left\{\sum_{q=1}^{n-1} P_qP_{q-1}+\half P_n P_{n-1} - \cos\sigma_0\sum_{q=0}^{n-1}P_q^2 + \half\right\}, \ \ \ n\neq 0
\eeq
From the fact that the Green function, and the Neumann coefficients, are real we obtain
\beq
 N^{11}_{-n,-m} = N^{11}_{nm}
\eeq
which allows us to compute the other components. 

Now we should compute $N^{21}_{nm}$. In order to do that we should expand $\tilde{G}(u,u')$ for $u\rightarrow\infty$ and $u'\rightarrow 0$. 
It can be easily done if one notices that
\beq
\tilde{G}(u,u') = \frac{1+\ei}{2} \ln |u| +(1+\ei)\ln(2\sin\sigma_0) + \ln\left|1-\frac{u'}{u}\right|
                  -\ei\left(\tilde{G}(\frac{1}{u},\bar{u}')-\ln\left|\frac{1}{u}-\bar{u}'\right|\right) 
\eeq
 When expanding the last term we get the coefficients $N^{11}_{nm}$ that we already computed. Including the contribution from expanding
$\ln\left|1-\frac{u'}{u}\right|$ we obtain
\beqa
 N^{21}_{nm} &=& -\ei N^{11}_{nm} + \half \delta_{n+m}, \ \ \ \ \mbox{except} \ \ \ n=m=0 \\
 N^{21}_{00} &=& (1+\ei) \ln(2\sin\sigma_0) -\ei N^{11}_{00} = \ln(2\sin\frac{\sigma_0}{2})+\ei \ln 2 
\eeqa
Finally it is easy to see that 
\beqa
 N^{22}_{mn} &=& N^{11}_{nm}\\
 N^{12}_{mn} &=& N^{21}_{nm} 
\eeqa
for all $n$, $m$. Notice also that we can write
\beq
N^{rs}_{00} = \frac{1-\ei}{2}\ln\sin\frac{\sigma_0}{2} + \delta^{rs} (1+\ei) \ln\cos\frac{\sigma_0}{2} 
             + \frac{1+\ei}{2} \ln\left(4\sin\frac{\sigma_0}{2}\right)
\eeq
The last term can be dropped by the reason we already discussed. Namely it goes away for $\ei=-1$ and for Neumann boundary conditions, \ie\ 
$\ei=+1$ it goes away in $\Pp$ by momentum conservation. So we prefer to write
\beq
N^{rs}_{00} = \frac{1-\ei}{2}\ln\sin\frac{\sigma_0}{2} + \delta^{rs} (1+\ei) \ln\cos\frac{\sigma_0}{2} 
\eeq
as we had found in \pone. To complete the calculation, we still need to compute $N^{11}_{m0}$. The result is actually contained in the expansions
we performed, the only point is that, for $n=0$, the terms depending on $u$, $u'$ and $u$, $\bar{u}'$ both contribute. Using the results we have 
(and remembering we defined $P_{-1}(\cos\sigma_0)=0$) we easily obtain
\beq
N^{11}_{m0} = \frac{1}{4m}\left(P_m(\cos\sigma_0)-\ei P_{m-1}(\cos\sigma_0)\right), \ \ \ m>0
\eeq
We also have $N^{11}_{m0}=N^{11}_{-m,0}$. 

To summarize we obtained the expressions:
\beqa
N^{rs}_{00} &=&  \frac{1-\ei}{2}\ln\sin\frac{\sigma_0}{2} + \delta^{rs} (1+\ei) \ln\cos\frac{\sigma_0}{2}  \\ 
N^{11}_{m0} &=&  N^{11}_{-m,0}  = \frac{1}{4m}\left(P_m(\cos\sigma_0)-\ei P_{m-1}(\cos\sigma_0)\right), \ \ \ m>0 \\
N^{11}_{nm} &=&  N^{11}_{-n,-m} = \frac{1}{4}\frac{P_nP_m-P_{n-1}P_{m-1}}{n+m} , \ \ \ \ n,m>0, \ \ n+m\neq 0           \label{N11pp} \\
N^{11}_{n,-m}&=& N^{11}_{-n,m}  = \frac{\ei}{4}\frac{P_nP_{m-1}-P_{n-1}P_{m}}{n-m} , \ \ \ \ n,m>0, \ \ n-m\neq 0 \label{N11pm}\\ 
N^{11}_{n,-n}&=&N^{11}_{-n,n}   =-\frac{\ei}{2n}\left\{\sum_{q=1}^{n-1} P_qP_{q-1}+\half P_n P_{n-1}              
                                 - \cos\sigma_0\sum_{q=0}^{n-1}P_q^2 + \half\right\}, \ \ \ n\neq 0    \nonumber                  \\ 
N^{21}_{nm} &=& -\ei N^{11}_{nm} + \half \delta_{n+m}, \ \ \ \ \mbox{except} \ \ \ n=m=0                          \\
N^{22}_{mn} &=& N^{11}_{nm}                                                                                              \\
N^{12}_{mn} &=& N^{21}_{nm} 
\eeqa
which determine all Neumann coefficients. As we show below following \pone, the case $n+m\neq0$ can also be summarized as
\beq
N^{rs}_{mn} = \frac{1}{(m+n)\sin\sigma_0} \Ima\left(f^r_m f^s_n\right)
\eeq
where $f^1_{m>0}=-\bar{f}_m$, $f^1_{m<0}=-\ei f_{-m}$, $f^2_{m\neq 0} = -\ei f^1_m$ with
\beqa
f_{m>0}= -\frac{i}{2} \left[e^{i\frac{\sigma_0}{2}}P_m(\cos\sigma_0) - e^{-i\frac{\sigma_0}{2}}P_{m-1}(\cos\sigma_0)\right]
\eeqa 
and
\beqa
 f^1_0 &=& \frac{1+\ei}{2}\left(1-\sin\frac{\sigma_0}{2}\right)-i \frac{1-\ei}{2} \cos\frac{\sigma_0}{2} \\
 f^2_0 &=& \frac{1+\ei}{2}\left(1+\sin\frac{\sigma_0}{2}\right)-i \frac{1-\ei}{2} \cos\frac{\sigma_0}{2} \\
\eeqa
 In the following we compare these results to our previous calculations in \pone\ and also to the expressions in \cite{GW}. Here let us just
note that what we just found are much simpler expressions than those we knew from before. 

\subsection{Comparison with previous results}

 In \pone\ we have already computed the Neumann coefficients in terms of finite sums of associated Legendre polynomials. In \cite{GW} Green and Wai had
previously discussed a similar calculation for the bosonic string for the case of Dirichlet boundary conditions in all directions (including $X^\pm$). 
They found the Neumann coefficients in terms of (sometimes infinite) double sums. Here we show briefly that the results have to agree since they
come from expanding in series the same Green function. We also make explicit the different expressions for the coefficients so one can compute them
for particular values of the subindices and see that they indeed agree. 

In \pone\ we expanded the Green function
\beq
 G^{(I)}(u,u') = \ln|w(u)-w(u')| + \ei \ln|w(u)-\overline{w(u')}|
\label{GreenI}
\eeq
 with 
\beq
 w(u) = -\frac{i}{u-1}\frac{1}{1+\sin\frac{\sigma_0}{2}} \left\{(1+u)\sin\frac{\sigma_0}{2} + \sqrt{(u-e^{i\sigma_0})(u-e^{-i\sigma_0})} \right\}
\eeq
where the square root has a cut on the slit. Replacing in eq.(\ref{GreenI}) and after some algebra we find
\beqa
  G^{(I)}(u,u') &=& \tilde{G}(u,u') +(1+\ei) \ln\frac{8\cos\frac{\sigma_0}{2}}{1+\sin\frac{\sigma_0}{2}} \\
                &&                  - (1+\ei) \ln|\sqrt{u-\mu}+i\sqrt{\mu u-1}|
                                    - (1+\ei) \ln|\sqrt{u'-\mu}+i\sqrt{\mu u'-1}| \nonumber
\eeqa
with $\mu=e^{i\sigma_0}$ as before. The first term, $\tilde{G}(u,u')$ is the same one as in eq.(\ref{tildeG}) and the others are irrelevant
for the same reason we already explained (Namely they do not appear in $\Pp$). Therefore the Neumann coefficients are proved to be the same since
they are the coefficients of the series expansion of the same function. In \pone we wrote the result as
\beq
N^{rs}_{mn} = -\frac{i}{8}\frac{(1+\ei)}{m+n}\left(a^r_m \delta_{n0} + a^s_n \delta_{m0}\right) 
              +\frac{1}{(m+n)\sin\sigma_0} \Ima\left(f^r_m f^s_n\right)
\eeq
where $f^1_{m>0}=-\bar{f}_m$, $f^1_{m<0}=-\ei f_{-m}$, $f^2_{m\neq 0} = -\ei f^1_m$ with
\beq
f_{m>0} = -\frac{i}{m} e^{i\frac{\sigma_0}{2}} \sum_{l=1}^m \frac{(-i)^lm!}{(m+l)!} l P^l_m(\cos\sigma_0)
\eeq
where $P^l_m$ are associated Legendre polynomials defined, in terms of regular Legendre polynomials $P_m$  as\footnote{Care should be taken when checking
these formulas with computer algebra programs since definitions of $P^l_m$ can differ between them.} 
\beq
P^l_m(x) = (-1)^l(1-x^2)^\frac{l}{2}\frac{d^l}{dx^l} P_m(x)
\eeq
 Comparing with the result we obtained in the previous subsection one finds that, agreement of the Neumann coefficients implies the following 
interesting identity between Legendre polynomials
\beqa
f_{m>0} &=& -\frac{i}{m} e^{i\frac{\sigma_0}{2}} \sum_{l=1}^m \frac{(-i)^lm!}{(m+l)!} l P^l_m(\cos\sigma_0) \\
        &=& -\frac{i}{2} \left[e^{i\frac{\sigma_0}{2}}P_m(\cos\sigma_0) - e^{-i\frac{\sigma_0}{2}}P_{m-1}(\cos\sigma_0)\right]
\eeqa 
 Although this is proven because it equates the coefficients of the series expansion of the same function, it is reassuring to compute the functions
$f_m(\sigma_0)$ for several explicit values of $m$ and see that they indeed agree. 

 We can also compare explicitly with the results of Green and Wai. Before doing that, however, let us dwell a little into the relation 
between that calculation and the one presented in \pone and here. In modern language, Green and Wai look at a $D$-instanton (although it is in Minkowski 
space making the interpretation less clear). In any case the coefficients they find should agree with ours for $\ei =-1$ giving us an extra check
of the calculations. It should be noted that, in our case, we have Neumann boundary conditions in $X^{\pm}$ which corresponds to the usual
$D$-brane interpretation (where the D-brane extends in the time direction). To see how this comes about we write one of the conformal
constraints as
\beq
 \partial_\tau X^+ \partial_\sigma X^- +  \partial_\tau X^- \partial_\sigma X^+  = -  \partial_\tau X^\perp\partial_\sigma X^\perp 
\eeq
 Since the slit is parallel to the $\sigma$ axis, Neumann boundary conditions are given by $\partial_\tau X=0$ and Dirichlet by $\partial_\sigma X=0$.
On the slit, each of the directions $X^\perp$ perpendicular to the light-cone  satisfies either Dirichlet or Neumann boundary conditions and therefore
the right  hand side of the equation vanishes implying
\beq
\left. \partial_\tau X^+ \partial_\sigma X^- +  \partial_\tau X^- \partial_\sigma X^+ \right|_{\mbox{bdy.}} = 0
\label{lcaa}
\eeq
 We use a gauge $X^+=\sigma$ which we call $\sigma$-gauge. It obviously means that 
$\partial_\tau X^+=0$, namely $X^+$ satisfies Neumann boundary conditions. Moreover, replacing $X^+=\sigma$ in 
eq.(\ref{lcaa}) we obtain $\partial_\tau X^-=0$, so both $X^\pm$ are Neumann. In \cite{GW} the authors considered a gauge 
$X^+=\tau$ implying $X^{\pm}$ are Dirichlet. 

Having said this, let us consider the Green function appearing in eq.(5.7) of \cite{GW}:
\beqa
\hat{G}_D(u,u') &=& \ln\left| i\sqrt{\frac{a^2}{4}-1-ia\frac{u+1}{u-1}}- i\sqrt{\frac{a^2}{4}-1-ia\frac{u'+1}{u'-1}}\right| \\
                & & - \ln\left| i\sqrt{\frac{a^2}{4}-1-ia\frac{u+1}{u-1}}+ i\sqrt{\frac{a^2}{4}-1+ia\frac{\bar{u}'+1}{\bar{u}'-1}}\right|
\eeqa
where $a= 2\tan\frac{L}{2}$ with $L=\sigma_0$. Again simple algebra reveals that
\beq
 \hat{G}_D(u,u') = \tilde{G}(e^{i\sigma_0} u, e^{i\sigma_0} u')
\label{relGW}
\eeq
where the extra phase factors appear because Green and Wai have the slit between $0\leq\sigma\leq2\sigma_0$, namely shifted by $\sigma_0$. This is 
of course irrelevant  for us since we integrate over the position of the slit when defining $\Pp$. In doing the comparison we should also
rescale $\sigma$ by two since in \cite{GW}, $0<\sigma<\pi$, instead of $-\pi<\sigma<\pi$.  Again, this implies equality of the Neumann coefficients 
up to a phase factor. The Neumann coefficients in \cite{GW} can be extracted from eq.(5.9) in that paper resulting in
\beqa
\tilde{N}^{11}_{pq} &=& \sum_{n,m=1}^{\infty}\frac{u_m u_n}{n+m}\binomial{m+p-1}{p}\binomial{n+q-1}{q}\left(1-e^{2i\sigma_0}\right)^{m+n} \label{GW1}\\
            &=& \sum_{m=1}^p \sum_{n=1}^q  \frac{u_m u_n}{n+m} \binomial{p-1}{p-m}\binomial{q-1}{q-n}
                   \left(e^{-2i\sigma_0}-1\right)^{m+n} \label{GW1b} \\
\tilde{N}^{11}_{p,-q} &=& \frac{1}{4} \sum_{m=1}^{\infty}\sum_{n=1}^q \frac{u_m u_n}{n+m} 
                                           (-1)^n \binomial{m+p-1}{p}\binomial{q-1}{q-n} \left(1-e^{-2i\sigma_0}\right)^{m+n} \label{GW2}
\eeqa
for $p,q>0$. Notice that in \cite{GW} two different but equivalent expressions are given for $N^{11}_{pq}$, $p,q>0$.
As we already said, it follows from eq.(\ref{relGW}) that 
\beq
 N^{11}_{mn} = e^{i\sigma_0 (m+n)} \tilde{N}^{11}_{mn}
\label{phasefactor}
\eeq
 It is reassuring to compute several of the coefficients, as functions of $\sigma_0$, and check that this relation is correct. Namely, check the agreement
(up to the phase factor in eq.(\ref{phasefactor})) between eqns. (\ref{GW1}), (\ref{GW1b}) and (\ref{N11pp}) and also (\ref{GW2}) and (\ref{N11pm}).

\section{Massless string scattering from D-branes}
\label{scattering}

As explained in section \ref{Dscattering} the operator $\Pp_S$ summarizes the scattering amplitudes for closed string states from a D3-brane. This includes
massive and massless states. Since scattering amplitudes for massless states are known \cite{Klebanov:1995ni,MG} their explicit computation provides a 
check for the calculation. This was already explained in section \ref{Dscattering} were we sketched the main steps of the calculation. Since 
the results depends on several details, it is a very useful check and deserves to be spelled out in more detail which we do in this appendix. 
First we explain the derivation of eqs.\eqn{HZM0}-\eqn{HZM8} which are the zero mode part of $\Pp_S$. It is the only relevant part since massless states, 
in the Green-Schwarz formalism are given by the different vacua, namely they contain no oscillator excitations.  After that we use the result to compute
the scattering amplitudes concentrating on bosonic states. Finally we compare with known results. The comparison fixes the overall normalization of
the hole insertion operator, namely the constant $\alpha_3$ in eq.\eqn{Pdef}.

\subsection{Zero modes Hamiltonian}

 The zero mode part of the operator $\Pp_S$, namely the part that contains no oscillators is given by
\beq
 \left. \Pp_S \right|_{\mbox{zero modes}} = \int_{0}^{\pi} d\sigma_0 \left(\sin\frac{\sigma_0}{2}\right)^{q^2}\left(\cos\frac{\sigma_0}{2}\right)^{4k^2}
                                              \hat{H}_{\mbox{zero modes}}
\eeq
where $\hat{H}_{\mbox{zero modes}}$ is the zero mode part of the insertions computed in \eqn{HZM} and the other factors come from the exponential
\beq
 e^{\sum_{rs,imn}N^{rs}_{i,00} a^\dagger_{ir0}a^\dagger_{is0}} = e^{q^2 \ln\sin\frac{\sigma_0}{2}+4k^2\ln\cos\frac{\sigma_0}{2}} 
     = \left(\sin\frac{\sigma_0}{2}\right)^{q^2}\left(\cos\frac{\sigma_0}{2}\right)^{4k^2}
\eeq
 From eq.\eqn{HZM} we can write $\hat{H}_{\mbox{zero modes}}$ separating the terms with different number of fermions:
\beqa
\hat{H}_{\mbox{z.m.}} &=& \frac{1}{2\pi^2} \frac{\sin^2\frac{\sigma_0}{2}}{\sin\sigma_0} k^Lk^L \nonumber\\
                            & & -\frac{1}{16\pi^2 \sqrt{2}}\frac{\sst\cst}{\sin^2\sz} k^L\left(y\sq y+\bar{y}\sq\bar{y} \right)
                                -\frac{1}{8\pi^2} \frac{\sin^2\frac{\sz}{2}}{\sin^2\sz} k^Lk^R (y^4+\bar{y}^4) \nonumber \\  
                          & & +\frac{1}{2^8\pi^2} \frac{\cos^2\frac{\sz}{2}}{\sin^3\sz} y\sq y \, \bar{y}\sq\bar{y} 
                             +\frac{1}{2^7\pi^2\sin^3\sz}\left(y^4+\bar{y}^4\frac{1}{4}\varepsilon_{ABCD}y^Ay^B\bar{y}^C\bar{y}^D\right) \nonumber \\
                          & &+ \frac{1}{2^6\pi^2\sqrt{2}}\frac{\sst\cst}{\sin^4\sz} k^R \left(\bar{y}^4 y\sq y + y^4 \bar{y} \sq \bar{y}\right) \nonumber\\
                            & & +\frac{1}{2^5\pi^2}\frac{\sin^2\frac{\sz}{2}}{\sin^5\sz} k^Rk^R y^4 \bar{y}^4   \label{Hzma}
\eeqa
where we use the notation
\beq
 y^4 = \frac{1}{24} \varepsilon_{ABCD}y^A y^By^Cy^D, \ \ \ y\sq y = q^I \rho^{I}_{AB} y^A y^B
\eeq
and the same for $\bar{y}$. The next step is to replace, from eq.\eqn{ydef},
\beq
y^A = \bar{a}^A_0 \sst + i \bar{b}_0^A \cst, \ \ \ \ \bar{y}^A =  \bar{a}^A_0 \sst - i \bar{b}_0^A \cst,
\eeq
in the previous equation. By expanding the corresponding terms and taking into account that
\beq
\bar{a}_0^A \bar{a}_0^B \bar{a}_0^C \bar{a}_0^D = \varepsilon^{ABCD} \bar{a}_0^4, \ \ \ \
\bar{b}_0^A \bar{b}_0^B \bar{b}_0^C \bar{b}_0^D = \varepsilon^{ABCD} \bar{b}_0^4
\eeq
which result in formulas such as 
\beqa
 y^4 + \bar{y}^4 + \frac{1}{4} \epsilon_{ABCD} y^Ay^B\bar{y}^C\bar{y}^D &=& 8 \sin^4\frac{\sz}{2} \bar{a}^4_0 + 8 \cos^4\frac{\sz}{2} \bar{b}^4_0 \\
 \epsilon_{ABCD} \bar{b}^A_0 \bar{a}^B_0\bar{a}^C_0\bar{a}^D_0\,\, \bar{a}_0 \sq \bar{b}_0 &=& -6 \bar{a}_0^4\, \bar{b}_0\sq \bar{b}_0 \\
 \epsilon_{ABCD} \bar{a}^A_0 \bar{a}^B_0\bar{b}^C_0\bar{b}^D_0\,\, \bar{a}_0 \sq \bar{a}_0 &=& 4 \bar{a}_0^4\, \bar{b}_0\sq \bar{b}_0 \\
\eeqa
we can easily find that $\hat{H}_{\mbox{z.m.}}$ reduces to
\beqa
\hat{H}_{\mbox{z.m.}}&=& \frac{1}{4\pi^2} k^Lk^L \frac{\sst}{\cst} \nonumber\\
          & & -\frac{k^L}{2^5\pi^2\sqrt{2}}\left(\bar{a}_0\sq\bar{a}_0 \tan\frac{\sz}{2} - \bar{b}_0\sq\bar{b}_0\cotan\frac{\sz}{2}\right)\nonumber\\
          & & -\frac{1}{2^8\pi^2}\left(\frac{\sst}{\cos^3\hsz}\bar{a}_0^4+\frac{\cst}{\sin^3\hsz}\bar{b}_0^4\right)
                                 \left(4k^2\sin^2\frac{\sz}{2}+q^2\cos^2\hsz-2\right) \nonumber\\
          & &                    - \frac{1}{2^9\pi^2}\cotan\hsz(\bar{a}_0\sq\bar{a}_0)(\bar{b}_0\sq\bar{b}_0)  
              + \frac{1}{2^{10}\pi^2} \epsilon_{ABCD}\bar{a}_0^A\bar{a}_0^B\bar{b}_0^C\bar{b}_0^D \left(4k^2\tan\hsz-q^2\cotan\hsz\right) \nonumber\\
   & &   -\frac{k^R}{2^7\pi^2\sqrt{2}}\left(\tan\hsz \bar{a}_0^4\, \bar{b}_0\sq\bar{b}_0 - \cotan\hsz \bar{b}_0^4\, \bar{a}_0\sq\bar{a}_0\right)\nonumber\\
          & &   \frac{1}{2^6\pi^2}\tan\hsz\, k^Rk^R \bar{a}_0^4\,\bar{b}_0^4
\eeqa
Now we proceed to replace in eq.\eqn{Hzma} and perform the $\sz$ integration using that
\beq
\int_0^\pi \left(\sst\right)^\alpha\left(\cst\right)^\beta d\sz = B\left(\frac{\alpha+1}{2},\frac{\beta+1}{2}\right)
\eeq
where $B(a,b)=\frac{\Gamma(a)\Gamma(b)}{\Gamma(a+b)}$ is Euler's Beta function. There is an important point that we should mention here which is that
the integral in $\sz$ can be divergent. We define it by analytic continuation to be equal to the Beta function. This is the usual procedure
when computing scattering amplitudes in string theory. Namely to assume we do the calculation in a region of momenta where all integrals are convergent
and continue to other regions assuming analyticity of the scattering amplitudes in the external momenta ($q^2$, $k^2$). To write the final expression 
we find convenient to define $s=-q^2$, $t=-4k^2$ and a function
\beq
 A(s,t) = \frac{\Gamma\left(-\frac{s}{2}\right)\Gamma\left(-\frac{t}{2}\right)}{\Gamma\left(1-\frac{s}{2}-\frac{t}{2}\right)} =
          \frac{\Gamma\left(2k^2\right)\Gamma\left(\frac{q^2}{2}\right)}{\Gamma\left(2k^2+\frac{q^2}{2}+1\right)}
\label{Adef}
\eeq
to which all Beta functions that appear can be reduced to using the properties of the Gamma functions. The result of the $\sz$ integration is a Hamiltonian
to which $\Pp_S$ reduces after eliminating all oscillators. It can better be written split into terms of different number of fermions as was
presented in eqs.\eqn{HZM0}-\eqn{HZM8}.
\beqa
H^{(0)}_{[0]} &=& -\frac{1}{8\pi^2}\,s\, k^Lk^L\, A(s,t), \label{HZM0B}\\
H^{(0)}_{[2]} &=& -\frac{1}{2^6\pi^2\sqrt{2}} k^L\left(t\,\bar{b}_0 \sq \bar{b}_0-s\,\bar{a}_0\sq\bar{a}_0\right)\,A(s,t), \label{HZM2B}\\
H^{(0)}_{[4]} &=& -\frac{1}{2^{8}\pi^2}\left\{\bar{a}_0^4s^2+\bar{b}_0^4 t^2
                  -\frac{t}{4}\bar{a}_0\sq\bar{a}_0 \bar{b}_0\sq\bar{b}_0\right\}\,A(s,t), \label{HZM4B}\\
H^{(0)}_{[6]} &=& -\frac{1}{2^8\pi^2\sqrt{2}} k^R
                   \left(t\,\bar{a}_0\sq\bar{a}_0\,\bar{b}_0^4-s\,\bar{b}_0\sq\bar{b}_0\,\bar{a}_0^4\right)\,A(s,t), \label{HZM6B}\\
H^{(0)}_{[8]} &=&  -\frac{1}{2^7\pi^2}\,s\,k^Rk^R\,\bar{a}_0^4\,\bar{b}_0^4\,A(s,t) \label{HZM8B} ,
\eeqa

\subsection{External states}

 Since we are going to consider only massless scattering the only states we need to consider for the Green-Schwarz closed IIB string 
are the vacuum states. One simple way to describe them is to define a state 
\beq
 \vl,\ \ \ \ \mbox{such that}\ \ \ \ \lambda_{0A} \vl=0, \ \ \tilde{\lambda}_{0A}\vl=0
\eeq
and act on it with the $\theta^A_0$, $\tilde{\theta}^A_0$. Since there are four $\theta^A_0$ and four $\tilde{\theta}^A_0$ we have $2^8$ states.
From the $2^8$ vacua, for simplicity, we will only consider the $2^6=64$ NS-NS states obtained by acting an even number of times with the $\theta^A_0$ 
and an even number of times with the $\tilde{\theta}_0^A$.  It is therefore convenient to define the following normalized states
\beq
\ket{R}=\vl, \ \ \ \ket{I}=\frac{1}{2\sqrt{2}}\rho^I_{AB}\theta^A\theta^B\vl, \ \ \ket{L} = \theta^4\vl
\label{states}
\eeq
Also, from now on, we do not put the subindex $0$ since in this section all fields are
reduced to their zero modes. We are going to define a polarization state as
\beq
 \ket{\xi} = \xi_L \ket{L} + \xi_I \ket{I} + \xi_R \ket{R}
\label{stat}
\eeq
 with $\xi_R=\xi_L^*$. The corresponding bra we define as
\beq
 \bra{\xi} = \bra{L} \xi_L + \bra{I} \xi_I + \bra{R} \xi_R
\eeq
which implies we use the notation $(\ket{L})^\dagger = \bra{R}$ and $(\ket{R})^\dagger = \bra{L}$ and therefore the identity is written as
\beq
 \mathbb{I} = \ket{R}\bra{L}+\ket{L}\bra{R}+\ket{I}\bra{I}
\label{idzm}
\eeq
 The state of a closed string is determined
by the product of two polarization states, one for the left and one for the right movers. Both should be transverse to the momentum $p$ of
the particle, \ie\ $(p\xi)=0$. For convenience we define 
\beq
 \theta_3 = \tilde{\theta}_1, \ \ \ \theta_4 = \tilde{\theta}_2
\eeq
so that the polarizations are $\epsilon^{(1)} = \xi_1 \otimes \xi_4$, $\epsilon^{(2)} = \xi_2 \otimes \xi_3$. 
While a general polarization tensor cannot be always factorized in that way, it certainly can be written as a linear combination of such 
factorized terms. The hole insertion operator is written, in section \ref{Pcomp} in a vertex state representation. This means that it is written
as a state in the product space of the Hilbert spaces of the initial and final strings. Formally, if we label the basis with an index $\nu$, we
get
\beq
 \ket{P} = \sum_{\nu_1\nu_2} \bra{\nu_1} P \ket{\nu_1} \ket{\nu_1}\otimes\ket{\nu_2}
\eeq
If we want to apply the operator to the state $\ket{\nu_1}$ we do
\beq
 \braket{\nu_1}{P} = \sum_{\nu_2} \ket{\nu_2}\bra{\nu_2} P \ket{\nu_1}= P\ket{\nu_1}
\eeq
One further point we need to make is that the vacuum we used in defining the hole insertion operator satisfies, according to eqs.\eqn{veq1},\eqn{veq2},
\beq
 a_A \ket{0} = 0,\ \ \  b_A \ket{0}=0, \ \ \ (\chi_A-\bar{\Xi}_A)\ket{0}=0, \ \ \ (\Xi^A-\bar{\chi}^A)\ket{0} = 0
\label{vdef}
\eeq
These are linear combinations of the $\lambda$'s except the last one. This means that the state $\ket{0}\neq\vl$. In fact, the state $\vl$
satisfies the same properties \eqn{vdef}, except that the last one is replaced by $(\chi_A+\bar{\Xi}_A)\ket{0}=0$. This means that, if we define
\beq
 \beta = \frac{1}{\sqrt{2}}(\Xi^A-\bar{\chi}^A) = \frac{1}{2}\left(\theta_1^A+\tilde{\theta}_2^A-\theta_2^A-\tilde{\theta}_1^A\right)
\label{betadef}
\eeq
we obtain
\beq
 \ket{0} = \beta^4 \vl
\eeq
which is a useful relation to write the zero mode hole insertion operator as linear combination of the states \eqn{stat}. 

 A final point has to do with the way in which we define the identity operator as a vertex state in the subspace of zero modes. 
It should impose the conditions: 
\beqa
 (\theta_1-\theta_2) \ket{\mathbb{I}} &=&0, \\ 
 (\tilde{\theta}_1 - \tilde{\theta}_2) \ket{\mathbb{I}} &=& 0 , \\
 (\lambda_1+\lambda_2) \ket{\mathbb{I}} &=&0, \\ 
 (\tilde{\lambda}_1 + \tilde{\lambda}_2) \ket{\mathbb{I}} &=& 0 , \\
\eeqa
which identifies both strings. This conditions are solved by
\beq
 \ket{\mathbb{I}} = (\theta_1-\theta_2)^4 (\tilde{\theta}_1-\tilde{\theta}_2)^4 \vl
\eeq
 To understand what it does we can consider a single fermion $\theta$ and states $\ket{0}$ and $\ket{1}=\theta\ket{0}$.
The condition 
\beq
 (\theta_1-\theta_2)\ket{\mathbb{I}} = 0, 
\eeq
imposes
\beq
 \ket{\mathbb{I}}=\ket{0}\otimes\ket{1}+\ket{1}\otimes\ket{0}
\label{vrep}
\eeq
whereas the identity operator is
\beq
 \mathbb{I}= \ket{0}\bra{0} + \ket{1}\bra{1}
\label{oprep}
\eeq
So, in converting a vertex representation as in \eqn{vrep} into and operator representation \eqn{oprep}  we should flip
$\ket{0}\rightarrow\bra{1}$ and $\ket{1}\rightarrow\bra{0}$. In our case, for the left moving fermions we have
\beq
 (\theta_1-\theta_2)^4 \vl \rightarrow (\theta_1^4  + \frac{1}{4}\epsilon_{ABCD}\theta_1^A\theta_1^B\theta_2^C\theta_2^D  + \theta_2^4)\vl =  
 \ket{L}\otimes\ket{R} -\ket{I}\otimes\ket{I} + \ket{R}\otimes\ket{L}
\eeq
Where the arrow indicates that we keep only the NS states, even in all fermion variables. We also made use of the formula
\beq
 \epsilon_{ABCD} \theta_1^A \theta_1^B \theta_2^C\theta_2^D \vl = -4 \rho^I_{AB} \rho^I_{CD} \theta_1^A\theta_1^B\theta_2^C\theta_2^D \vl 
 = - 4\ket{I}\otimes\ket{I}
\label{f2}
\eeq 
Comparing with \eqn{idzm} implies that we should flip $\ket{R}\rightarrow\bra{R}$, $\ket{L}\rightarrow\bra{L}$, $\ket{I}\rightarrow-\bra{I}$.
The final upshot is that we should simply interpret the states as polarizations but flip the sign for each perpendicular polarization in the
initial state. Finally notice that we can use $\beta$ from eq.\eqn{betadef} and $\bar{b}_0$ from eq.\eqn{abdef} to write
\beq
 \bar{b}_0^4 \beta^4 \vl =\frac{1}{4} \ket{\mathbb{I}} 
\label{1/4}
\eeq
meaning that we should understand $\bar{b}_0^4$ in the vertex representation of the Hamiltonian $H$ as $\frac{1}{4}$ times the identity. 
It can be seen that this matters only for the overall normalization and does no affect the relative coefficients of the terms in $H$.

\subsection{Massless scattering}

We need to expand the zero mode Hamiltonian in powers of $\theta_{r=1\cdots4}^A$. As mentioned in the previous subsection, for convenience we define 
\beq
 \theta_3 = \tilde{\theta}_1, \ \ \ \theta_4 = \tilde{\theta}_2
\eeq
which then gives
\beq
\beta=\half\left(-\thon-\thfo+\thth+\thtw\right), \ \ \ \bar{a}_0=\frac{1}{\sqrt{2}}\left(\thon+\thtw-\thth-\thfo\right),
\ \ \ \bar{b}_0=\frac{1}{\sqrt{2}}\left(\thon+\thth-\thtw-\thfo\right)
\eeq
We should replace in the Hamiltonian and expand in powers of the $\theta_r$. Since we are interested in the scattering of NS-NS states,
only terms which contain an even number of each $\theta_r$ are kept. It is not difficult to obtain that
\beqa
\beta^4 &=& \frac{1}{2^4} \left(-\thon-\thfo+\thth+\thtw\right)^4 \non\\
        &\rightarrow& \frac{1}{2^4}\left[\thon^4+\thtw^4+\thth^4+\thfo^4\right]
               +\frac{1}{2^6} \left[ \varepsilon\thon^2\thtw^2+\varepsilon\thon^2\thth^2+\varepsilon\thon^2\thfo^2+
                                     \varepsilon\thtw^2\thth^2+\varepsilon\thtw^2\thfo^2+\varepsilon\thth^2\thfo^2\right] \non\\
\bar{a}_0\sq \bar{a}_0\,\beta^4 &\rightarrow& \frac{1}{2^4}\left[ \aab{\uno}{\dos} + \aab{\uno}{\cuatro}+\aab{\dos}{\uno}+\aab{\dos}{\tres}\right.\non \\
                 &&           \left.\         +\aab{\tres}{\dos} + \aab{\tres}{\cuatro} + \aab{\cuatro}{\uno} + \aab{\cuatro}{\tres} \right] \non\\
         && +\frac{1}{2^6}\left[ \aabc{\dos}{\uno}{\tres} + \aabc{\uno}{\dos}{\cuatro} +\aabc{\cuatro}{\uno}{\tres} +\aabc{\tres}{\dos}{\cuatro}\right] \non\\ 
\bar{b}_0\sq \bar{b}_0\, \beta^4 &\rightarrow& \frac{1}{2^4}\left[ \aab{\uno}{\tres}+\aab{\uno}{\cuatro}+\aab{\tres}{\uno}+\aab{\tres}{\dos}\right.\non\\
                   && \left.\                  +\aab{\dos}{\tres} + \aab{\dos}{\cuatro} + \aab{\cuatro}{\uno} + \aab{\cuatro}{\dos} \right] \non\\
         && +\frac{1}{2^6}\left[ \aabc{\tres}{\uno}{\dos} + \aabc{\uno}{\tres}{\cuatro} +\aabc{\cuatro}{\uno}{\dos} +\aabc{\dos}{\tres}{\cuatro}\right] \non\\ 
\bar{a}_0^4\beta^4 &=& \frac{1}{16}(\thtw-\thfo)^4(\thon-\thth)^4 \non\\
                  &\rightarrow& \frac{1}{2^4}\left[\thon^4\thtw^4+\thon^4\thfo^4+\thtw^4\thth^4+\thth^4\thfo^4\right] 
                                + \frac{1}{2^8} \epsilon\thon^2\thth^2\,\epsilon\thtw^2\thfo^4    \non\\        
                  &&            + \frac{1}{2^6}\left[\aad{\uno}{\dos}{\cuatro}+\aad{\dos}{\uno}{\tres}
                                                 +\aad{\tres}{\dos}{\cuatro}+\aad{\cuatro}{\uno}{\tres}\right] \non\\
\bar{b}_0^4\beta^4 &=& \frac{1}{2^4}(\thon-\thtw)^4(\thth-\thfo)^4 \non\\
                  &\rightarrow& \frac{1}{2^4}\left[\thon^4\thth^4+\thon^4\thfo^4+\thtw^4\thth^4+\thtw^4\thfo^4\right] 
                                 + \frac{1}{2^8} \epsilon\thon^2\thtw^2\,\epsilon\thth^2\thfo^4 \non\\ 
                  &&             + \frac{1}{2^6}\left[\aad{\uno}{\tres}{\cuatro}+\aad{\tres}{\uno}{\dos}
                                                 +\aad{\dos}{\tres}{\cuatro}+\aad{\cuatro}{\uno}{\dos}\right] \non\\          
\bar{a}_0\sq\bar{a}_0\,\bar{b}_0\sq\bar{b}_0\,\beta^4 &\rightarrow& -\half q^2\left[\thon^4\thfo^4+\thtw^4\thth^4\right] 
                   +\frac{1}{2^4}\left[ \thon^4\left(\aaf{\dos}{\tres}+\aaf{\tres}{\cuatro}+\aaf{\dos}{\cuatro}\right) \right. \non\\
                  &&  +                    \thtw^4\left(\aaf{\uno}{\cuatro}+\aaf{\tres}{\cuatro}+\aaf{\uno}{\tres}\right) \non\\
                  &&                        +\thth^4\left(\aaf{\dos}{\cuatro}+\aaf{\uno}{\dos}+\aaf{\uno}{\cuatro}\right)  \non\\
                  &&  \left.                +\thfo^4\left(\aaf{\uno}{\tres}+\aaf{\uno}{\dos}+\aaf{\dos}{\tres}\right) \right] \non\\
                  && +\frac{1}{2^6}\left[ q^2\aag{\uno}{\dos}{\tres}{\cuatro}+ q^2\aag{\uno}{\tres}{\dos}{\cuatro} 
                     - q^2\aag{\uno}{\cuatro}{\dos}{\tres}\right. \non\\
                  &&            + \aah{\uno}{\dos}{\tres}{\cuatro}+\aah{\uno}{\tres}{\dos}{\cuatro}+\aah{\tres}{\cuatro}{\uno}{\dos} \non\\
                  && +\left. \aah{\dos}{\cuatro}{\uno}{\tres}-\aah{\dos}{\tres}{\uno}{\cuatro}-\aah{\uno}{\cuatro}{\dos}{\tres}\right]\non\\
\bar{a}_0\sq\bar{a}_0\bar{b}_0^4\beta^4 &\rightarrow& \frac{1}{2^4}\left[\aaj{\uno}{\tres}{\dos}{\cuatro}+\aaj{\dos}{\cuatro}{\uno}{\tres}\right.\non\\
                  && \left.                                            +\aaj{\uno}{\cuatro}{\dos}{\tres}+\aaj{\dos}{\tres}{\uno}{\cuatro} \right]\non\\
                  && +\frac{1}{2^6}\left[\aak{\uno}{\dos}{\tres}{\cuatro}+\aak{\dos}{\uno}{\tres}{\cuatro}+
                                        \aak{\tres}{\cuatro}{\uno}{\dos}+\aak{\cuatro}{\tres}{\uno}{\dos} \right] \non\\
\bar{b}_0\sq\bar{b}_0\bar{a}_0^4\beta^4 &\rightarrow& \frac{1}{2^4}\left[\aaj{\uno}{\dos}{\tres}{\cuatro}+\aaj{\tres}{\cuatro}{\uno}{\dos} \right.\non\\
                  &&\left.                       +\aaj{\uno}{\cuatro}{\tres}{\dos}+\aaj{\tres}{\dos}{\uno}{\cuatro} \right]\non\\
                  && +\frac{1}{2^6}\left[\aak{\uno}{\tres}{\dos}{\cuatro}+\aak{\tres}{\uno}{\dos}{\cuatro}+
                                        \aak{\dos}{\cuatro}{\uno}{\tres}+\aak{\cuatro}{\dos}{\uno}{\tres} \right] \non\\
\bar{a}_0^4\bar{b}_0^4 \beta^4 &\rightarrow& \frac{1}{2^4}\left[\aal{\uno}{\dos}{\tres}+\aal{\uno}{\dos}{\cuatro}
                                                               +\aal{\uno}{\tres}{\cuatro}+\aal{\dos}{\tres}{\cuatro}\right]  \non\\
                &&    +\frac{1}{2^6} \left[\aam{\uno}{\dos}{\tres}{\cuatro}+\aam{\uno}{\tres}{\dos}{\cuatro}+\aam{\dos}{\cuatro}{\uno}{\tres}\right.\non\\
                &&    \left.               +\aam{\tres}{\cuatro}{\uno}{\dos}+\aam{\uno}{\cuatro}{\dos}{\tres}+\aam{\dos}{\tres}{\uno}{\cuatro} \right]
\eeqa 
where the arrow indicates that we only keep terms even in all fermionic variables $\theta_{r=1\ldots4}$. We also used the following notation
\beq
 \theta_r^4 = \frac{1}{24}\varepsilon_{ABCD} \theta_r^A\theta_r^B\theta_r^C\theta_r^D, \ \ \ 
\varepsilon\theta_r^2\theta_s^2 = \varepsilon_{ABCD} \theta_r^A\theta_r^A\theta_s^C\theta_s^D, \ \ \ 
\theta_r\sq\theta_r = q^I\rho^I_{AB}\theta_r^A\theta_r^B
\eeq
Having expanded in the zero modes, we can proceed to apply the operators to the vacuum state. Using the formulas in eqns.\eqn{states} and \eqn{f2} 
it easily follows that
\beqa
 \beta^4 \vl &\rightarrow& \frac{1}{2^4}\left(\kets{L}{R}{R}{R}+\kets{R}{L}{R}{R}+\kets{R}{R}{L}{R}+\kets{R}{R}{R}{L}\right) \non\\ 
        && -\frac{1}{2^4}\left(\kets{K}{K}{R}{R}+\kets{K}{R}{K}{R}+\kets{K}{R}{R}{K}+\kets{R}{K}{K}{R}+\kets{R}{K}{R}{K}+\kets{R}{R}{K}{K}\right) \non\\
 \bar{a}_0\sq\bar{a}_0 \beta^4 \vl &\rightarrow& \frac{\sqrt{2}}{2^2} \left(\kets{L}{q}{R}{R}+\kets{L}{R}{R}{q}+\kets{q}{L}{R}{R}+\kets{R}{L}{q}{R}
                                      +\kets{R}{q}{L}{R}+\kets{R}{R}{L}{q}+\kets{q}{R}{R}{L}+\kets{R}{R}{q}{L}\right) \non\\
                 && -\frac{\sqrt{2}}{2^2} \left(\kets{K}{q}{K}{R}+\kets{q}{K}{R}{K}+\kets{K}{R}{K}{q}+\kets{R}{K}{q}{K}\right) \non\\
 \bar{b}_0\sq\bar{b}_0 \beta^4 \vl &\rightarrow& \frac{\sqrt{2}}{2^2} \left(\kets{L}{R}{q}{R}+\kets{L}{R}{R}{q}+\kets{q}{R}{L}{R}+\kets{R}{q}{L}{R}
                                      +\kets{R}{L}{q}{R}+\kets{R}{L}{R}{q}+\kets{q}{R}{R}{L}+\kets{R}{q}{R}{L}\right) \non\\
             && -\frac{\sqrt{2}}{2^2} \left(\kets{K}{K}{q}{R}+\kets{q}{R}{K}{K}+\kets{K}{K}{R}{q}+\kets{R}{q}{K}{K}\right) \non\\
 \bar{a}_0^4 \beta^4 \vl &\rightarrow& \frac{1}{2^2}\left(\kets{L}{L}{R}{R}+\kets{L}{R}{R}{L}+\kets{R}{L}{L}{R}+\kets{R}{R}{L}{L}\right) \non\\
             && -\frac{1}{2^2} \left(\kets{L}{K}{R}{K}+\kets{K}{L}{K}{R}+\kets{R}{K}{L}{K}+\kets{K}{R}{K}{L}\right) +\frac{1}{2^2}\kets{K}{I}{K}{I} \non\\
 \bar{b}_0^4 \beta^4 \vl &\rightarrow& \frac{1}{2^2}\left(\kets{L}{R}{L}{R}+\kets{L}{R}{R}{L}+\kets{R}{L}{L}{R}+\kets{R}{L}{R}{L}\right) \non\\
             && -\frac{1}{2^2} \left(\kets{L}{R}{K}{K}+\kets{K}{K}{L}{R}+\kets{R}{L}{K}{K}+\kets{K}{K}{R}{L}\right) +\frac{1}{2^2}\kets{K}{K}{I}{I} \non\\
 \bar{a}_0\sq\bar{a}_0\,\bar{b}_0\sq\bar{b}_0\,\beta^4 &\rightarrow& -2 q^2 \left(\kets{L}{R}{R}{L}+\kets{R}{L}{L}{R}\right) + 
                q^2 \left(\kets{I}{I}{K}{K}+\kets{I}{K}{I}{K}-\kets{I}{K}{K}{I}\right) \non\\
             && + 2 \left(\kets{L}{q}{q}{R}+\kets{L}{q}{R}{q}+\kets{L}{R}{q}{q}+\kets{q}{L}{q}{R}+\kets{q}{L}{R}{q}+\kets{R}{L}{q}{q}\right.  \non\\
             &&         \left.+\kets{q}{q}{L}{R}+\kets{q}{R}{L}{q}+\kets{R}{q}{L}{q}+\kets{q}{q}{R}{L}+\kets{q}{R}{q}{L}+\kets{R}{q}{q}{L}\right) \non\\
             && - 2 \left(\kets{q}{q}{K}{K}+\kets{q}{K}{q}{K}+\kets{K}{K}{q}{q}+\kets{K}{q}{K}{q}-\kets{K}{q}{q}{K}-\kets{q}{K}{K}{q}\right) \non\\
\bar{b}_0\sq\bar{b}_0\,\bar{a}_0^4\beta^4\vl &\rightarrow& \sqrt{2} 
           \left(\kets{L}{L}{R}{q}+\kets{L}{L}{q}{R}+\kets{q}{R}{L}{L}+\kets{R}{q}{L}{L}
                +\kets{L}{q}{R}{L}+\kets{L}{R}{q}{L}+\kets{q}{L}{L}{R}+\kets{R}{L}{L}{q}\right) \non\\
             && - \sqrt{2} \left(\kets{L}{K}{q}{K}+\kets{q}{K}{L}{K}+\kets{K}{L}{K}{q}+\kets{K}{q}{K}{L}\right) \non\\
\bar{a}_0\sq\bar{a}_0\,\bar{b}_0^4\beta^4\vl &\rightarrow& \sqrt{2} 
                      \left(\kets{L}{R}{L}{q}+\kets{L}{q}{L}{R}+\kets{q}{L}{R}{L}+\kets{R}{L}{q}{L}
                           +\kets{L}{R}{q}{L}+\kets{L}{q}{R}{L}+\kets{q}{L}{L}{R}+\kets{R}{L}{L}{q}\right) \non\\
              &&  - \sqrt{2} \left(\kets{L}{q}{K}{K}+\kets{q}{L}{K}{K}+\kets{K}{K}{L}{q}+\kets{K}{K}{q}{L}\right) \non\\
\bar{a}_0^4\bar{b}_0^4\beta^4\vl &\rightarrow& \left(\kets{L}{L}{L}{R}+\kets{L}{L}{R}{L}+\kets{L}{R}{L}{L}+\kets{R}{L}{L}{L}\right) \non\\
              && -  \left(\kets{L}{L}{K}{K}+\kets{L}{K}{L}{K}+\kets{K}{L}{K}{L}+\kets{K}{K}{L}{L}+\kets{L}{K}{K}{L}+\kets{K}{L}{L}{K}\right)
\eeqa
Now we can write the same information in terms of polarizations using that $\epsilon^{(1)}=\xi_1\otimes\xi_4$ and $\epsilon^{(2)}=\xi_2\otimes\xi_3$
and what we found before, for each perpendicular polarization in the initial state we should flip the sign:
\beq
\ket{abcd} \rightarrow (-)^{n_a+n_d} \polab{a}{b}{d}{c}
\eeq
where $n_L=n_R=0$ and $n_I=1$. We obtain the contribution of each term in the Hamiltonian to the polarization part of 
the NSNS scattering amplitude:
\beq
 H \ket{0} = -\frac{1}{2^8\pi^2} A(s,t) K, \ \ \ \ K=K_{[0]}+K_{[2]}+K_{[4]}+K_{[6]}+K_{[8]}
\eeq
 where, for convenience, we extracted a common factor, lumping all the polarization dependence in $K$ which has the following contributions from each term
in the Hamiltonian:
\beqa 
 K_{[0]} &=& 2s k^Lk^L\left[\left(\polab{L}{R}{R}{R}+\polab{R}{L}{R}{R}+\polab{R}{R}{L}{R}+\polab{R}{R}{R}{L}\right)\right. \non\\ 
         &&   \left.+\left(\polab{K}{K}{R}{R}-\polab{K}{R}{K}{R}+\polab{K}{R}{R}{K}
                                     +\polab{R}{K}{K}{R}-\polab{R}{K}{R}{K}+\polab{R}{R}{K}{K}\right)\right] \non\\
 K_{[2]} &=&  k^L \left\{ t \left[ \left(-\polab{L}{R}{q}{R}+\polab{L}{R}{R}{q}-\polab{q}{R}{L}{R}+\polab{R}{q}{L}{R}
                                      -\polab{R}{L}{q}{R}+\polab{R}{L}{R}{q}-\polab{q}{R}{R}{L}+\polab{R}{q}{R}{L}\right)\right.\right. \non\\
             && \left. +  \left(-\polab{K}{K}{q}{R}-\polab{q}{R}{K}{K}+\polab{K}{K}{R}{q}+\polab{R}{q}{K}{K}\right)\right] \non\\
             &&   - s  \left[ \left(\polab{L}{q}{R}{R}+\polab{L}{R}{R}{q}-\polab{q}{L}{R}{R}-\polab{R}{L}{q}{R}
                                      +\polab{R}{q}{L}{R}+\polab{R}{R}{L}{q}-\polab{q}{R}{R}{L}-\polab{R}{R}{q}{L}\right)\right. \non\\
                 &&\left.\left. +  \left(-\polab{K}{q}{K}{R}+\polab{q}{K}{R}{K}-\polab{K}{R}{K}{q}+\polab{R}{K}{q}{K}\right)\right]\right\} \non\\
 K_{[4]} &=&  \left\{ \frac{s^2}{4} \left[\left(\polab{L}{L}{R}{R}+\polab{L}{R}{R}{L}+\polab{R}{L}{L}{R}+\polab{R}{R}{L}{L}\right)\right.\right. \non\\
             && \left.- \left(\polab{L}{K}{R}{K}+\polab{K}{L}{K}{R}+\polab{R}{K}{L}{K}+\polab{K}{R}{K}{L}\right) +\polab{K}{I}{K}{I} \right]\non\\
         && +\frac{t^2}{4}\left[\left(\polab{L}{R}{L}{R}+\polab{L}{R}{R}{L}+\polab{R}{L}{L}{R}+\polab{R}{L}{R}{L}\right) \right.\non\\
             && \left. + \left(\polab{L}{R}{K}{K}+\polab{K}{K}{L}{R}+\polab{R}{L}{K}{K}+\polab{K}{K}{R}{L}\right)+\polab{K}{K}{I}{I} \right]\non\\
         && +\frac{st}{4}\left[ -2  \left(\polab{L}{R}{R}{L}+\polab{R}{L}{L}{R}\right) + 
                 \left(\polab{I}{I}{K}{K}+\polab{I}{K}{I}{K}-\polab{I}{K}{K}{I}\right)\right] \non\\
         && +\frac{t}{2} \left[ \left(\polab{L}{q}{q}{R}-\polab{L}{q}{R}{q}+\polab{L}{R}{q}{q}
                                -\polab{q}{L}{q}{R}+\polab{q}{L}{R}{q}+\polab{R}{L}{q}{q}\right.\right.  \non\\
         &&      \left.+\polab{q}{q}{L}{R}+\polab{q}{R}{L}{q}-\polab{R}{q}{L}{q}+\polab{q}{q}{R}{L}-\polab{q}{R}{q}{L}+\polab{R}{q}{q}{L}\right) \non\\
         && \left.\left.+\left(\polab{q}{q}{K}{K}+\polab{q}{K}{q}{K}+\polab{K}{K}{q}{q}
                              +\polab{K}{q}{K}{q}-\polab{K}{q}{q}{K}-\polab{q}{K}{K}{q}\right)\right]\right\} \non\\
 K_ {[6]} &=&  k^R \left\{ t\left[
                      \left(\polab{L}{R}{L}{q}+\polab{L}{q}{L}{R}-\polab{q}{L}{R}{L}-\polab{R}{L}{q}{L}
                           -\polab{L}{R}{q}{L}+\polab{L}{q}{R}{L}-\polab{q}{L}{L}{R}+\polab{R}{L}{L}{q}\right) \right.\right.\non\\
              && \left. + \left(\polab{L}{q}{K}{K}-\polab{q}{L}{K}{K}+\polab{K}{K}{L}{q}-\polab{K}{K}{q}{L}\right) \right]\non\\
              && - s\left[
           \left(\polab{L}{L}{R}{q}-\polab{L}{L}{q}{R}-\polab{q}{R}{L}{L}+\polab{R}{q}{L}{L}
                +\polab{L}{q}{R}{L}-\polab{L}{R}{q}{L}-\polab{q}{L}{L}{R}+\polab{R}{L}{L}{q}\right) \right.\non\\
             && \left.\left.+  \left(\polab{L}{K}{q}{K}+\polab{q}{K}{L}{K}-\polab{K}{L}{K}{q}-\polab{K}{q}{K}{L}\right) \right]\right\}\non\\
 K_{[8]} &=&  2 s k^Rk^R \left[\left(\polab{L}{L}{L}{R}+\polab{L}{L}{R}{L}+\polab{L}{R}{L}{L}+\polab{R}{L}{L}{L}\right) \right.\non\\
              && \left.+  \left(\polab{L}{L}{K}{K}-\polab{L}{K}{L}{K}-\polab{K}{L}{K}{L}+\polab{K}{K}{L}{L}+\polab{L}{K}{K}{L}+\polab{K}{L}{L}{K}\right)\right]
\label{Kh}
\eeqa
where $K_{[r]}$ is the polarization factor corresponding to $H_{[r]}$ ($r=0,2,4,6,8$).
The scattering amplitude is then simply
\beq
 \cA = -\frac{2\pi\alpha_3}{2^8\pi^2} A(s,t) K = -\frac{\alpha_3}{2^7\pi} A(s,t) K,
\eeq
where the $2\pi$ comes from the $\sigma$ integral in eq.\eqn{Pdef} (since the zero modes do not depend on $\sigma$) and $\alpha_3$ is the
overall constant introduced in the same equation (and which we are going to determine later in the appendix). 

\subsection{Comparison with known results}

 The scattering of massless strings from D-branes has been studied in detail \cite{Klebanov:1995ni,MG}. Here we follow the work of Myers and Garousi 
who give, for NS-NS states scattering from a $p$-brane the amplitude
\beq
 \cA = -i\frac{\kappa T_p}{2} A(s,t) K 
\eeq
 where $A(s,t)$ is the same function defined in eq.\eqn{Adef}, $\kappa$ the closed string coupling constant and $T_p$ is the D-brane tension. 
The kinematic factor $K$ is given by
\beqa
 K &=& \frac{t^2}{4} \tr(\eone^T \etwo) + \frac{s^2}{4} \tr(\eone D)\tr(\etwo D) 
     + \frac{st}{4}\left[\tr(\eone^T\etwo) + \tr(\eone D)\tr(\etwo D) -\tr(D\eone D\etwo) \right] \non \\
   & & +\frac{s}{2} \left[ \tr(\eone D)\left(p_1\etwo Dp_2+p_2D\etwo p_1+p_2D\etwo Dp_2\right) + p_1D\eone D\etwo Dp_2-p_2D\etwo\eone^TDp_1 \right.\non\\
   & &\ \ \, +\left. \tr(\etwo D)\left(p_1 D \eone  p_2+p_2\eone D p_1+p_1D\eone Dp_1\right) + p_2D\etwo D\eone Dp_1-p_1D\eone^T\etwo Dp_2 \right]\non \\
   & & -\frac{t}{2} \left[\tr(\eone D) p_1\etwo p_1-p_1\etwo D\eone p_2-p_1\etwo \eone^TDp_1-p_1\etwo^T\eone Dp_1-p_1\etwo \eone^Tp_2 \right.\non\\
   & &\ \ \, +\left.\tr(\etwo D) p_2\eone p_2-p_2\eone D\etwo p_1-p_2\eone \etwo^TDp_2-p_2\eone^T\etwo Dp_2-p_2\eone^T \etwo p_1\right]
\label{K1}
\eeqa
where $p_{1,2}$ are the momenta of the initial and final particles, $D_{\mu\nu}$ is a diagonal matrix with diagonal element equal to 1 or $-1$ according
if $\mu$ is parallel or perpendicular to the $D$-brane. Finally $(\eone)_{\mu\nu}$, $(\etwo)_{\mu\nu}$ are the polarizations of the initial and final 
particles. The result \eqn{K1} can be obtained\footnote{This is not completely true. We replaced two terms 
$p_1\etwo\eone^Tp_2\rightarrow p_2\eone^T\etwo p_1$ and $p_2D\etwo\eone^TDp_\rightarrow p_1D\eone^T\etwo Dp_2$ since \cite{MG} seems to contain a typo.}
 from equations (11), (12) and (13) of \cite{MG} after replacing, according to our conventions 
$q^2\rightarrow -\frac{t}{4}$ and $t\rightarrow s$. Another, alternative way to compute the kinematic factor $K$ is, as also shown in \cite{MG}, 
to start from the four open string kinematic factor \cite{GSW}
\beqa
 K &=& -(p_2p_3)(p_2p_4)(\zeta_1\zeta_2)(\zeta_3\zeta_4) -(p_2p_3)(p_3p_4)(\zeta_1\zeta_3)(\zeta_2\zeta_4)
       -(p_3p_4)(p_2p_4)(\zeta_1\zeta_4)(\zeta_2\zeta_3)  \label{K2}\\
   & & -(p_1p_2) \left[ (p_4\zeta_1)(p_2\zeta_3)(\zeta_2\zeta_4)+(p_3\zeta_2)(p_1\zeta_4)(\zeta_1\zeta_3) +
                        (p_3\zeta_1)(p_2\zeta_4)(\zeta_2\zeta_3)+(p_4\zeta_2)(p_1\zeta_3)(\zeta_1\zeta_4)\right] \non\\
   & & -(p_1p_3) \left[ (p_4\zeta_1)(p_3\zeta_2)(\zeta_3\zeta_4)+(p_2\zeta_3)(p_1\zeta_4)(\zeta_1\zeta_2) +
                        (p_2\zeta_1)(p_3\zeta_4)(\zeta_2\zeta_3)+(p_4\zeta_3)(p_1\zeta_2)(\zeta_1\zeta_4)\right] \non\\
   & & -(p_1p_4) \left[ (p_2\zeta_1)(p_4\zeta_3)(\zeta_2\zeta_4)+(p_3\zeta_4)(p_1\zeta_2)(\zeta_1\zeta_3) +
                        (p_3\zeta_1)(p_4\zeta_2)(\zeta_3\zeta_4)+(p_2\zeta_4)(p_1\zeta_3)(\zeta_1\zeta_2)\right] \non
\eeqa
 and replace\footnote{Here we have some factors of two discrepancies with \cite{MG} which we attribute to some typos in \cite{MG}.} 
$p_4\rightarrow Dp_1$, $p_3\rightarrow Dp_2$, $\zeta_1 \otimes \zeta_4 \rightarrow \eone D$ and $\zeta_2 \otimes \zeta_3 \rightarrow \etwo D$.
To convert to the $SU(4)\times U(1)$ notation we should now replace in \eqn{K1} or alternatively \eqn{K2}
\beq
 p_1 = (k,q_1), \ \ p_2 = (-k,q_2), \ \ p_3=(-k,-q_2), \ \ p_4=(k,-q_1)
\eeq
and the scalar products
\beq
(q_1\zeta_1)=-(k\zeta_1), \ \ (q_1\zeta_4)=(k\zeta_4), \ \ (q_2\zeta_2)=(k\zeta_2), \ \ (q_2\zeta_3)=-(k\zeta_3)
\eeq
which follows from $(p_i\zeta_i)=0$, $p_i^2=0$ and $q=q_1+q_2=p_1+p_2$.  The result is then expanded using that
\beq
(vw) = v_R w_L+v_Lw_R+v_I w_I
\eeq
which gives
\beqa
K &=&   \frac{t^2}{4}  \left[\pola{I}{I}{K}{K}+\pola{L}{R}{K}{K}+\pola{R}{L}{K}{K}+\pola{K}{K}{L}{R}+\pola{K}{K}{R}{L}
                           +\pola{L}{R}{L}{R}+\pola{L}{R}{R}{L}+\pola{R}{L}{L}{R}+\pola{R}{L}{R}{L}\right] \non\\
  &&   +\frac{s^2}{4}  \left[\pola{I}{K}{K}{I}-\pola{L}{K}{K}{R}-\pola{R}{K}{K}{L}-\pola{K}{R}{L}{K}
                           -\pola{K}{L}{R}{K}+\pola{L}{R}{L}{R}+\pola{L}{L}{R}{R}+\pola{R}{R}{L}{L}+\pola{R}{L}{R}{L}\right]\non \\
  &&   +\frac{st}{4}   \left[\pola{I}{I}{K}{K}+\pola{I}{K}{K}{I}-\pola{I}{K}{I}{K}-2\pola{R}{L}{R}{L}-2\pola{L}{R}{L}{R}  \right] \non \\
  &&   +\frac{t}{2}    \left[\pola{q}{q}{K}{K}-\pola{q}{K}{q}{K}-\pola{K}{q}{K}{q}+\pola{q}{K}{K}{q}+\pola{K}{q}{q}{K}+\pola{K}{K}{q}{q} 
                            +\pola{q}{q}{R}{L}+\pola{q}{R}{q}{L}+\pola{R}{q}{L}{q}\right. \non\\
  &&\phantom{\frac{t}{2}}\left.-\pola{q}{R}{L}{q}-\pola{R}{q}{q}{L}+\pola{R}{L}{q}{q}+\pola{q}{q}{L}{R}+\pola{q}{L}{q}{R}
                           +\pola{L}{q}{R}{q}-\pola{q}{L}{R}{q}-\pola{L}{q}{q}{R}+\pola{L}{R}{q}{q}  \right] \non\\
  &&   +2sk_Lk_L       \left[\pola{R}{R}{K}{K}+\pola{R}{K}{R}{K}+\pola{K}{R}{K}{R}-\pola{R}{K}{K}{R}-\pola{K}{R}{R}{K}+\pola{K}{K}{R}{R} \right. \non\\
  &&\phantom{+2sk_Lk_L}\left.+\pola{L}{R}{R}{R}+\pola{R}{L}{R}{R}+\pola{R}{R}{L}{R}+\pola{R}{R}{R}{L}\right] \label{Ko}\\
  &&   +2sk_Rk_R       \left[\pola{L}{L}{K}{K}+\pola{L}{K}{L}{K}+\pola{K}{L}{K}{L}-\pola{L}{K}{K}{L}-\pola{K}{L}{L}{K}+\pola{K}{K}{L}{L} \right. \non\\
  &&\phantom{+2sk_Rk_R}\left.+\pola{R}{L}{L}{L}+\pola{L}{R}{L}{L}+\pola{L}{L}{R}{L}+\pola{L}{L}{L}{R}\right] \non\\
  &&   +sk_L           \left[-\pola{q}{K}{K}{R}-\pola{R}{K}{K}{q}+\pola{K}{q}{R}{K}+\pola{K}{R}{q}{K} \right.\non\\
  &&\phantom{+sk_L}    \left.+\pola{q}{R}{L}{R}+\pola{q}{L}{R}{R}+\pola{R}{R}{L}{q}+\pola{R}{L}{R}{q}
                           -\pola{R}{q}{R}{L}-\pola{L}{q}{R}{R}-\pola{L}{R}{q}{R}-\pola{R}{R}{q}{L}\right] \non\\
  &&   +sk_R           \left[-\pola{q}{K}{K}{L}-\pola{L}{K}{K}{q}+\pola{K}{q}{L}{K}+\pola{K}{L}{q}{K} \right.\non\\
  &&\phantom{+sk_R}    \left.+\pola{q}{L}{R}{L}+\pola{q}{R}{L}{L}+\pola{L}{L}{R}{q}+\pola{L}{R}{L}{q}
                           -\pola{L}{q}{L}{R}-\pola{R}{q}{L}{L}-\pola{R}{L}{q}{L}-\pola{L}{L}{q}{R}\right] \non\\
  &&   -tk_L           \left[\pola{q}{R}{K}{K}-\pola{R}{q}{K}{K}+\pola{K}{K}{R}{q}-\pola{K}{K}{q}{R} \right.\non\\
  &&\phantom{-tk_L}    \left.+\pola{q}{R}{L}{R}+\pola{q}{R}{R}{L}-\pola{L}{R}{q}{R}-\pola{R}{L}{q}{R}
                           -\pola{R}{q}{L}{R}-\pola{R}{q}{R}{L}+\pola{R}{L}{R}{q}+\pola{L}{R}{R}{q}\right] \non\\
  &&   -tk_R           \left[\pola{q}{L}{K}{K}-\pola{L}{q}{K}{K}+\pola{K}{K}{L}{q}-\pola{K}{K}{q}{L} \right.\non\\
  &&\phantom{-tk_R}    \left.+\pola{q}{L}{R}{L}+\pola{q}{L}{L}{R}-\pola{R}{L}{q}{L}-\pola{L}{R}{q}{L}
                           -\pola{L}{q}{R}{L}-\pola{L}{q}{L}{R}+\pola{L}{R}{L}{q}+\pola{R}{L}{L}{q}\right] \non
\eeqa
where, by a slight abuse of notation, we used the subindex $q$ to denote contraction with $q^I$, for example, 
$ \epsilon^{(1)}_{qR} = q^I \epsilon^{(1)}_{IR}$, etc. 
Comparison with eq.\eqn{Kh} shows a perfect agreement of the polarization factor $K$. The overall coefficient would agree if we choose
\beq
 \frac{\alpha_3}{2^7\pi} = \half \kappa T_3
\eeq
 If we use from \cite{MG} and \cite{Polchinski} that
\beq
 T_3 = \sqrt{\pi}, \ \ \ \kappa^2=\half(2\pi)^7 g^2 \alpha'{}^4
\eeq
we get
\beq 
\kappa T_3 = 2^5 \pi^4 g_s
\eeq
where we set $\alpha'=2$ as used here and in \cite{MG}. This fixes the normalization coefficient 
\beq
 \alpha_3 = 2^{11} \pi^5
\label{a31}
\eeq
  which is the correct value for $\alpha_3$ as we will discuss in more detail in the next subsection. Notice also that, in our case, 
the string coupling constant $g_s$ appears in $\lambda=g_sN$ which multiplies $P$.

\subsection{Overall normalization}

 The overall normalization of the amplitude determines the coefficient $\alpha_3$ in eq.\eqn{Pdef}. The calculation is tantamount to
determine the tension of the D-brane and requires some extra material to which we devote this subsection. The final result is that 
$\alpha_3=2^{11}\pi^5$ in units where $\alpha'=2$. This is the same as we obtained in the previous section using the results of \cite{MG} and
\cite{Polchinski}. Here we summarize how it is obtained, namely by normalizing the amplitude to agree, at large distances, with the classical 
scattering from the corresponding D3-brane supergravity background. Notice that this is the only point where we make any reference to 
the existence of such background. All the rest we derived from 
summing the planar diagrams of the open string theory. In principle it should be possible to determine the correct normalization, namely 
the tension of the D-brane, from an open string argument alone. It appears a difficult task however and we do not attempt to do so here. 

 The supergravity solution we use in paper \pone for the D3-brane has a metric
\beq
 ds^2 = \frac{1}{\sqrt{f}}\left(dX^+dX^- + dX^2\right) + \sqrt{f} dY^2, \ \ \ \ f=1+4\pi\alpha'{}^2\frac{g_sN}{Y^4}
\eeq
 Far from the D-brane, the metric is
\beq
 g_{\mu\nu} = \eta_{\mu\nu} - 2\pi \alpha'{}^2 \frac{g_sN}{Y^4} D_{\mu\nu}
\label{htilde}
\eeq
where $D_{\mu\nu}$ is a diagonal matrix with diagonal elements equal to one for directions parallel to the brane and minus one for directions
perpendicular. Consider now the action of a B-field in this background\footnote{As in \cite{MG} we consider a B-field for simplicity.}:
\beq
S = -\frac{3}{2} \int d^{10} x \sqrt{-g} H^2 e^{-\sqrt{2}\kappa\phi}
\eeq
where 
\beq
 H_{\mu\nu\rho} = \frac{1}{3} \left(\partial_{\mu} B_{\nu\rho} + \partial_{\rho} B_{\mu\nu} + \partial_{\nu} B_{\rho\mu} \right)
\eeq
 Expanding for small $h_{\mu\nu}$ and setting the dilaton $\phi$ to zero we get for the term linear in $h_{\mu\nu}$
\beq
 S = -\frac{3}{2} \int d^{10}x \left[\half \eta^{\mu\nu}h_{\mu\nu} H^2 - 3 h^{\mu\nu} H_{\mu\alpha\beta} H_{\nu}^{\phantom{\nu}\alpha\beta}\right]
\eeq
Now we are going to consider, for simplicity, only transverse B-fields and also, since we are interested in the D-brane, 
assume that $h_{\mu\nu}=h(y^2) D_{\mu\nu}$.
We obtain
\beq
 S = -3 \int d^{10}x h(y^2) H_{IJK} H_{IJK}
\eeq
where $I,J,K=1\ldots 6$ denote transverse indices. This vertex corrects the propagator of the B-field by an amount
\beq
 \cA_{BB}^\perp =  6 \tilde{h}(q) \tilde{H}_{IJK} \tilde{H}_{IJK}
\eeq
where there is a factor of two from the two different ways to contract the fields to the external states and we have introduced
the Fourier transforms
\beq
 \tilde{h}(q) = \int d^6y h(y) e^{iqy}, \ \ \tilde{H}_{IJK} = \frac{i}{3} \left(p_{I} B_{JK}+p_{K}B_{IJ}+p_{J}B_{KI}\right)
\label{hqq}
\eeq
with $B_{IJ}$ the B-field in momentum representation. The matrix element of the perturbation $\cA_{BB}^\perp$ determines the scattering
amplitude of transverse B-fields by the D-brane. Using, from eq.\eqn{htilde}, that $h(y) = - 2\pi \alpha'{}^2 \frac{g_sN}{Y^4}$ we get
\beq
  \cA_{BB}^\perp = - 6 g_s N \frac{8\pi^4\alpha'{}^2}{s} \tilde{H}_{IJK} \tilde{H}_{IJK}, \ \ \ \ \ \ s=-q^2
\eeq
as the supergravity result. In our previous calculation, the kinematic factor $K$, for this situation reduces to
\beq
 K = t \left[\left(\frac{s}{2}+\frac{t}{4}\right) \epsilon^{(1)}_{IK}\epsilon^{(2)}_{IK} + 2\epsilon^{(1)}_{qK} \epsilon^{(2)}_{qK}\right]
\eeq
Using that $H_{IJK}=\frac{i}{3}\left(p_{I} \epsilon_{JK}+p_{K}\epsilon_{IJ}+p_{J}\epsilon_{KI}\right)$ we can write
\beq
K = 3 t H_{IJK} H_{IJK}
\eeq
Far from the brane, in momentum space, means that the momentum transfer $q^2=-s$ is small. In that limit we have
\beq
 A(s,t)= \frac{\Gamma\left(-\frac{s}{2}\right)\Gamma\left(-\frac{t}{2}\right)}{\Gamma\left(1-\frac{s}{2}-\frac{t}{2}\right)} 
        \simeq_{s\rightarrow 0} \frac{4}{st}
\eeq
 which gives, from Garousi and Myers paper
\beq
 \cA^\perp_{BB} \simeq -\frac{6i}{s} \kappa T_3 H_{IJK} H_{IJK}
\eeq
and from our calculation
\beq
 \cA^\perp_{BB} \simeq -\frac{6i}{s} \frac{\alpha_3}{2^6\pi} H_{IJK} H_{IJK}
\eeq
 Therefore, matching all the results, including the supergravity one requires that we set
\beq
 \kappa T_3 = g_s N \frac{\alpha_3}{2^6\pi} = 8\pi^4 \alpha'{}^2 g_s N
\eeq
 which, with $\alpha'=2$,  fixes $\alpha_3=2^{11}\pi^5$ and $\kappa T_3= 2^5\pi^4 g_sN$. This verifies the result \eqn{a31}.
Finally note that inverting eq.\eqn{hqq} gives
\beq
 h(y) =\int \frac{d^6q}{(2\pi)^6} e^{-iqy} \tilde{h}(q)
\eeq
which shows that, in these conventions, the integration measure for $q$ contains $(2\pi)^{-6}$

\end{document}